\documentclass[aps,prd, reprint,twocolumn,nofootinbib,superscriptaddress,showpacs, amsmath,floats,floatfix]{revtex4-1}
\usepackage{graphicx}
\usepackage{amsmath,amssymb}
\usepackage{amsfonts}
\usepackage{xspace} 
\usepackage[usenames]{color}
\usepackage{dcolumn}
\usepackage{bm}
\usepackage{mathrsfs}
\usepackage[colorlinks=true]{hyperref}
\usepackage[all]{hypcap} 
\usepackage[utf8]{inputenc} 

\def\be{\begin{equation}}
\def\ee{\end{equation}}
\def\beq{\begin{eqnarray}}
\def\eeq{\end{eqnarray}}

\newcommand{\tn}{\textnormal}

\begin{document}
%
%

\title{Dark stars: gravitational and electromagnetic observables}

\author{Andrea Maselli}
	\email{andrea.maselli@uni-tuebingen.de}
\affiliation{Theoretical Astrophysics, IAAT, University of T\"uebingen, 
T\"uebingen 72076, Germany}

\author{Pantelis Pnigouras}
\email{pantelis.pnigouras@uni-tuebingen.de}
\affiliation{Theoretical Astrophysics, IAAT, University of T\"uebingen, 
T\"uebingen 72076, Germany}

\author{Niklas Gr\o nlund Nielsen}
	\email{ngnielsen@cp3.sdu.dk}
	\affiliation{CP$^3$-Origins, Centre for Cosmology and Particle Physics Phenomenology University of Southern Denmark, Campusvej 55, 5230 Odense M, Denmark}

\author{Chris Kouvaris}
\email{kouvaris@cp3.sdu.dk}
	\affiliation{CP$^3$-Origins, Centre for Cosmology and Particle Physics Phenomenology University of Southern Denmark, Campusvej 55, 5230 Odense M, Denmark}

\author {Kostas D. Kokkotas}
	\email{kostas.kokkotas@uni-tuebingen.de}

\affiliation{Theoretical Astrophysics, IAAT, University of T\"uebingen, 
T\"uebingen 72076, Germany}

\date{\today}


\begin{abstract}
Theoretical models of self-interacting dark matter represent a promising answer to a series of open problems within the so-called collisionless cold dark matter (CCDM) paradigm. In case of asymmetric dark matter, self-interactions might facilitate gravitational collapse and potentially lead to formation of compact objects predominantly made of dark matter. Considering both fermionic and bosonic equations of state, we construct the equilibrium structure of rotating dark stars, focusing on their bulk properties, and comparing them with baryonic neutron stars. We also show that these dark objects admit the $I$-Love-$Q$ universal relations, which link their moments of inertia, tidal deformabilities, and quadrupole moments. Finally, we prove that stars built with a dark matter equation of state are not compact enough to mimic black holes in general relativity, thus making them 
distinguishable in potential events of gravitational interferometers.
\end{abstract}



\maketitle 

\section{Introduction}
It is quite probable that if dark matter (DM) exists in the form of particles, it might experience non-negligible self-interactions. 
This is highly motivated both theoretically and observationally. From a theoretical point of view, if the dark sector is 
embedded in a unification scheme in a theory beyond the Standard Model, it is hard to imagine DM particles that 
do not interact among themselves via some gauge bosons. In addition, DM self-interactions might be a desirable 
feature due to the fact that the CCDM paradigm seems to be currently at odds with 
observations. There are three main challenges that CCDM faces today. The first one is related to the flatness 
of the DM density profile at the core of dwarf galaxies~\cite{Moore:1994yx,Flores:1994gz}. The latter are dominated 
by DM and, although numerical simulations of CCDM~\cite{Navarro:1996gj} predict a cuspy profile for the DM density 
at the core of these galaxies, measurements of the rotation curves suggest that the density profile is flat. A second 
issue is that numerical simulations of CCDM also predict a larger number of satellite galaxies in the Milky Way than 
what has been observed so far~\cite{Klypin:1999uc,Moore:1999nt,Kauffmann:1993gv}. Finally a third serious problem 
for CCDM is the ``too big to fail'' ~\cite{BoylanKolchin:2011de}, i.e. CCDM numerical simulations predict massive 
dwarf galaxies that are too big to not form visible and observable stars. These discrepancies between numerical 
simulations of CCDM and observations could be alleviated by taking into account DM-baryon interactions~\cite{Oh:2010mc,Brook:2011nz,Pontzen:2011ty,Governato:2012fa}. 
In addition, the satellite discrepancy could be attributed to Milky Way being a statistical fluctuation~\cite{Liu:2010tn,Tollerud:2011wt,Strigari:2011ps}, 
thus deviating from what numerical simulations predict. Apart from these explanations, another possible solution is 
the existence of substantial DM self-interactions, which can solve all three aforementioned problems~\cite{Vogelsberger:2012ku,Rocha:2012jg,Zavala:2012us,Peter:2012jh}. It is not hard for example to see that DM self-interactions would lead to increased rates of self-scattering in high 
DM density regions, thus flattening out dense dwarf galaxy cores.

In this picture, DM self-interactions have been thoroughly studied in the literature in different 
contexts~\cite{Spergel:1999mh,Wandelt:2000ad,Faraggi:2000pv,Mohapatra:2001sx,Kusenko:2001vu,Loeb:2010gj,Kouvaris:2011gb,
Rocha:2012jg,Peter:2012jh,Vogelsberger:2012sa,Zavala:2012us,Tulin:2013teo,Kaplinghat:2013xca,Kaplinghat:2013yxa,Cline:2013pca,
Cline:2013zca,Petraki:2014uza,Buckley:2014hja,Boddy:2014yra,Schutz:2014nka}. Although depending on the type of 
DM self-interactions, the general consensus is that DM interactions falling within the range  $0.1~\text{cm}^2/\text{g}< \sigma_{XX}/m_{X}<10~\text{cm}^2/\text{g}$ ($\sigma_{XX}$ and $m_\tn{X}$ being the  DM self-interaction cross section and DM particle mass respectively) are sufficient to resolve the 
CCDM problems. If DM is made of one species, DM self-interactions cannot be arbitrarily strong because in this 
case they could destroy the ellipticity of spiral galaxies~\cite{Feng:2009mn,Feng:2009hw}, dissociate the bullet 
cluster~\cite{Markevitch:2003at} or destroy old neutron stars (NSs) by accelerating the collapse of captured DM at the 
core of the stars leading to formation of black holes that could eat up the star (thus imposing constraints due to 
observation of old NSs)~\cite{Kouvaris:2014uoa}.

Apart from the associated problems of CCDM, there is another orthogonal scenario where DM self-interactions 
might be needed. The supermassive black hole at the center of the Milky Way seems to be too big to have grown within 
the lifetime of the galaxy from collapsed baryonic stars. One possible solution is to envision a strongly self-interacting 
subdominant component of DM that collapses via a gravothermal process, providing the seeds for the black hole to 
grow to today's mass within the lifetime of the Milky Way~\cite{Pollack:2014rja}.

As a desirable feature of DM, self-interactions may assist DM clumping together and forming compact objects\footnote{DM 
can also clump without self-interactions, e.g. when density perturbations fulfil the Jean's criterion or through 
gravothermal evolution.}, if they are dissipative or they speed up gravothermal evolution~\cite{Balberg:2002ue}. 
There could be two different possibilities here: i) DM with substantial amount of annihilations and ii) DM with 
negligible amount of annihilations. The Weakly Interacting Massive Particle (WIMP) paradigm belongs to the first 
category. In this case, the DM relic density in the Universe is determined by the DM annihilations. DM is in thermal 
equilibrium with the primordial plasma until the rate of DM annihilations becomes smaller than the expansion of the 
Universe. In this WIMP paradigm, particle and antiparticle populations of DM come with equal numbers. Gravitational 
collapse of such type of DM could create dark stars that oppose further gravitational collapse by radiation 
pressure~\cite{Spolyar:2007qv,Freese:2008hb,Freese:2008wh}. However, these types of stars cannot exist 
anymore, as DM annihilations would have already lead to the depletion of the DM population and therefore to 
the extinction of these dark stars long time ago. 

On the contrary, asymmetric DM can lead to the formation of compact star-like objects that can be stable today. The 
asymmetric DM scenario is a well-motivated alternative to the WIMP paradigm~\cite{Nussinov:1985xr,Barr:1990ca,Gudnason:2006yj,Foadi:2008qv,Dietrich:2006cm,Sannino:2009za,Ryttov:2008xe,Sannino:2008nv,Kaplan:2009ag,Frandsen:2009mi,MarchRussell:2011fi,
Frandsen:2011cg,Gao:2011ka,Arina:2011cu,Buckley:2011ye,Lewis:2011zb,Davoudiasl:2011fj,Graesser:2011wi,Bell:2011tn,Cheung:2011if}. 
In this case, there is a conserved quantum number, as, e.g., the baryon number. A mechanism similar 
to the one responsible for the baryon asymmetry in the Universe could also create a particle-antiparticle asymmetry in the 
DM sector. DM annihilations deplete the species with the smaller population, leaving at the end only the species in excess, 
to account for the DM relic density. One can easily see that, e.g., a DM particle of mass $\sim5$ GeV could account for the DM 
abundance, provided that a common asymmetry mechanism that creates simultaneously a baryon and a DM unit is in place. 
Obviously, in such an asymmetric DM scenario, there is no substantial amount of annihilations today due to lack of DM 
antiparticles. Therefore, provided that DM self-interactions can facilitate the collapse, asymmetric DM can form compact 
objects that can be stable, thus possibly detectable today via, e.g., gravitational wave (GW) emission in binary systems.

The possibility of asymmetric DM forming compact-like objects has been studied both in the case of 
fermionic~\cite{Narain:2006kx,Kouvaris:2015rea} as well as bosonic~\cite{Eby:2015hsq} DM. In both of these papers, 
the mass--radius relations, density profiles, and maximum ``Chandrasekhar'' 
mass limits were established for a wide range of DM particle masses and DM self-coupling, for both attractive and repulsive interactions. 
Bosonic DM forming compact objects has also been studied in other than asymmetric DM contexts, i.e., in the case DM is ultra 
light, e.g., axions~\cite{Kolb:1993zz,Kolb:1993hw,Chavanis:2011zi,Chavanis:2011zm,Eby:2014fya,Brito:2015yfh,Eby:2015hyx,Eby:2016cnq,Cotner:2016aaq,Davidson:2016uok,Chavanis:2016dab,Levkov:2016rkk,Hui:2016ltb,Bai:2016wpg,Eby:2017xaw} or other theoretically motivated bosonic candidates~\cite{Soni:2016gzf} 

These proposed dark stars, if in binary systems, can produce GW signals that could potentially distinguish 
them from corresponding signals of black hole binaries, as it was suggested in~\cite{Giudice:2016zpa,Cardoso:2017cfl}. 
Other probes of bosonic DM stars via GWs have been proposed in~\cite{Dev:2016hxv}. We remark that objects 
inconsistent with either black holes or NSs may suggest the existence of new particle physics. 
Therefore, compact binary mergers could become a search strategy for beyond-standard-model physics which is 
completely orthogonal to the LHC and (in-)direct DM searches.
Interesting scenario of compact objects made of asymmetric DM with a substantial baryonic component 
has also been studied~\cite{Leung:2011zz,Leung:2013pra,Tolos:2015qra,Mukhopadhyay:2015xhs}. 

One should mention that there are several  scenarios of how these dark stars can form in the first place. Gravothermal 
collapse is one option~\cite{LyndenBell:1968yw}. In this case, DM self-interactions facilitate the eviction of DM particles 
that acquire excessive energy from DM-DM collisions, thus leading to a lower energy DM cloud that shrinks gradually 
forming a dark star. Another possibility is by DM accretion in supermassive stars. Once the star collapses, DM is not 
necessarily carried by the supernova shock wave, leaving a highly compact DM population at the core~\cite{Kouvaris:2010vv}. Moreover, if DM interactions are dissipative, DM can clump via direct cooling \cite{Fan:2013yva}.

It is crucial to determine the most important features which characterize the bulk properties of 
dark compact objects, and form a set of suitable observables to be potentially constrained by gravitational 
and electromagnetic surveys.
In this paper, we investigate the structure of slowly-rotating and tidally-deformed stars, modeled 
with a DM equation of state (EoS), based on fermionic and bosonic DM particles. 
As far as rotation is concerned, we follow the approach developed in \cite{Hartle:1967he,1968ApJ...153..807H},
in which spin corrections are described as a small perturbation of a static, spherically symmetric spacetime. 
At the background level, the star structure is determined by solving the usual Tolman-Oppenheimer-Volkoff equations (TOV). 
Rotational terms are included up to second order in the angular momentum $J$, which allow to compute the moment 
of inertia $I$ and the quadrupole moment $Q$ of the star. Similarly, we model tidal effects through the relativistic perturbative 
formalism described in \cite{Hinderer:2007mb}. At leading order, this approach leads to the Love number $k_2$, 
or, equivalently, the tidal deformability $\lambda=2/3k_2 R^5$, 
which encodes all the properties of the star's quadrupolar deformations. We refer the reader to the references 
cited above for a detailed description of the equations needed to compute these quantities.

The plan of the paper is the following: In Sec.~\ref{Sec:eos} we describe the main properties of the two classes 
of the dark EoS considered. In Secs.~\ref{Sec:MR} and \ref{Sec:IL} we analyze the bulk properties of the stellar 
models, such as masses and radii, which can be potentially constrained through electromagnetic and GW 
observations. An explicit example of such constraints is discussed in Sec.~\ref{Sec:detectability}. In Sec.~\ref{Sec:ILQ} we investigate universal relations for dark stars. Finally, in Sec.~\ref{Sec:summary} 
we summarize our results.

Throughout the paper (with the exception of Sec.~\ref{Sec:eos} and the \hyperref[Sec:app]{Appendix}) we use geometrized units, in which $G=c=1$.

\section{The dark matter equation of state} \label{Sec:eos}

In this section we describe the most important features of the dark EoS used in this 
paper to model fermion and boson stars. Moreover, in our analysis we will also consider two standard EoS, \texttt{apr} 
\cite{Akmal:1998cf} and \texttt{ms1} \cite{Muller:1995ji}, which represent two extreme examples of soft 
and stiff nuclear matter, and will allow to make a direct comparison between the macroscopic features of 
baryonic and dark objects. 

 \subsection{Fermion star}

We consider a fermionic particle interacting via a repulsive Yukawa potential (e.g., due to a massive dark 
photon):
\begin{equation}
	V = \frac{\alpha_\tn{X} }{r} \exp \left(-\frac{\hbar m_\phi r}{ c} \right),
\end{equation}
where $\alpha_X$ is the dark fine structure constant and $m_\phi$ is the mass of the mediator. The mass of the DM fermion is 
denoted as $m_\tn{X}$. Models that interact through a Yukawa potential are useful in the context of self-interacting DM, because 
the scattering cross section is suppressed at large relative velocities~\cite{Tulin:2013teo}. As a result, the success of collisionless 
cold DM is left untouched at super-glactic scales, while sub-galactic structure is flattened. In the context of self-interacting DM, 
both attractive and repulsive interactions flatten structures. However, attractive interactions in a compact object will soften the EoS. 
Since we are interested in dense objects, we only consider repulsive interactions.

Pressure in fermion stars has two contributions: one from Fermi-repulsion and one due to the Yukawa-interactions. 
We calculate the energy density and pressure due to Yukawa interactions in the mean field approximation; in this case the 
EoS is given by two implicitly related equations (see~\cite{Kouvaris:2015rea} for further details):
\begin{subequations}
	\begin{align}
	\rho &= \frac{m_\tn{X}^4 c^3}{\hbar^3}\left[\xi (x) + \frac{2}{9\pi^3}\frac{\alpha_\tn{X}}{\hbar c}\frac{ m_\tn{X}^2}{m_\phi^2} x^6\right],\\
	P &= \frac{m_\tn{X}^4 c^5}{\hbar^3}\left[\chi (x) + \frac{2}{9\pi^3}\frac{\alpha_\tn{X}}{\hbar c}\frac{ m_\tn{X}^2}{m_\phi^2} x^6\right],
\end{align}
\label{Eq: Fermion EoS}
\end{subequations}
where $x\equiv p_\text{F}/(m_\tn{X} c)$ is a dimensionless quantity that measures the Fermi-momentum compared to the 
DM mass (note that the density is defined such that $\rho c^2$ is the total energy density). The functions 
$\xi$ and $\chi$ are the contributions from Fermi-repulsion \cite{1983bhwd.book.....S}, given by
\begin{align*}
	\xi(x) &= \frac{1}{8\pi^2}\left[x\sqrt{1+x^2}(2x^2+1)- \ln\left(x+\sqrt{1+x^2} \right) \right],\\
		\chi(x) &= \frac{1}{8\pi^2}\left[x\sqrt{1+x^2}(2x^2/3-1)+ \ln\left(x+\sqrt{1+x^2} \right) \right].
\end{align*}

Both pressure and density are smooth monotonic functions of the parameters $x$. 
At low density the EoS becomes an approximate polytrope $P = K \rho^\gamma$ with index 
$\gamma \simeq 5/3$, whereas at large density the index changes to $\gamma \simeq 1$. 
At large density the proportionality constant is $K=c^2/3$ or $K=c^2$, depending on whether the Fermi-repulsion 
or the Yukawa-interactions dominates, respectively.

\subsection{Boson star}

Boson stars are naturally much smaller than their fermionic counterparts, because they lack Fermi-pressure to 
balance their self-gravity~\cite{Kaup:1968zz,Ruffini:1969qy}. In the absence of self-interactions, bosons stars 
are stabilized by a quantum mechanical pressure due to the uncertainty principle. Unless the bosons are 
extraordinarily light, this pressure is inherently tiny and can only balance small lumps of matter. If the field 
self-interacts, the boson star would naturally be similar to a fermion star in size~\cite{Colpi:1986ye}.

A wide variety of boson stars have been investigated in the literature (see~\cite{Liebling:2012fv} for a comprehensive 
review). The most studied examples include a complex scalar field with a $U(1)$ symmetry and an associated 
Noether charge. Other solutions include: real scalar field oscillatons~\cite{Seidel:1991zh}, Proca stars~\cite{Brito:2015pxa}, 
and axion stars~\cite{Sikivie:2009qn}, to name a few. 
In this work, we consider a complex scalar field coupled to gravity with the action
\begin{equation}
	S_\text{BS} = \int d^4x \sqrt{-g}\left(\frac{c^4R}{16\pi G}-|\partial_\mu \phi|^2 - \frac{m_\tn{X}^2c^2}{\hbar^2} |\phi|^2- \frac{1}{2}\frac{\beta}{\hbar c} |\phi|^4 \right),
\end{equation}
where $m_\tn{X}$ is the boson mass and $\beta$ is a dimensionless coupling constant.
The energy-momentum tensor is not automatically isotropic for a boson star. As such, the TOV formalism does 
not always apply. However, if we choose a spherically symmetric ansatz for the metric and the field 
[$\phi = \varphi(r)e^{i\omega t}$], the energy-momentum tensor becomes approximately isotropic.\footnote{This is only 
true when $\beta/(4\pi \hbar c)\gg G m_\tn{X}^2$, in which case spatial derivatives of the field can be dropped such that the 
Klein-Gordon equation becomes algebraic.} The EoS for this boson star model was first derived in~\cite{Colpi:1986ye} 
and it is given by
\begin{equation}
	P = \frac{c^5}{9\hbar^3}\frac{m_\tn{X}^4}{\beta}\left(\sqrt{1+\frac{3\hbar^3}{c^3} \frac{\beta \rho}{m_\tn{X}^4}} -1\right)^2.
	\label{Eq: Boson star EoS}
\end{equation}
This EoS behaves as a $\gamma \simeq 2$ polytrope at low density, and smoothly softens to $\gamma \simeq 1$ at high density.

\section{Mass - radius profiles} \label{Sec:MR}

Masses and radii are the macroscopic quantities which immediately characterize astrophysical 
compact objects, and they represent the primary target of both gravitational and electromagnetic surveys. 
X-ray and radio observations of astrophysical binaries are expected to provide precise measurements of the 
mass components, as they develop multiple relativistic effects which can be used to independently constrain 
the stellar structure \cite{Will:2014kxa,Lattimer:2006xb}. However, an accurate estimate of the radius still 
represents a challenging task, mostly relied upon the observation of signals coming from the interaction of the star with its
surrounding environment, which strongly depends on the assumption employed to model the process. 
Coalescences of binary NSs are also among the most powerful sources of GWs for ground-based 
interferometers, like LIGO \cite{0264-9381-32-11-115012} and Virgo \cite{0264-9381-32-2-024001}, as the large 
number of cycles before the merger will allow to extract the mass of the objects with good accuracy 
\cite{Sathyaprakash:2009xs}. Moreover, unlike the EM bandwidth, GW observatories have access 
to another quantity, the tidal deformability \cite{Flanagan:2007ix,Hinderer:2007mb}, which offers complementary 
information about the stellar radius (see Sec.~\ref{Sec:IL}).

These considerations point out that it is crucial to understand how the values of $M$ and $R$ change according 
to the extra parameters which specify the DM sector, and to which extent they differ from ordinary NSs.\footnote{We note that in computing the star's quadrupole moment (Sec.~\ref{Sec:IL}), the rotation 
rate introduces a monopole correction which modifies the mass of 
the compact object, namely $\bar{M}=M+\delta M$, with $\delta M\ll M$, $M$ being the {\it bare} mass of the star. Therefore, in general, $M\ne\bar{M}$. However, for the sake of clarity, in the next sections we will mostly use $M$ 
as the fundamental parameter of our analysis.}
The first section of our analysis will be therefore devoted to investigate these features. 

\subsection{Fermion stars}

\begin{figure*}[ht]
\includegraphics[width=5.cm]{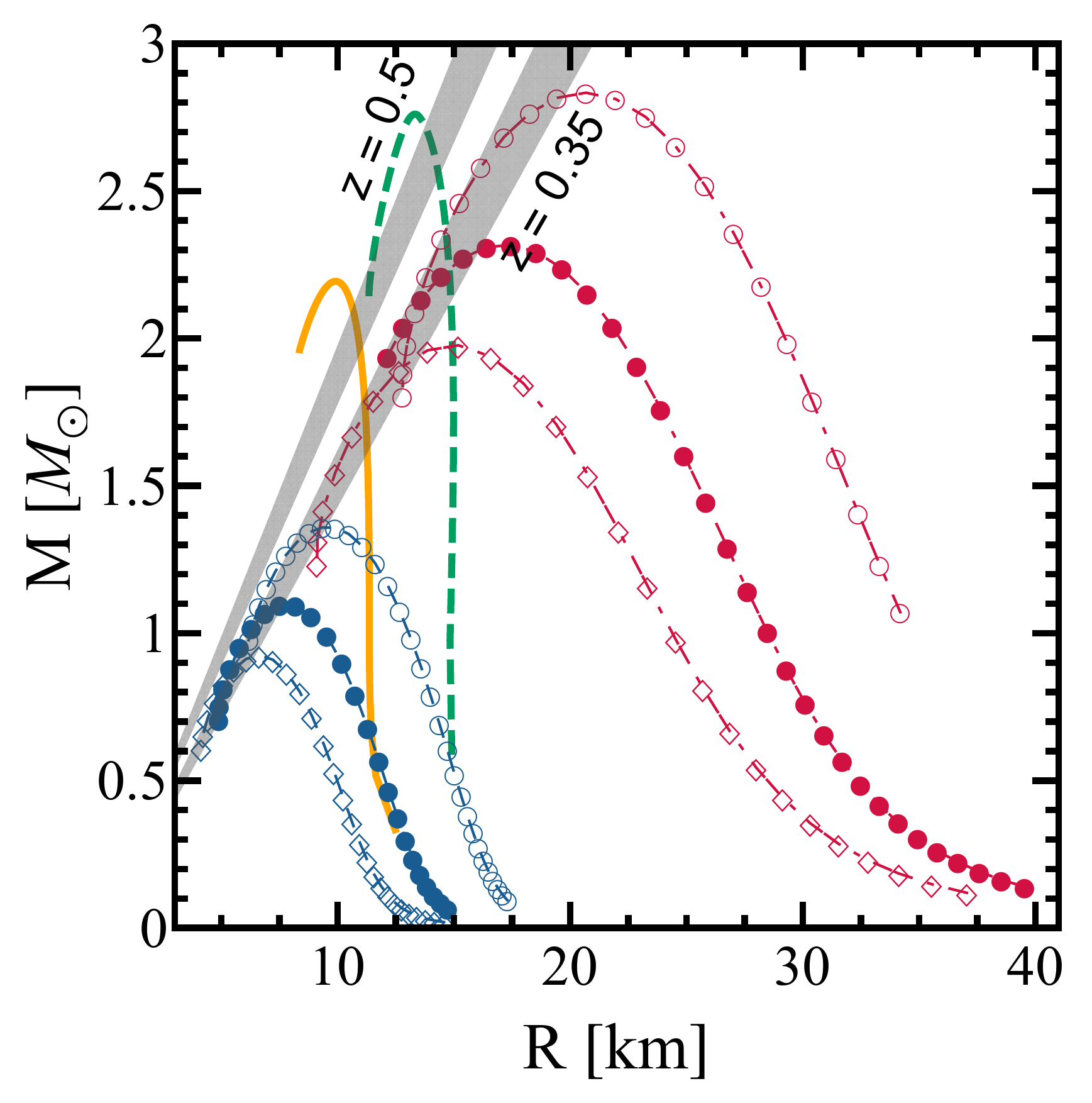}
\includegraphics[width=5cm]{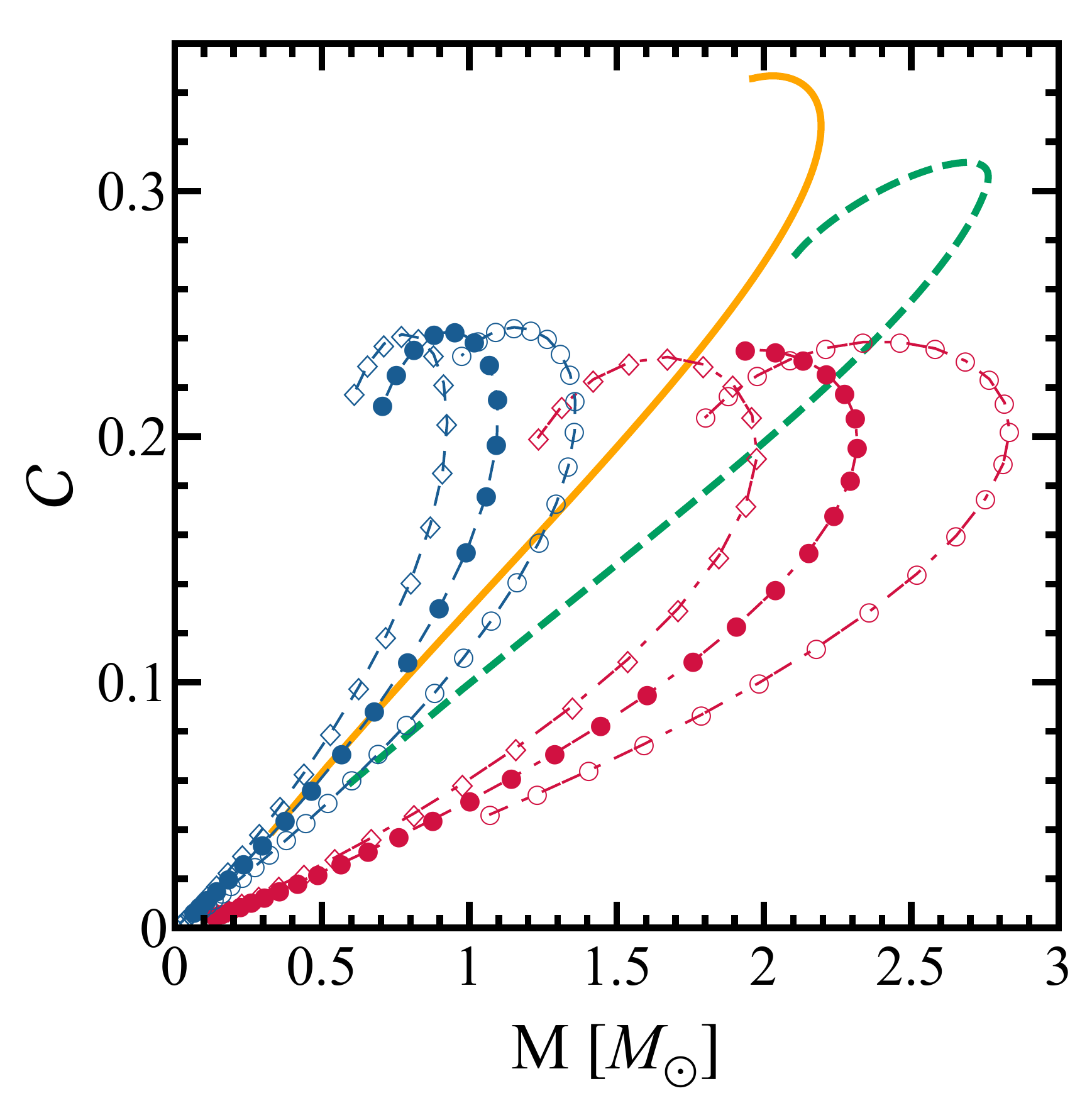}
\includegraphics[width=7.4cm]{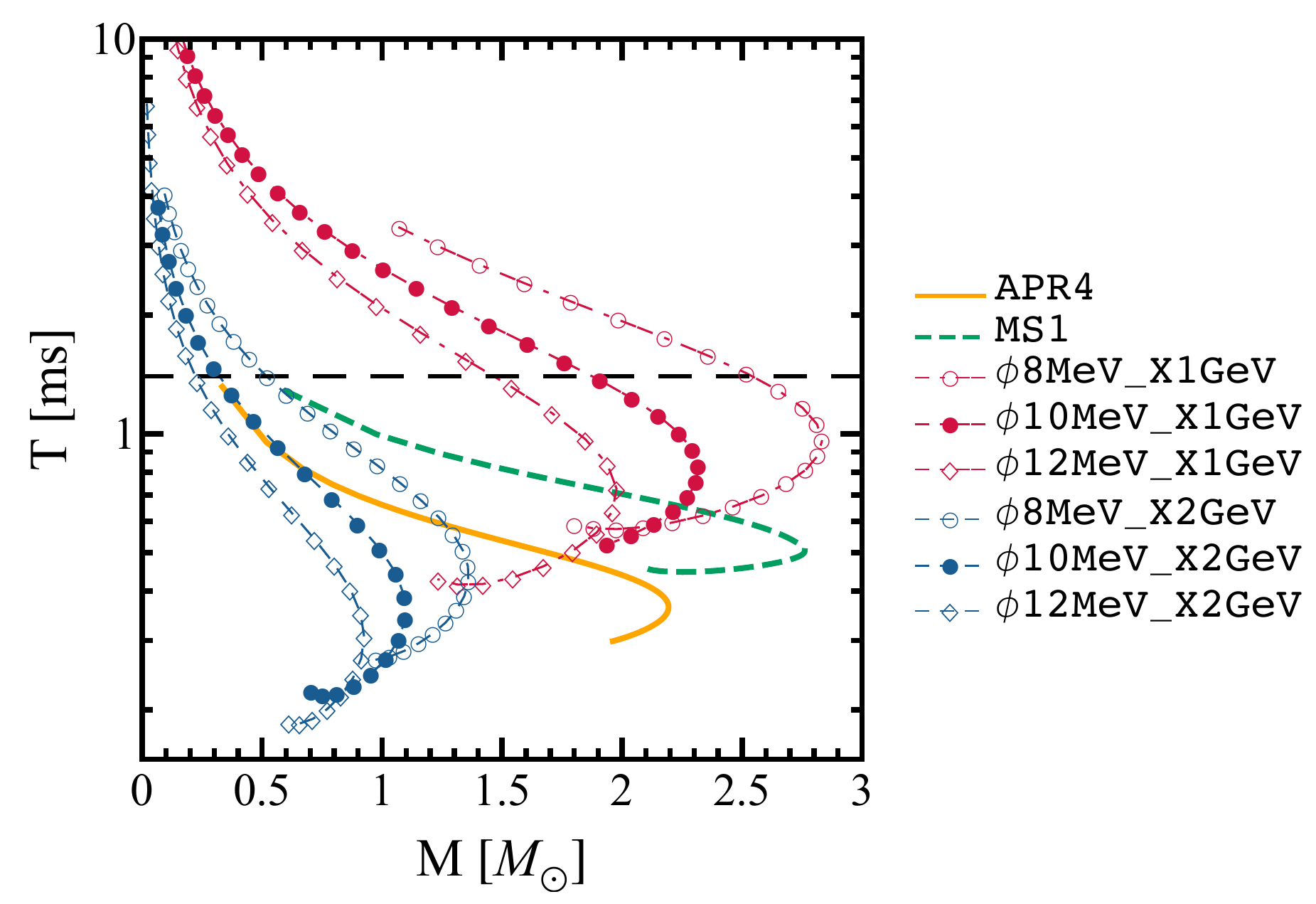}
\label{fig:MR} 
\caption{(Left) Mass-radius profiles for fermion stars with fixed $\alpha=10^{-3}$, dark particle mass $m_{X}=(1,2)$GeV, and mediator mass $m_\phi=(8,10,12)$MeV. The standard EoS \texttt{apr} and \texttt{ms1} are represented by the yellow solid and the dashed green curves respectively. The shaded regions correspond to contour regions of constant surface redshift $z=0.35$ and $z=0.5$ with 10\% of accuracy. (Center) Compactness ${\cal C}=M/R$ as a function of the stellar mass, for the models considered in the left panel. (Right) Minimum period, according to the Kepler frequency limit, $T=2\pi/\Omega_\tn{K}$. The dashed 
horizontal line represents the fastest known pulsar with $f=716$ Hz.}
\end{figure*}

The fermion stars described in this paper are fully specified by three parameters: the coupling constant 
$\alpha_\tn{X}$, the dark particle mass $m_\tn{X}$, and the mediator mass $m_\phi$. Hereafter, 
we will fix $\alpha_\tn{X}/\hbar c\equiv\alpha=10^{-3}$, varying the other two coefficients $m_\phi=(8,10,12)$MeV and 
$m_{X}=(1,2)$GeV. These values lead to mass--radius profiles comparable with those computed for 
$\texttt{apr}$ and $\texttt{ms1}$, and therefore will allow for a direct and more clear comparison with NSs. 
All our models, identified by the label $\phi\texttt{m}_\phi\_\texttt{X}\texttt{m}_\texttt{X}$, 
are presented in Fig.~\ref{fig:MR}. Each point of the plot is obtained, for a chosen EoS, by varying the star's 
central pressure.

The left panel of the plot shows how the dark sector parameters 
affect the stellar configurations. We note first that, for a fixed radius, larger values of $m_\tn{X}$ 
rapidly decrease the mass, therefore leading to less compact objects. This is more evident from the 
center panel in which we draw the compactness ${\cal C}=M/R$ for all of our models.
The mediator mass also provides large changes, still in the same direction as $m_\tn{X}$, as
less compact stars are obtained passing from $m_\phi=8$MeV to $m_\phi=12$MeV.
It is interesting to note that the two baryonic EoS considered are characterized by steeper 
slopes, which lead to larger mass/radius variations. Both left and center plots also show that overlapping regions 
do exist between fermion stars and baryonic matter profiles which yield the same 
configurations. This is particularly relevant from the experimental point of view, as mass and 
radius measurements lead to degeneracies which may prevent a clear identification of the nature 
of the compact object.

Mass-radius profiles are extremely useful, since they can exploit astrophysical observations to constrain 
the space of nuclear EoS. As an example, in the left plot of Fig.~\ref{fig:MR} we show two shaded regions 
corresponding to constant surface redshift 
\begin{equation}
z=\left(1-\frac{2M}{R}\right)^{-1/2}-1\ ,
\end{equation}
with two reference values, namely $z=0.35$ and $z=0.5$. The first one matches the data obtained for 
EXO 0748-676, a NS  showing repeated X-ray bursts \cite{Cottam:2002cu}. The width of the bands 
represents a 10\% of accuracy in the measurements. It is clear that, in the first case ($z=0.35$), the observed 
value is already inconsistent with all the stable branches of the models considered in this section. However, 
a potential observation of a surface redshift $z=0.5$ would 
set a tighter bound, ruling out the possibility that the source is a fermion star with one of the EoS we have used here.   
Using multiple observables, like the Eddington flux and the ratio between the thermal flux and the color 
temperature, would further reduce the parameter space of allowed configurations 

The center panel of Fig.~\ref{fig:MR} shows another interesting property of fermion stars. The 
EoS considered cover large changes in the mass--radius space, but the corresponding compactness never 
exceeds a threshold ${\cal C}\approx 0.22$, contrary to $\texttt{apr}$ and $\texttt{ms1}$, which can 
achieve values higher than $0.3$. Although significantly different from white dwarf and main-sequence stars 
(with ${\cal C}\lesssim 10^{-6}$), this indicates that fermion stars cannot act as black hole mimickers, 
i.e., compact objects with a compactness approaching the limit value ${\cal C}\rightarrow 0.5$ 
\cite{Cardoso:2017cfl,2017arXiv170308156V}. 

As a final remark, for each model it is useful to investigate the maximum rotation rate allowed by the stellar structure. 
This quantity is of particular interest for astrophysical observations, as spinning frequencies of isolated and binary 
NSs are measured with exquisite precision and they can be used to constrain the underlying EoS \cite{Lattimer:2006xb}. 
The right panel of Fig.~\ref{fig:MR} shows the minimum rotational period for fermion stars, derived from the Keplerian limit 
$T=2\pi/\Omega_\tn{K}$, where $\Omega_\tn{K}\approx\sqrt{M/R^3}$ (mass-shedding limit). Although this is a Newtonian approximation, it 
gives a good estimate of the order of magnitude of this quantity, and  provides an absolute upper limit on the 
spin. As a benchmark, we also draw (horizontal black line) 
the value corresponding to the maximum frequency observed for a spinning NS, $f=716$ Hz, i.e., $T\approx 1.4$ ms
\cite{Hessels:2006ze}. This constraint alone cannot rule out any theoretical model that we have used to describe both 
dark stars and regular NSs. Assuming a future dark star observation, a bound potentially able to exclude configurations with 
$m_\tn{X}\lesssim 1$GeV would require a much faster rotating object, 
with $f\approx 1800$Hz or, equivalently, $T\approx 0.55$ ms.

\subsection{Boson stars}

\begin{figure*}[ht]
\centering
\includegraphics[width=7.cm]{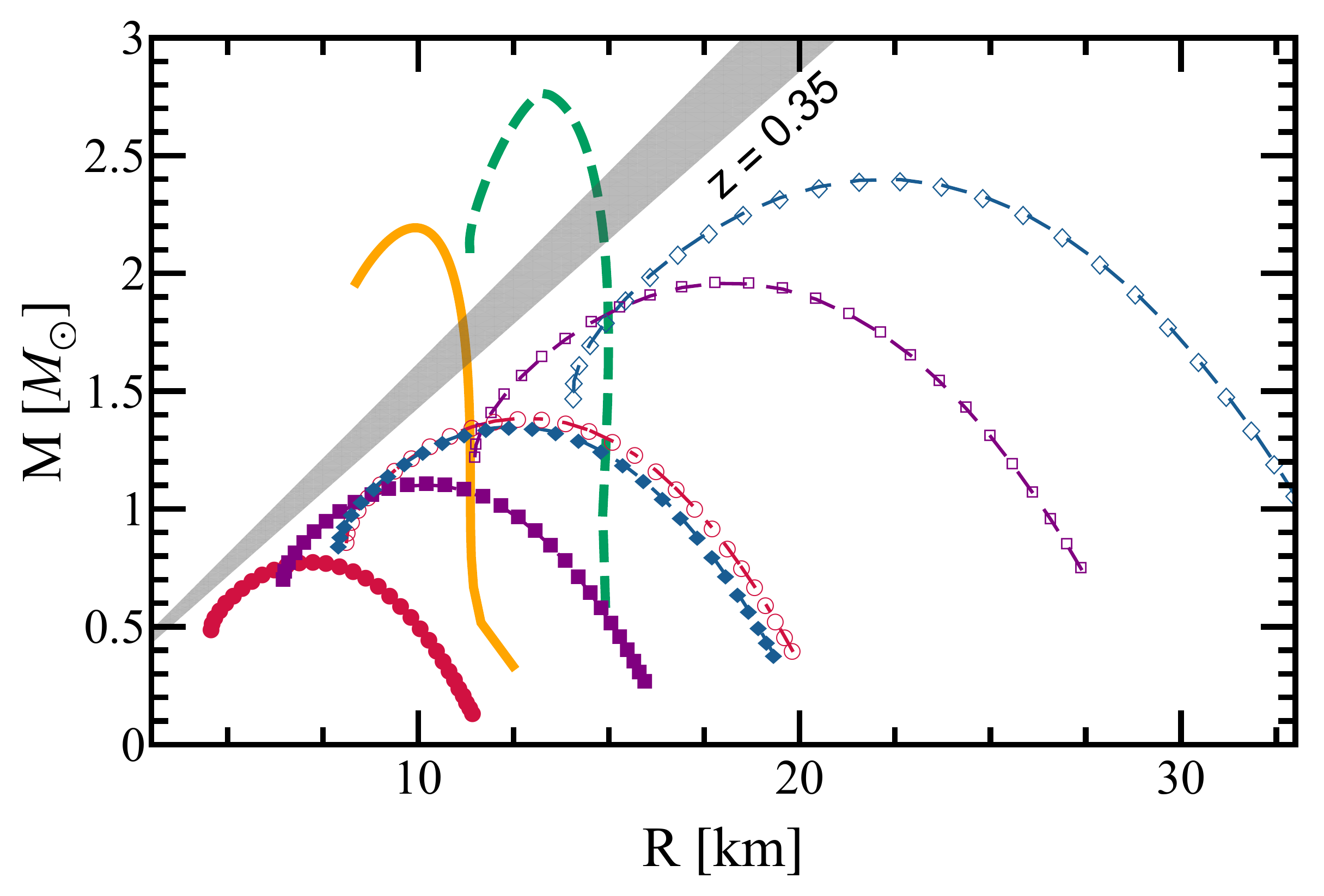}
\includegraphics[width=7.2cm]{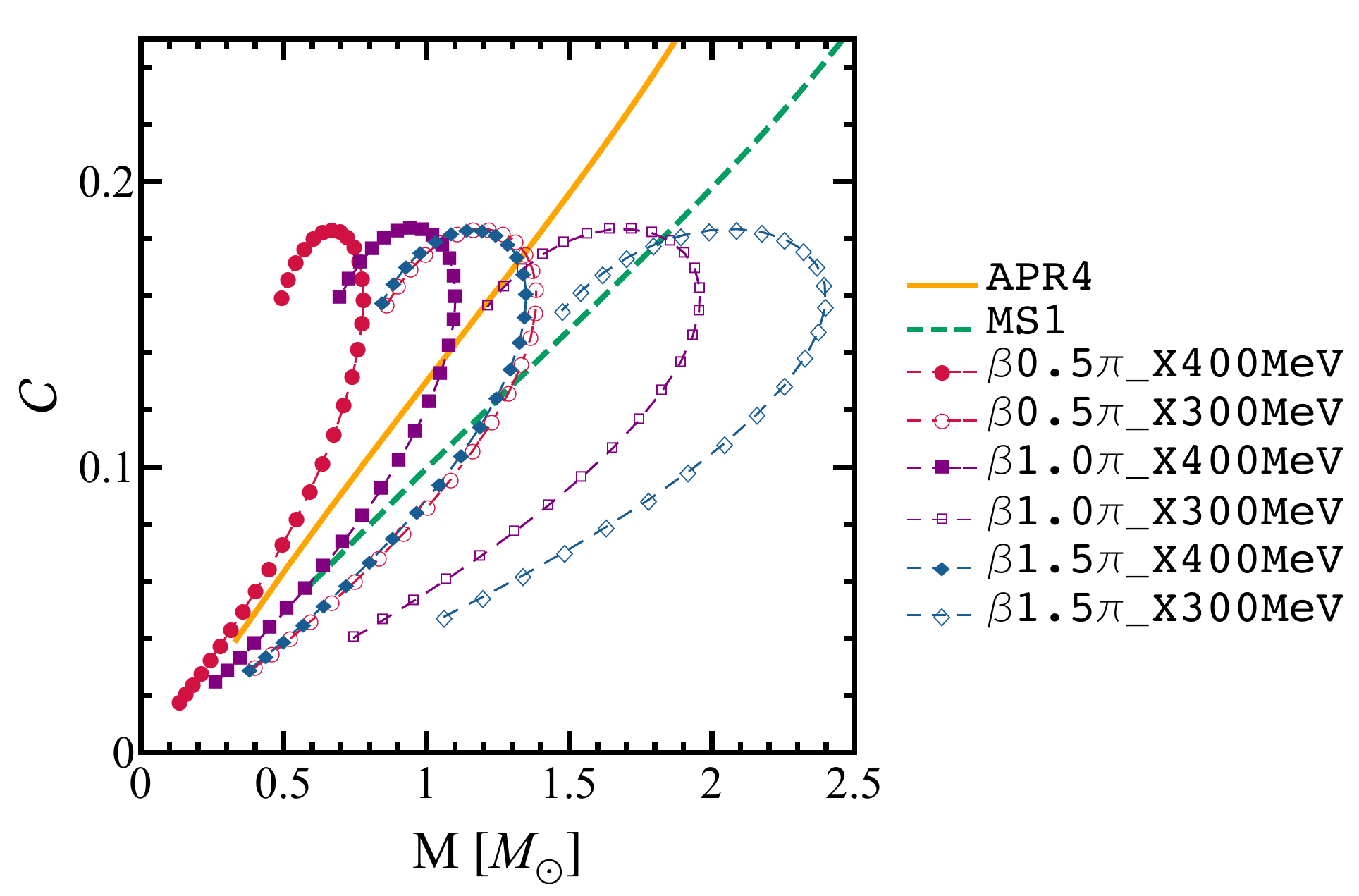}
\label{fig:MR2} 
\caption{Same as Fig.~\ref{fig:MR}, but for boson stars with different values of the coupling constant 
$\beta=(0.5,1,1.5)\pi$ and boson mass $m_\tn{X}=(300,400)$MeV.}
\end{figure*}

The mass--radius profiles for boson stars are shown in the left panel of Fig.~\ref{fig:MR2}. We consider three 
values of the coupling parameter\footnote{The values of $\beta$ are chosen to be less than 
$2\pi$, such that the interactions can be treated perturbatively \cite{Eby:2015hsq}.} $\beta=(0.5,1,1.5)\pi$ and two values of the boson mass $m_\tn{X}=(300,400)$MeV. Like the fermion case, these configurations are chosen to provide the stellar models 
closest to the standard NSs built with \texttt{apr} and \texttt{ms1}.
We label the boson EoS as $\beta\_\texttt{X} \texttt{m}_\texttt{X}$.

We first note that, for a given mass, stronger couplings stiffen the EoS, leading to larger radii and,
therefore, to less compact objects. At the same time, $\beta$ modifies the maximum mass of each model, 
shifting the value to the top end of the parameter space. The same trend, although with a major impact, 
occurs if we consider lighter dark particles.

Even for boson stars, the slope of the curves is smoother than that obtained for standard nuclear matter. 
This produces more pronounced changes in the radius distribution, as the central pressure of the star varies. 
The right panel of Fig.~\ref{fig:MR2} also shows the stellar compactness $M/R$. For all the considered 
models, we observe a maximum value ${\cal C}\approx 0.16$, well below the edge of the curve related to \texttt{apr} 
and \texttt{ms1} which, for a fixed radius, yield softer EoS and therefore larger masses.
This also excludes the chance to interpret boson stars as astrophysical objects compact enough to 
mimic black holes. Remarkably, we find that this peoperty holds in general for any class of boson EoS, 
independently from the coupling parameter. A mathematical proof of this feature is outlined in the 
\hyperref[Sec:app]{Appendix}.

Electromagnetic observations of the stellar spin frequency are still too weak to considerably 
narrow the star parameter space, as the maximum value observed so far leaves the EoS essentially 
unbound. Larger values of $f\approx 1400$ Hz (or, equivalently, $T\approx 0.7$ ms) would be required to 
exclude the presence of a dark star. On the other hand, as seen in the previous section, precise measurements of the surface redshift 
represent a powerful tool to constrain the stellar structure. As an example, the value $z=0.35$, derived for EXO 
0748-676, seems already to exclude the possibility that this object is built by one of the bosonic EoS considered.

\section{Moments of inertia, tidal Love numbers and quadrupole moments} \label{Sec:IL}

The moment of inertia represents another global feature of compact objects, potentially observable by 
electromagnetic surveys, which depends more on the compactness ${\cal C}$ rather than on the 
microphysical details of the EoS. The moment of inertia is found to be correlated through semi-analytical 
relations with different stellar parameters, scaling approximately as $R^2$. Therefore, any constraint 
on the stellar radius naturally provides a bound for $I$ \cite{0004-637X-550-1-426,0004-637X-771-1-51}. 
Moreover, this quantity affects different astrophysical processes, such as pulsar glitches, 
characterized by sudden increases of the stellar rotational frequency (of the order of $10^{-6}$). 
The relativistic spin-orbit coupling in compact binary systems also depends on the moment of inertia. 
In the near future, high precision pulsar timing could determine the periastron advance of such systems 
in order to provide an estimate of $I$ (and therefore of $R$) with an accuracy of $10\%$ \cite{Lattimer:2004nj}.

\begin{figure}[ht]
\centering
\includegraphics[width=8.5cm]{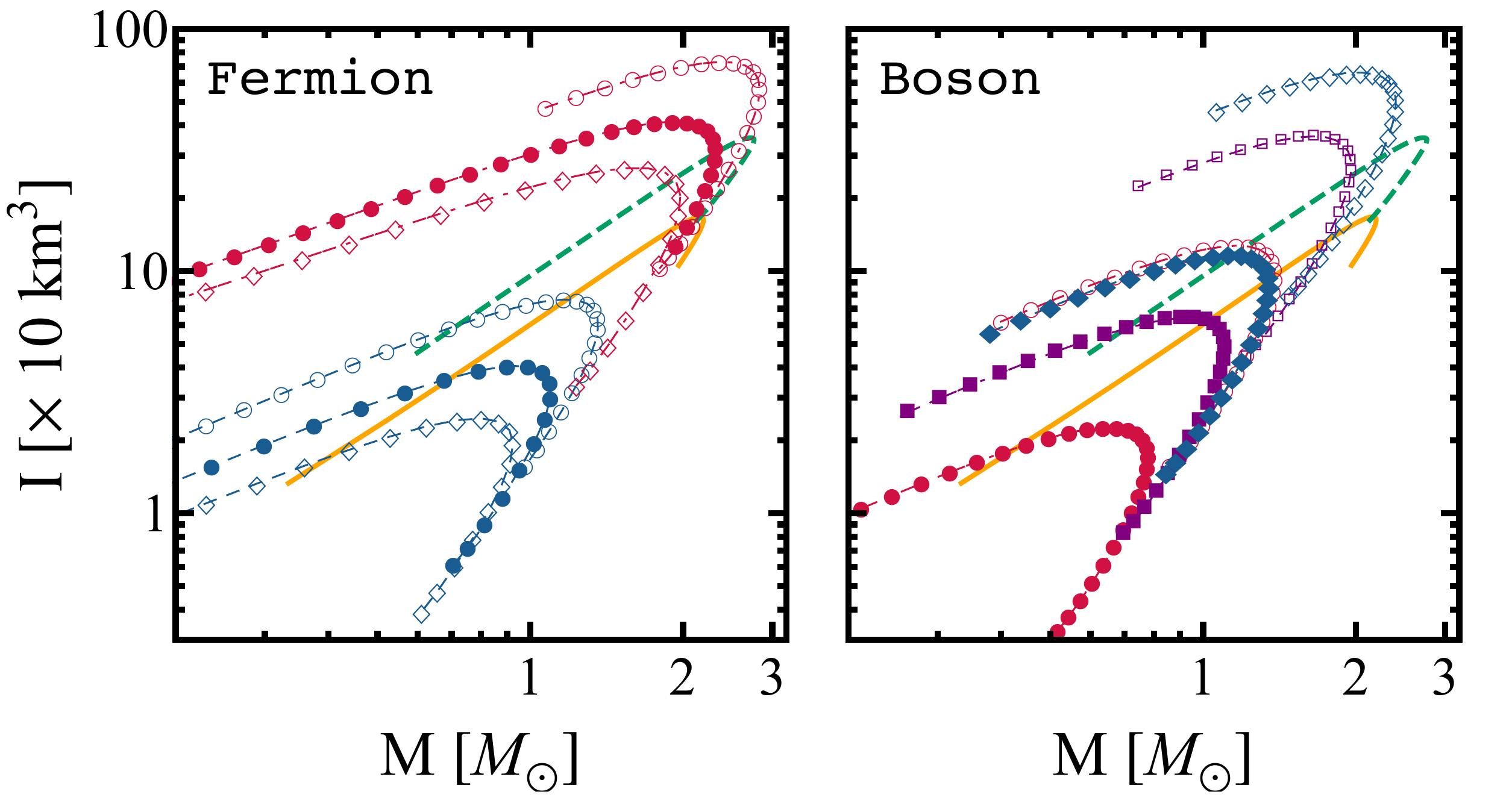}
\label{fig:inertia} 
\caption{Moment of inertia for fermion (left) and boson (right) stars, for the various EoS 
considered in Figs.~\ref{fig:MR}-\ref{fig:MR2}. The yellow solid and green dashed curve correspond to 
the \texttt{apr} and \texttt{ms1} EoS, respectively.}
\end{figure}

Motivated by these considerations, in this section we shall compare the values of the moment of inertia 
computed for DM and baryonic EoS. Our results are shown in Fig.~\ref{fig:inertia}.
As expected, for fermion stars and a fixed stellar mass, $I$ increases for smaller values of $m_\phi$ and 
$m_\tn{X}$, with the latter leading to the largest variations, as it mainly affects the stellar compactness. 
For the boson case, the largest moment of inertia results from light particles, with $m_\tn{X}<300$ MeV, 
and stronger repulsive interactions $\beta\gtrsim \pi$. 
We also note that for fermion stars and a fixed mass $m_\tn{X}$, the spread within the configurations 
given by the mediator $\phi$, typically $\Delta I\gtrsim 10\%$, is much larger than the gap between 
\texttt{apr} and \texttt{ms1}, which cover a rather wide range of standard EoS currently known. This 
is particularly relevant for future space observations, as measurements of the spin-orbit effect (previously 
described) would provide errors smaller or equal to $\Delta I$, and therefore would be able to set (at least) 
an upper bound on $m_\phi$. Similar considerations also apply to the boson sector, if we consider the 
deviations produced by the coupling parameter $\beta$.\\

Extracting information on the internal structure of compact objects is also a 
primary goal of current and future GW interferometers. 
The imprint of the EoS within the signals emitted during binary coalescences is mostly 
determined by adiabatic tidal interactions, characterized in terms of a set of coefficients, 
the {\it Love numbers}, which are computed assuming that tidal effects are produced by 
an external, time-independent gravitational field 
\cite{Hinderer:2007mb,Binnington:2009bb,Damour:2009vw}. The dominant 
contribution $k_2$, associated to a quadrupolar deformation, is defined by 
the relation 
\begin{equation}
{\cal Q}_{ij}=\frac{2}{3}k_2R^5{\cal E}_{ij}=\lambda {\cal E}_{ij}\ ,\label{QlLoveE}
\end{equation}
where ${\cal E}_{ij}$ is the external tidal tensor and ${\cal Q}_{ij}$ is the 
(tidally-deformed) star's quadrupole tensor.\footnote{Not to be confused with the spin-induced quadrupole 
moment introduced later.} The Love number $k_2$ or, equivalently, the tidal deformability 
$\lambda$, depends {\it solely} on the star's EoS. 
The inclusion of the Love number into semi-analytical templates for GW searches, 
and its detectability\footnote{We note that, so far, most of the works concerning tidal 
effects focused on NS binaries only, as in general relativity $k_2=0$ for black holes. However, 
the Love number formalism has been recently extended to exotic compact 
objects, showing that they represent a powerful probe to distinguish between such 
alternative scenarios and regular black holes \cite{Cardoso:2017cfl}.} by current 
and future detectors, have been deeply investigated in the literature \cite{Flanagan:2007ix,Damour:2009wj,Maselli:2013rza,Damour:2012yf,Hinderer:2009ca,Bernuzzi:2012ci,Read:2013zra,Baiotti:2010xh,Baiotti:2011am,Vines:2011ud,Pannarale:2011pk,Vines:2010ca,Lackey:2011vz,Lackey:2013axa,Yagi:2013baa}.
As an example, fully relativistic numerical simulations have shown that, for stiff EoS, the  
radius of a standard NS can be constrained within $\sim10\%$ of accuracy by advanced 
detectors, with the measurability rapidly getting worse for softer matter, i.e., for stellar 
configurations with larger compactness \cite{Hotokezaka:2016bzh}.
More recently, the effect of dynamic tides has been taken into account, 
proving that they also provide a significant contribution to the GW 
emission \cite{Hinderer:2016eia,Essick:2016tkn}.

\begin{figure}[ht]
\centering
\includegraphics[width=8.5cm]{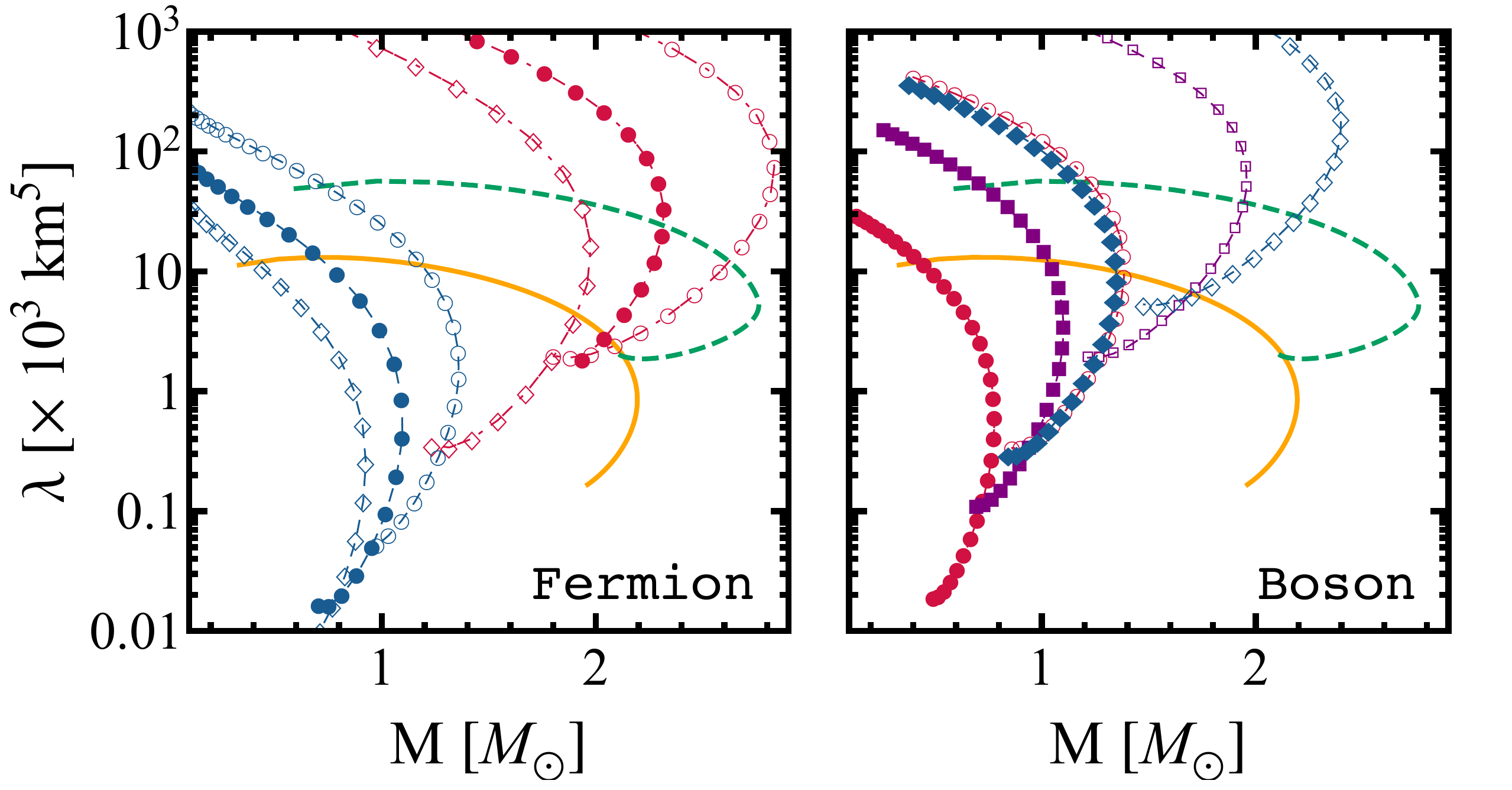}
\label{fig:Iambda} 
\caption{Same as Fig.~\ref{fig:inertia} but for the tidal deformability $\lambda$.}
\end{figure}

To this end, it is crucial to analyze how $\lambda$ behaves for dark stars, as they  
may lead to large signatures, potentially detectable by GW interferometers, to be used 
together with measurements of $I$ and $Q$ for multi-messenger constraints. 
Figure~\ref{fig:Iambda} shows the tidal deformability as a function of the stellar mass, for  
fermion and boson stars. 
For the former, different values of $m_\tn{X}$ and $m_\phi$ 
yield large variations of $\lambda$ within the parameter space. Such differences are mainly 
related to the strong dependence of the tidal deformability on the stellar radius, 
$\lambda\propto k_2 R^5$, which amplifies the discrepancies between the models. 
We also note that for boson stars a universal relation between $k_2$ and the compactness 
${\cal C}=M/R$ exists, which is independent of the specific choice of $m_X$ and $\beta$.

It is worth to remark that these features, which ultimately reflects the stellar compactness, may be 
a crucial ingredient for future GW detections, as $\lambda$ is the actual parameter entering 
the waveform. In this regard, dark stars with a lighter mediator $\phi$ would experience large 
deformations, improving our ability to constrain the tidal Love number.
On the other hand, EoS with $m_\phi\gg 1$GeV would provide smaller $\lambda$, 
leading to weaker effects within the signal and, hence, to 
looser bounds on the star's structure.

Similar considerations hold as far as boson interactions are taken into account. For a 
chosen mass, both larger couplings and lighter particles lead to larger values of 
the Love number. The right panel of Fig.~\ref{fig:Iambda} 
shows indeed that for a canonical $1.4M_\odot$ star, even a very stiff EoS like \texttt{ms1} would 
provide a tidal deformability more than two orders of magnitude smaller than those computed for
$\beta \gtrsim \pi$ and $m_\tn{X}\lesssim 300$MeV. Such enhancement would strongly improve 
the measurability of $\lambda$ from GW signals.\\

\begin{figure}[ht]
\centering
\includegraphics[width=8cm]{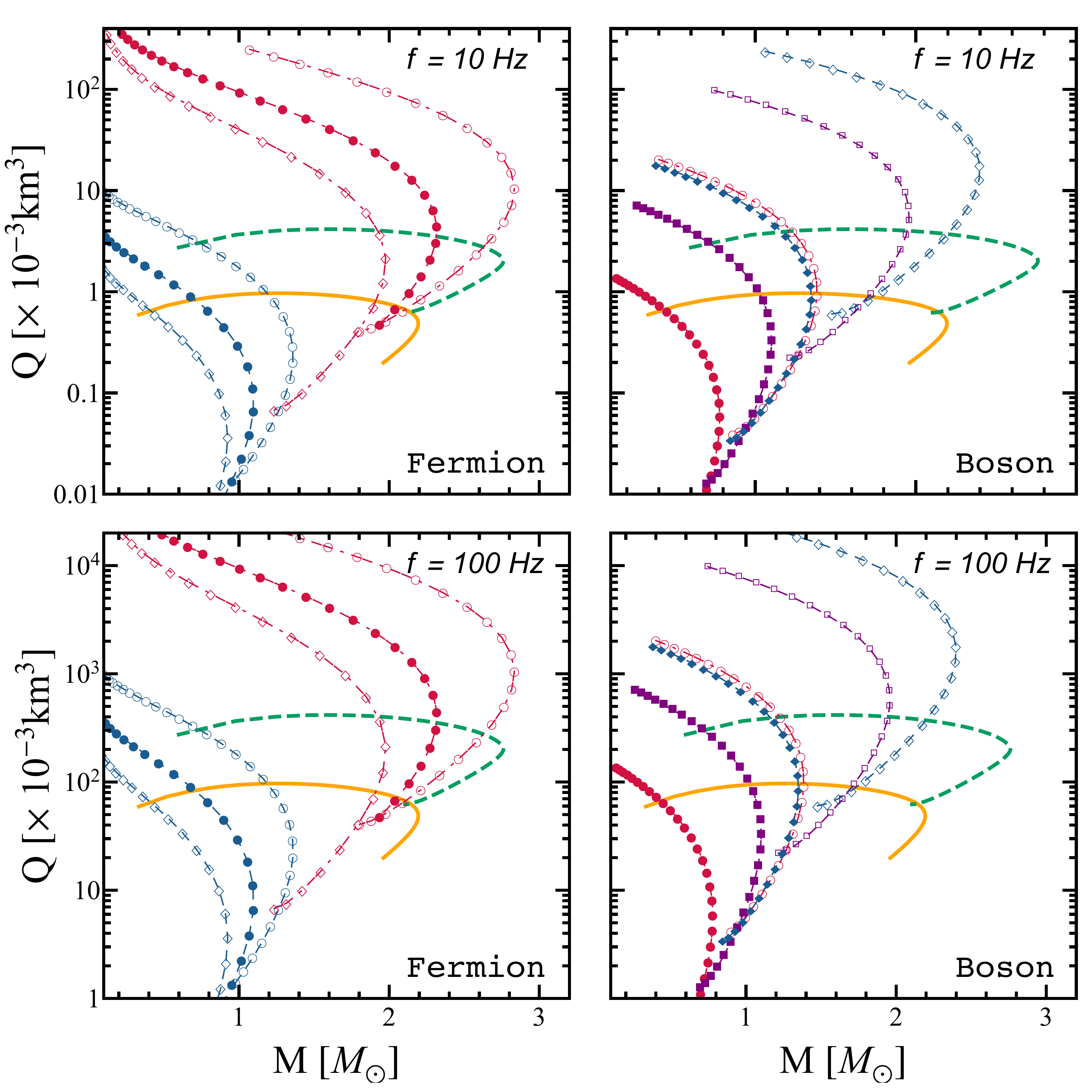}
\label{fig:quadrupole} 
\caption{Quadrupole moments as functions of the stellar 
mass $M$ for two values of the rotational frequency $f=(10,100)$Hz. The data for 
fermion and boson stars are compared against the results derived for the standard EoS 
\texttt{apr} and \texttt{ms1}.}
\end{figure}

As a final remark, in Fig.~\ref{fig:quadrupole} we show the spin-induced quadrupole moment 
for fermion and boson stars. Following \cite{1968ApJ...153..807H,Hartle:1967he}, the spacetime 
describing a spinning compact object can be obtained perturbing a spherical non-rotating metric, 
as a power series of the dimensionless spin variable $\chi=J/M^2$, $J$ being the star's intrinsic 
angular momentum. The quadrupole moment affects the perturbed metric at the second order 
in $\chi$. In our analysis we consider rotational frequencies $f=(10,100)$Hz, such that $\chi\ll1$, 
i.e., requiring that spin effects represent a small perturbation of the static, spherically symmetric 
background. Looking at the figure we immediately note that, for a fixed mass, the values 
of $Q$ for dark and neutron stars yield large differences, which can be potentially tested both by 
GW and electromagnetic observations. Indeed, the quadrupole moment modifies the 
gravitational waveform produced by binary coalescences, leading to signatures detectable by 
terrestrial interferometers \cite{Krishnendu:2017shb}. Moreover, $Q$ is expected to affect the 
location of the innermost stable circular orbit, and therefore to influence the geodesic motion 
around the star \cite{Pappas:2012nt}. 
The latter plays a crucial role in several astrophysical 
phenomena related to accretion processes, which produce characteristic signals (quasi-periodic
oscillations), that have been proven to be a powerful diagnostic tool of the nature of gravity 
in the strong-field regime.

\subsection{Constraining the bulk properties: a practical example} \label{Sec:detectability}

Before further discussing the basic properties of boson and fermion stars, it is useful to provide an explicit 
example of how future observations will constrain the bulk properties described in the previous sections. 
For the sake of simplicity, we shall consider one particular quantity, the tidal deformability $\lambda$, which 
affects the GW signals emitted by binary systems. We consider indeed the coalescence of two 
non-spinning dark stars with masses $m_1,m_2$ and the same EoS. The emitted sky-averaged waveform 
in the frequency domain
\begin{equation}
\tilde{h}(f)={\cal A}e^{i\psi(f)}\ ,
\end{equation}
is specified by the overall amplitude ${\cal A}$ and the phase $\psi(f)$, which depends on the GW frequency $f$ 
and the physical parameters $\bm{\theta}=({\cal A},\ln {\cal M},\ln \nu,t_c,\phi_c,\Lambda)$, where 
${\cal M}=(m_1+m_2)\nu^{3/5}$ and $\nu=m_1m_2/(m_1+m_2)^2$ are the chirp mass and the symmetric 
mass ratio, while $(t_c,\phi_c)$ the time and phase at the coalescence. The parameter $\Lambda$ 
is an {\it average} tidal deformability: 
\begin{equation}
\Lambda=\frac{1}{26}\left[(1+12/q)\lambda_1+(1+12q)\lambda_2\right]\ ,
\end{equation}
where $q=m_1/m_2\geq1$, related to the $\lambda_{1,2}$ of the single objects \cite{Vines:2011ud,Damour:2012yf}. 
Equal-mass binaries, with $q=1$, yield $\Lambda=\lambda_1=\lambda_2$. For strong signals, with a 
large signal-to-noise ratio, the errors on the parameters $\sigma_{\bm{\theta}}$ can be estimated using a 
Fisher matrix approach (see \cite{Vallisneri:2007ev} and references therein). In this framework, 
the covariance matrix of $\bm{\theta}$, $\Sigma_{ij}$, is given by the inverse of the Fisher matrix 
\begin{equation}\label{covariance}
\Sigma_{ij}=(\Gamma_{ij})^{-1}\quad \ , \quad \Gamma_{ij}=\left(\frac{\partial h}{\partial \theta_i}\bigg\vert 
\frac{\partial h}{\partial \theta_j}\right)_{\bm{\theta}=\bar{\bm{\theta}}}\ ,
\end{equation}
which contains the derivatives with respect to the binary parameters computed around the true values 
$\bar{\bm{\theta}}$, and we have defined the scalar product $(\cdot\vert \cdot)$ between two waveforms
\begin{equation}
(h_1\vert h_2)=2\int_{f_\tn{min}}^{f_\tn{max}}df\frac{\tilde{h}_1(f)\tilde{h}^\star_2(f)
+\tilde{h}^\star_1(f)\tilde{h}_2(f)}{S_n(f)}\ ,
\end{equation}
weighted with  the detector noise spectral density $S_n(f)$. In the following, we consider one single interferometer, 
Advanced LIGO, with the sensitivity curve provided in \cite{zerodet}, numerically computing Eqs.~(\ref{covariance}) 
between $f_\tn{min}=20$Hz and $f_\tn{max}=f_\tn{ISCO}$, the latter being the frequency at the innermost stable 
circular orbit for the Schwarzschild spacetime, i.e., $f_\tn{ISCO}=[\pi6^{3/2}(m_1+m_2)]^{-1}$. 
We also assume sources at distance $d=100$ Mpc, with masses $m_{1,2}=(1.2,1.4)M_\odot$ and $q=1$.

\begin{table}[ht]	
  \centering
  \begin{tabular}{ccccc}
     \hline
    \hline
\texttt{EoS} &  $\ln \Lambda_{1.4}$ & $\sigma_\Lambda/\Lambda_{1.4}$ &  $\ln \Lambda_{1.2}$ & $\sigma_\Lambda/\lambda_{1.2}$\\
     \hline
\texttt{ms1} & 10.88 & 20.55 & 10.93 & 19.41 \\
\texttt{apr} &  9.16 & 117.6 & 9.321 & 99.93 \\
$\phi8\texttt{MeV}\_\texttt{X}1\texttt{GeV}$  &14.98 & 0.6746 & 15.22 & 0.7248 \\
$\phi10\texttt{MeV}\_\texttt{X}1\texttt{GeV}$ &  13.73 & 0.7228 & 14.06 & 0.5555 \\
$\phi12\texttt{MeV}\_\texttt{X}1\texttt{GeV}$  &12.63 & 2.986  & 13.08 & 1.683\\
$\phi8\texttt{MeV}\_\texttt{X}2\texttt{GeV}$ & - & - & 9.272 & 105 \\
$\beta0.5\pi\_\texttt{X}300\texttt{MeV}$ & -  & - & 11.05 & 17.17\\
$\beta1.0\pi\_\texttt{X}300\texttt{MeV}$ & 13.49 & 0.9701 & 13.85 & 0.6402\\
$\beta1.5\pi\_\texttt{X}300\texttt{MeV}$ & 14.95 & 0.6677 & 15.18 & 0.717\\
$\beta1.5\pi\_\texttt{X}400\texttt{MeV}$ & - & - & 10.77 & 22.92\\
   \hline
  \hline
\end{tabular}
  \caption{Average tidal deformability $\Lambda$ and corresponding relative percentage errors, computed for 
  boson and fermion dark star binaries, with $m_{1,2}=1.4M_\odot$ (second and third columns) and 
  $m_{1,2}=1.2M_\odot$ (fourth and fifth columns). We assume non-spinning sources at a distance 
  $d=100$ Mpc. For some EoS no model with $1.4M_\odot$ exists.}
  \label{table:errors}
\end{table}

Table~\ref{table:errors} shows the relative percentage errors $\sigma_\Lambda/\Lambda$ for the 
binary systems considered, and different EoS, together with the values of $\Lambda$. The third and fifth 
columns immediately show how, for a fixed mass, the uncertainties change among all the models.  
As described in the previous sections, for fermion stars, small values of the mediator mass $\phi$ lead 
to larger tidal deformations, which drastically improve the errors on $\Lambda$, around $1\%$. 
These numbers have to be compared against the results for standard nuclear matter, which provide 
much looser bounds. The same trend is observed for boson EoS with $\beta\gtrsim\pi$ and 
$m_\tn{X}\lesssim 300$MeV.

These data can be combined with other information, coming from different experiments and/or 
bandwidths to further constrain the stellar EoS. A more detailed analysis on this topic, focused on how 
to join the results from both electromagnetic and GW surveys, will be presented in a 
forthcoming publication.

\section{Universal relations} \label{Sec:ILQ}

Astrophysical observations of compact objects both in the electromagnetic and gravitational bandwidth are 
limited by our ignorance on their internal structure. As discussed in the previous sections, macroscopic 
quantities, such as masses and radii, strictly depend on the underlying EoS, and their measurement is strongly 
affected by the behavior of matter at extreme densities. This lack of information can be mitigated by exploiting 
the recently discovered $I$-Love-$Q$ universal relations \cite{Yagi:2013bca,Yagi:2013awa}, which relate the 
moment of inertia, the tidal Love number, and the spin-induced quarupole moment of slowly-rotating compact 
objects through semi-analytical relations, that are almost insensitive to the stellar composition and accurate within 
$1\%$. The $I$-Love-$Q$ have several applications, as they can be used to break degeneracies between astrophysical 
parameters and make redundancy tests of general relativity (GR) \cite{Baubock:2013gna,Psaltis:2013zja}. These relations have 
been extensively investigated in the literature so far, extending their domain of validity to binary coalescence 
\cite{Maselli:2013mva}, fast-rotating bodies \cite{Doneva:2013rha,Chakrabarti:2013tca,Pappas:2013naa,Stein:2013ofa}, 
magnetars \cite{Haskell:2013vha}, and proto-NS \cite{PhysRevD.90.064026}. This analysis also led to the discovery of
new universal relations, both in GR 
\cite{PhysRevD.89.064056,Pappas:2013naa,Stein:2013ofa,Yagi:2014bxa,Chatziioannou:2014tha,Yagi:2015pkc,Majumder:2015kfa} 
and in alternative theories of gravity \cite{Sham:2013cya,Pani:2015tga,Doneva:2014faa,Doneva:2015hsa}.

The $I$-Love-$Q$ relations are described by semi-analytic fits of the following form:
\begin{equation}
\ln y=a+b\ln x+c(\ln x)^2+d(\ln x)^3+e(\ln x)^4\,, \label{ILQ}
\end{equation}
where $(a\ldots e)$ are numerical coefficients (provided in \cite{Yagi:2013awa}), while $(y,x)$ correspond 
to the trio $(\bar{I},\bar{Q},\bar{\lambda})$, normalized such that 
\begin{equation}
\bar{I}=\frac{I}{M^3}\quad\ ,\quad \bar{Q}=-\frac{Q}{M^3\chi^2}\quad\ ,
\quad \bar{\lambda}=\frac{\lambda}{M^5}\ ,
\end{equation}
where $M$ is the mass of the non-rotating configuration. 
Although the reason of the universality is not completely clear, several works have already provided 
interesting proofs to support the discovery, which can be classified into three main arguments: (i) an 
approximate version of the no-hair theorem which holds for isolated black holes in GR \cite{Hawking:1973uf}, 
(ii) the assumption that NS are modeled by isodensity contours which are self-similar ellipsoids, with large 
variations of the eccentricity being able to destroy the universality 
\cite{PhysRevD.90.063010,Yagi:2015hda,Yagi:2016ejg,PhysRevD.90.064026}, 
(iii) the stationarity of $I$-Love-$Q$ under perturbations of the EoS around the incompressible limit, suggesting 
that EoS-independence could be related to the proximity of NSs to incompressible objects 
\cite{Sham:2014kea,Chan:2015iou}.

In this regard, it is extremely interesting to analyze the validity of the $I$-Love-$Q$ relations for non-ordinary 
NSs. The next sections will be devoted to test our results for fermion and boson stars against the original 
relations derived in \cite{Yagi:2013awa}. 

\subsection{Fermion stars}

\begin{figure*}[ht]
\includegraphics[width=5.9cm]{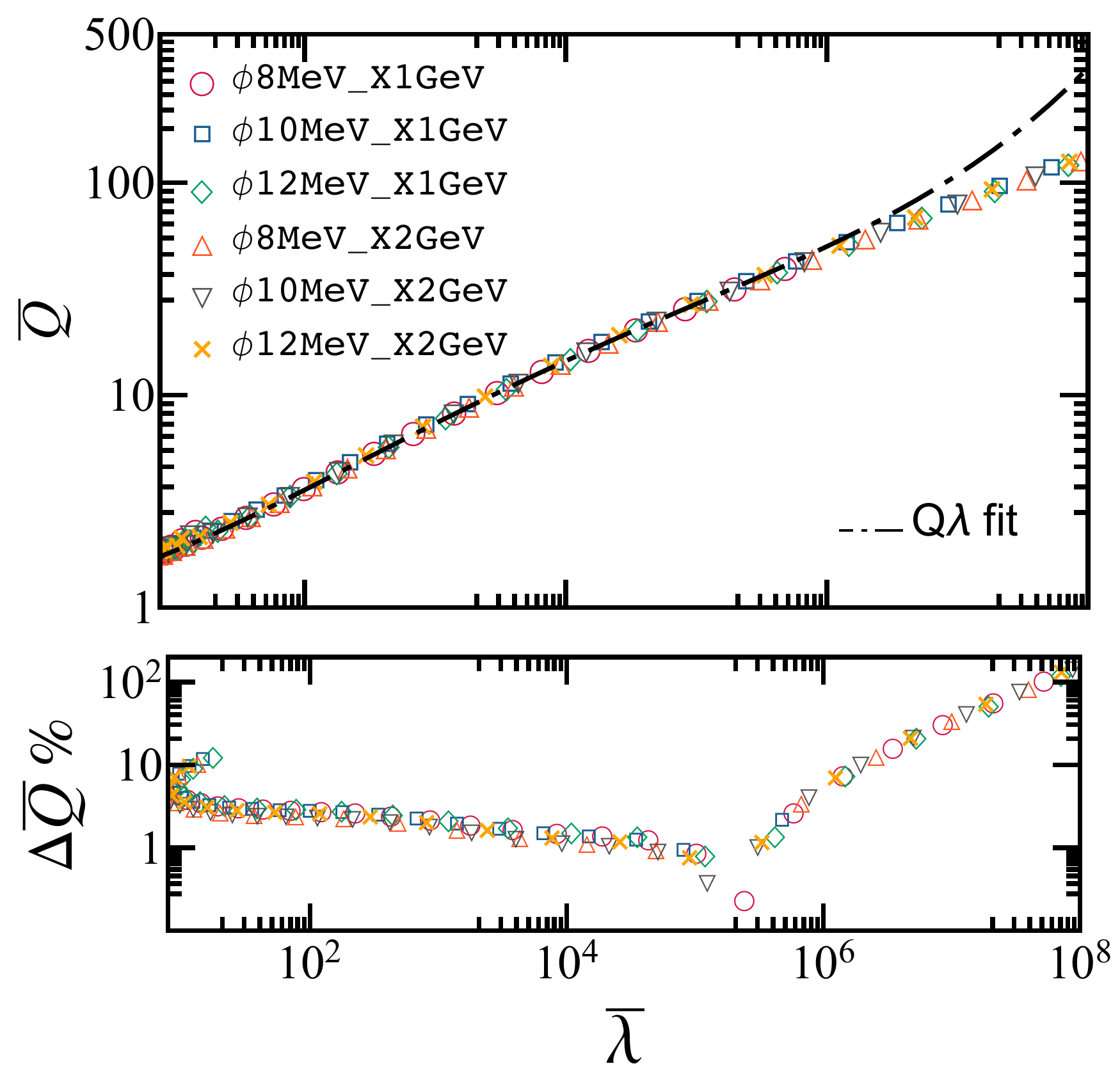}
\includegraphics[width=5.9cm]{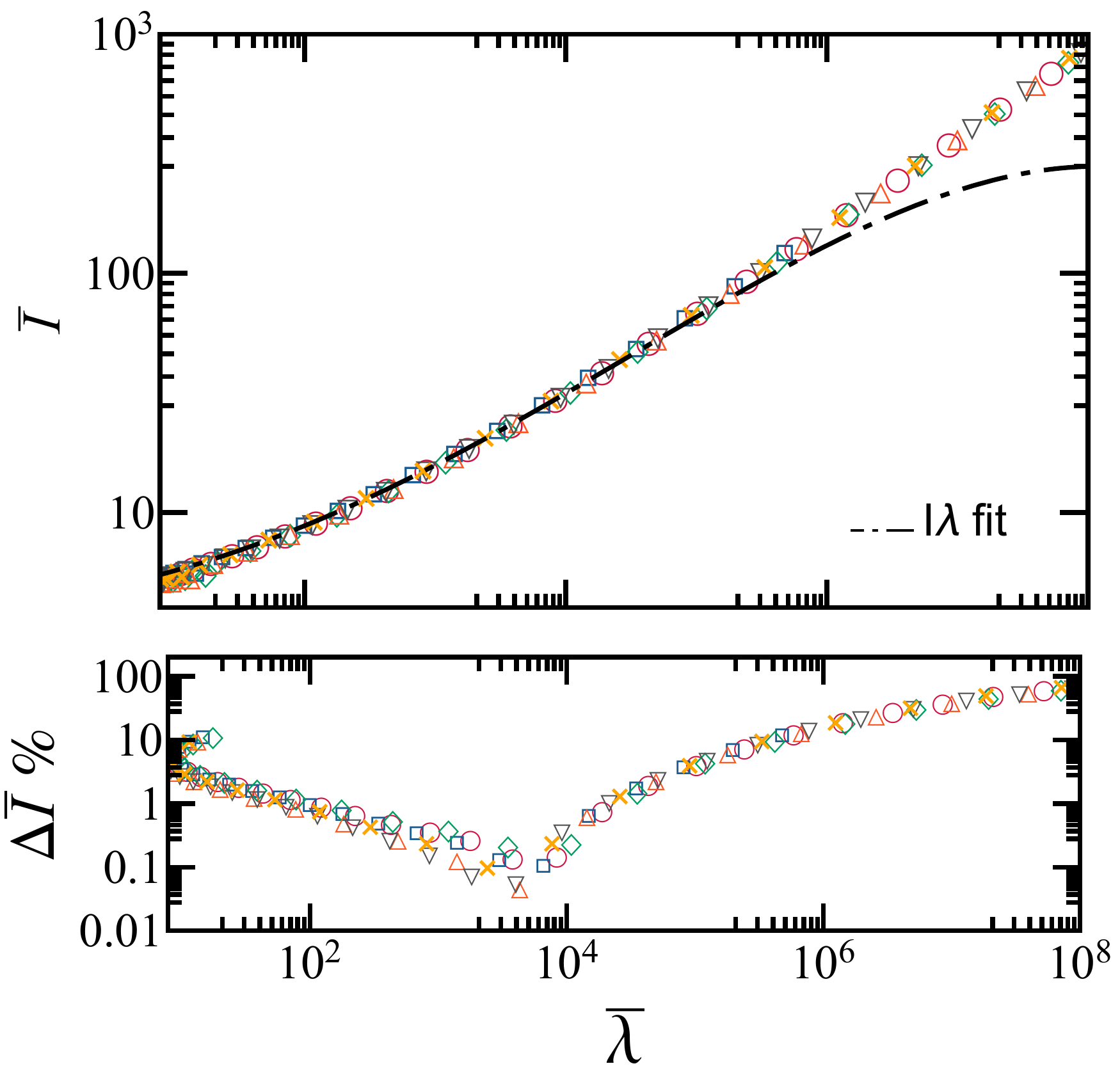}
\includegraphics[width=5.9cm]{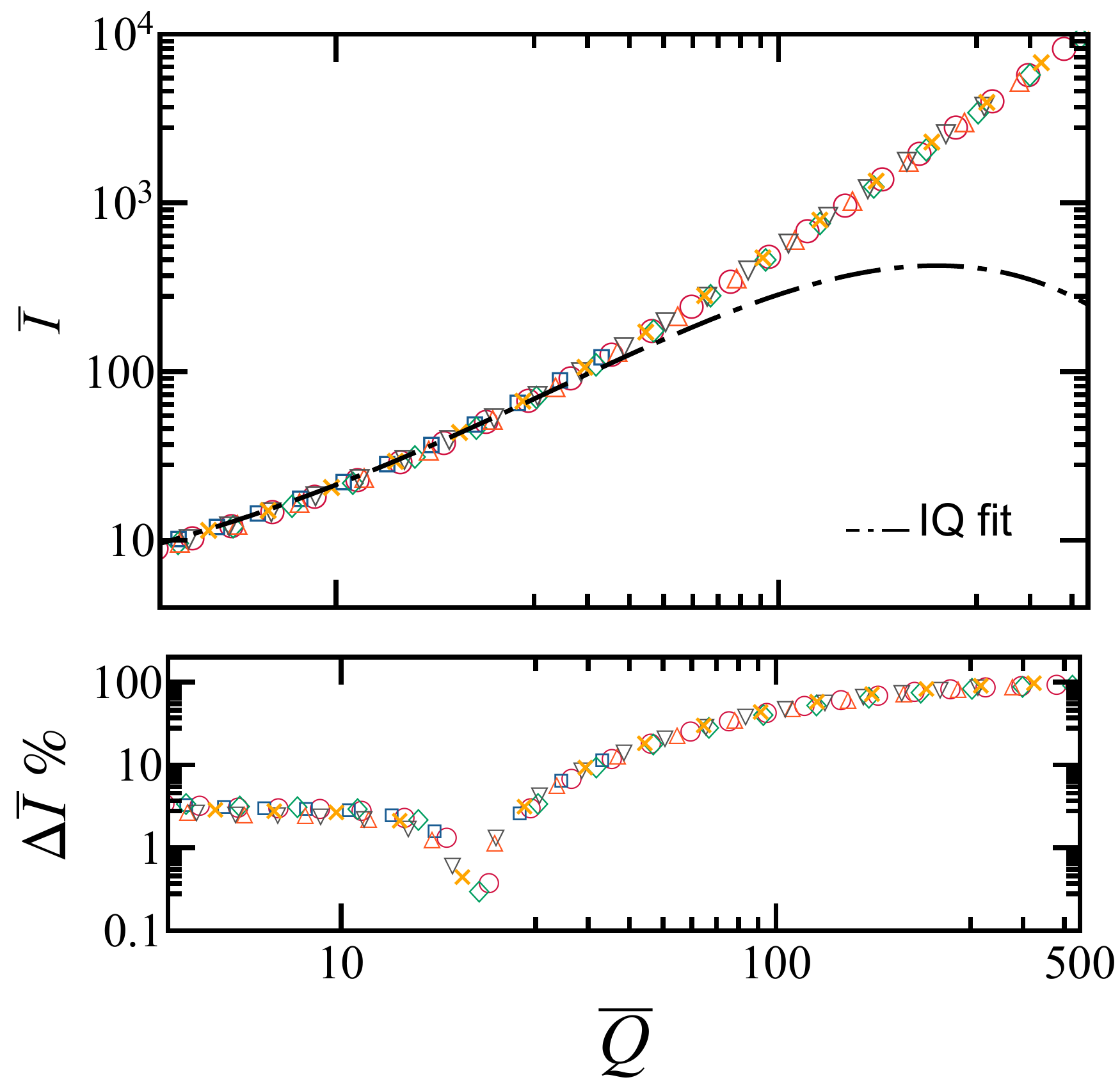}
\label{fig:ILQ1} 
\caption{$I$-Love-$Q$ relations for fermion stars with a fixed coupling constant $\alpha=10^{-3}$ and different 
values for $m_\phi$ and $m_\tn{X}$. The bottom panels show the relative percentage 
errors between the numerical data and the universal relation \eqref{ILQ} (dashed black curve).}
\end{figure*}

Figure \ref{fig:ILQ1} shows the universal relations among the $(\bar{I},\bar{\lambda},\bar{Q})$ trio for the 
fermion stars considered in this paper. Colored dots represent our results, obtained by solving the TOV 
equations for different values of $m_\phi$ and $m_\tn{X}$, while the dashed black curve refers to the 
semi-analytical fit \eqref{ILQ}. The bottom panel of each plot also shows the relative errors between the 
latter and the numerical values. We note that, in all the three cases, the original universal relations seem to 
accurately describe the data for $\bar{\lambda}\lesssim 10^5$ and $\bar{Q}\lesssim 50$, with errors less than 
$10\%$. Although these values are larger compared with those obtained for standard NSs, which are of order
$1\%$, it is still notable that the $I$-Love-$Q$ relations are able to describe such exotic objects with reasonable accuracy. 
However, larger values of the tidal deformability and quadrupole moment, corresponding to less compact stars, 
rapidly deteriorate the agreement. Nevertheless, Fig.~\ref{fig:ILQ1} also shows that it is still 
possible to interpolate the points with one single curve, which better approximates the data. 
Indeed, fitting our results with the same functional form of Eq.~\eqref{ILQ}, we find new universal relations, 
which reproduce the numerical values with accuracy better than $10\%$ within the entire spectrum 
of models. The fitting coefficients of these relations are listed in Table~\ref{table:coeffILQFS}.

\begin{table}[ht]
	\renewcommand*{\arraystretch}{1.2}
  \centering
  \begin{tabular}{ccccccc}
   \hline
    \hline
     $y$ & $x$ & $a$ & $b$ & $c$ & $d$ & $e$\\  
     \hline
 $\bar{I}$ & $\bar{\lambda}$ & 1.38 & 0.0946 & 0.0184 & -0.000514 & 5.51$\times 10^{-6}$\\
 $\bar{I}$ & $\bar{Q}$ & 1.27 & 0.632 & -0.0118 & 0.0383 &-0.0031\\
  $\bar{Q}$ & $\bar{\lambda}$ & 0.00796 & 0.272 & 0.00526 & -0.00046 &8.53$\times 10^{-6}$\\
  \hline
  \hline
  \end{tabular}  
  \caption{Best-fit coefficients of the $I$-Love-$Q$ relations for fermion stars, with the same functional 
  form of Eqs.~\eqref{ILQ}.}\label{table:coeffILQFS}
\end{table}

\subsection{Boson stars}

\begin{figure}[ht]
\includegraphics[width=8cm]{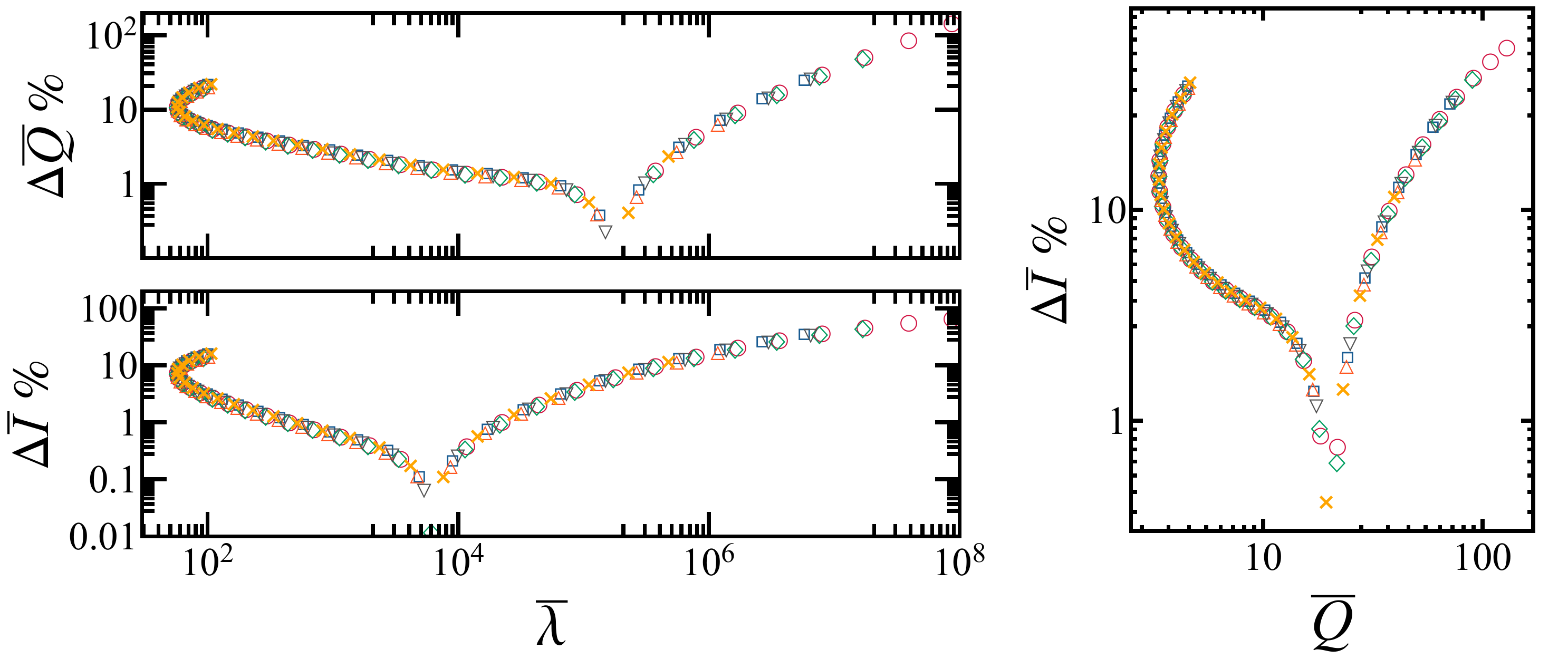}
\label{fig:ILQ2} 
\caption{Relative percentage errors between the universal relations \eqref{ILQ} and the 
$(\bar{I},\bar{Q},\bar{\lambda})$ trio computed for the boson star models considered in 
this paper.}
\end{figure}

Universal relations between $\bar{I},\bar{Q}$, and $\bar{\lambda}$ also exist for boson stars. 
As a first comparison, it is useful to analyze the agreement between our numerical results for the dark EoS 
described in Sec.~\ref{Sec:IL} and the original fits \eqref{ILQ}. Figure~\ref{fig:ILQ2} shows indeed the 
relative percentage errors between the semi-analytical predictions and the actual data. As for the fermion case, 
the largest differences occur for higher values of the quadrupole moment and of the tidal deformability. 
More precisely, for $100\lesssim \lambda \lesssim10^4$, the $I$-Love-$Q$ relations are as accurate as for standard NSs, 
with both $\Delta \bar{I}$ and $\Delta \bar{Q}$ being of the order of $1\%$ (or even less). For the $\bar{I}$--$\bar{Q}$ 
pair a reasonable agreement holds for $2\lesssim Q\lesssim 30$, with relative errors smaller than $10\%$. Outside these ranges, the discrepancies increase monotonically with $\bar{\lambda}$ and $\bar{Q}$, up to 
$100\%$. 

\begin{table}[ht]
	\renewcommand*{\arraystretch}{1.2}
  \centering
  \begin{tabular}{ccccccc}
   \hline
    \hline
     $y$ & $x$ & $a$ & $b$ & $c$ & $d$ & $e$\\  
     \hline
 $\bar{I}$ & $\bar{\lambda}$ & 0.967 & 0.245 & -0.00146 &0.000622&-0.0000181\\
 $\bar{I}$ & $\bar{Q}$ & 1.03 & 0.719 & 0.031 & 0.0153 & -0.000443\\
  $\bar{Q}$ & $\bar{\lambda}$ & 0.618 & 0.0218 & 0.0429 & -0.00284 & 0.0000615\\
  \hline
  \hline
  \end{tabular}  
  \caption{Best-fit coefficients of the new $I$-Love-$Q$ relations for boson stars. The semi-analytic relations follow 
  the form given by Eqns.~\eqref{ILQ}.}\label{table:coeffILQFS2}
\end{table}

However, as described in the previous section, it is still possible to fit all the data to obtain new 
universal relations with improved accuracy, which reproduce the numerical results of our boson stars 
with relative errors smaller than $1\%$ for $\bar{\lambda}\gtrsim 10^2$ and $\bar{Q}\gtrsim5$. 
The coefficients of these relations are listed in Table~\ref{table:coeffILQFS2}.

\section{Conclusions} \label{Sec:summary}

Self-interacting DM particles represent a well-motivated theoretical and observational scenario, potentially 
able to solve a number of important problems, currently unresolved by the CCDM paradigm. In this picture, asymmetric DM, may cluster to form stable astrophysical objects, compact enough to mimic regular NSs. 
If such objects form in nature, they offer the unique chance to explore the dark sector in extreme 
physical conditions, characterized by the strong-gravity regime. Since the  only dark matter property which is known with certainty is that it gravitates,  the existence of dark stars could reveal particle properties of DM without any non-gravitational coupling to the Standard Model.

In this paper we have investigated compact stars modeled with fermionic and bosonic DM EoS, as viable candidates 
to be tested with future electromagnetic and GW observations. By solving the stellar 
structure equations for slowly-rotating and tidally deformed bodies, we have derived the most important 
bulk properties of such dark stars, namely, their moment of inertia, tidal deformability, and quadrupole moment. Together 
with the mass and the radius, these quantities specify (at leading order) the shape of the compact object and  
its external gravitational field, and therefore they affect the orbital motion 
and the astrophysical phenomena in its close surroundings. 
We have compared these results with two extreme 
cases of soft and stiff standard nuclear EoS, showing that dark objects may cover large portions of the parameter space 
close to standard NSs. Moreover, as an explicit example, we have computed the constraints that current 
GW interferometers may already be able to set from signals emitted by binary systems 
composed of two dark stars.

Our results also show that universal relations for both the fermion and the boson 
case do exist, which connect the $(I,\lambda,Q)$ trio regardless of the specific EoS. These relations could be 
extremely useful in the near future to combine multiple observations and perform redundancy tests of the stellar model. 
The validity of the $I$-Love-$Q$ relations for dark stars seems to also confirm that besides the particle content, 
the universality may be related to the ellipsoidal isodensity contours used to model the spinning and 
tidally-deformed stars.

Finally, a simple analysis of the mass--radius profiles shows that the stellar compactness of all the considered 
models never exceeds the threshold ${\cal C}\approx 0.24$. Interestingly, we find that this is actually a more 
general result, holding for all fermionic and bosonic EoS for which a self-similar symmetry 
exists, such that the mass and the radius scale identically. We prove this statement 
analytically in the \hyperref[Sec:app]{Appendix}. As a consequence, bosonic EoS lead to stellar configurations with 
a maximum compactness ${\cal C}_\tn{max}\simeq 0.16$, independently from the coupling. In the fermionic 
case, self-similarity can be proven for non-interacting $(\alpha_\tn{X}=0)$ or strong-interacting particles, 
leading to ${\cal C}_\tn{max}\simeq0.15$ and ${\cal C}_\tn{max}\simeq0.22$, respectively. 
This result clearly indicates that dark  stars are not enough compact to act as black hole mimickers.


\begin{acknowledgments}
This work was partially supported from 'NewCompStar', COST Action MP1304.
CK is partially funded by the Danish Council for Independent Research, grant number DFF 4181-00055
\end{acknowledgments}


\renewcommand{\appendixname}{APPENDIX} 

\appendix*

\let\oldsection=\section
\def\section#1{\oldsection{\uppercase{#1}}} 

\section{Self-similarity} \label{Sec:app}

The solutions to the TOV equations in this paper exhibit self-similar symmetries, i.e., the shape (not the scale) 
of the mass--radius relation is independent of the EoS parameters (such as the particle mass or the 
interaction strength). In this Appendix we discuss the features of the EoS which leads to self-similarity, so 
that general statements about our models can be made, without scanning the entire space of parameters. 

We first define a dimensionless mass and radius:
\begin{equation}
	M_* \equiv \frac{GM}{c^2\ell}\quad
	,\quad r_* \equiv \frac{r}{\ell},
\end{equation}
where $\ell$ is a length scale. Note that ${\cal C}=M_*/r_*$ independent of $\ell$.
If we further use the scaling $\ell$ to define a dimensionless density $\rho_*$ and pressure $P_*$, the TOV 
equations can be cast in a dimensionless form, again independent of the choice of $\ell$. The dimensionless 
density and pressure are
\begin{equation}
	\rho_* \equiv \frac{\ell^2 G }{c^2}\rho\quad ,\quad P_* \equiv \frac{\ell^2G}{c^4}P.
	\label{Eq: density pressure rescaling}
\end{equation}
Using these variables, the TOV equations read
	\begin{align}
	\frac{dP_*}{dr_*} &= -\frac{M_*\rho_*}{r_*^2}\left(1+ \frac{P_*}{\rho_*} \right)\left(1+ \frac{4\pi r_*^3 P_*}{M_*} \right)\left(1- \frac{2M_*}{r_*} \right)^{-1},\notag\\
	\frac{dM_*}{dr_*} &= 4\pi r_*^2\rho_*.
\end{align}

To solve these equations, we must specify the central density $\rho_*(0)$ and the EoS $P_*=P_*(\rho_*)$. The 
mass--radius relation follows from scanning over all values of $\rho_*(0)$. The parameters of the model can only 
affect the differential equations if they enter through the EoS. Therefore, if we can choose the scaling parameter 
$\ell$, such that the EoS is independent of model parameters (when written in dimensionless variables), 
then the solution will be self-similar.

For fermionic EoS, it is in general \emph{not} possible to rescale pressure and density such that the EoS 
[Eq.~(\ref{Eq: Fermion EoS})] is independent of the model parameters $m_\tn{X}$, $\alpha_\tn{X}$ and $m_\phi$. 
However, a scaling does exist if the interactions vanish ($\alpha_\tn{X}=0$), and we may choose
\begin{equation}
	\ell_\text{FS}=	\sqrt{\frac{\hbar^3}{G c}}\frac{1}{m_\tn{X}^2}.
\end{equation}
In this case, the dimensionless EoS is $P_* = \chi\left[\xi^{-1}(\rho_*)\right]$ and the maximum compactness is 
${\cal C}\simeq 0.15$. If the interaction term in Eq.~\eqref{Eq: Fermion EoS} is so large that Fermi-repulsion is negligible, 
the EoS can also be rescaled to dimensionless form with $P_*\simeq \rho_*$. However, this limit cannot be satisfied everywhere 
in the star, since interactions are always subdominant near the surface.\footnote{Because $x\to 0$ and $x^6$ vanishes faster than 
the functions $\chi(x)$ and $\xi(x)$.} Still, in the regime of large interactions $(\alpha_\tn{X}/\hbar c) m_\tn{X}^2/m_\phi^2\gg 1 $ we find that 
the mass--radius relations become approximately self-similar, with a maximum compactness around ${\cal C}\simeq 0.22$.
Whereas, in the range of intermediate interaction strength, $(\alpha_\tn{X}/\hbar c) m_\tn{X}^2/m_\phi^2\sim 1 $, the solutions are \emph{not} 
self-similar.

Unlike the fermion star EoS with non-zero $\alpha_\tn{X}$, the boson star EoS with self-interactions in Eq.~(\ref{Eq: Boson 
star EoS}) produces exactly self-similar mass--radius relations. The relevant rescaling is
\begin{equation}
	\ell_\text{BS} = \sqrt{\frac{3\hbar^3\beta}{Gc}}\frac{1}{m_\tn{X}^2},
\end{equation}
and the dimensionless EoS is given by
\begin{equation}
	P_* = \frac{1}{3}\left(\sqrt{1+\rho_*} -1\right)^2.
\end{equation}
We find the maximum compactness with this EoS to be ${\cal C}\simeq 0.16$.

%
\bibliography{bibnote}

\begin{thebibliography}{167}%
\makeatletter
\providecommand \@ifxundefined [1]{%
 \@ifx{#1\undefined}
}%
\providecommand \@ifnum [1]{%
 \ifnum #1\expandafter \@firstoftwo
 \else \expandafter \@secondoftwo
 \fi
}%
\providecommand \@ifx [1]{%
 \ifx #1\expandafter \@firstoftwo
 \else \expandafter \@secondoftwo
 \fi
}%
\providecommand \natexlab [1]{#1}%
\providecommand \enquote  [1]{``#1''}%
\providecommand \bibnamefont  [1]{#1}%
\providecommand \bibfnamefont [1]{#1}%
\providecommand \citenamefont [1]{#1}%
\providecommand \href@noop [0]{\@secondoftwo}%
\providecommand \href [0]{\begingroup \@sanitize@url \@href}%
\providecommand \@href[1]{\@@startlink{#1}\@@href}%
\providecommand \@@href[1]{\endgroup#1\@@endlink}%
\providecommand \@sanitize@url [0]{\catcode `\\12\catcode `\$12\catcode
  `\&12\catcode `\#12\catcode `\^12\catcode `\_12\catcode `\%12\relax}%
\providecommand \@@startlink[1]{}%
\providecommand \@@endlink[0]{}%
\providecommand \url  [0]{\begingroup\@sanitize@url \@url }%
\providecommand \@url [1]{\endgroup\@href {#1}{\urlprefix }}%
\providecommand \urlprefix  [0]{URL }%
\providecommand \Eprint [0]{\href }%
\providecommand \doibase [0]{http://dx.doi.org/}%
\providecommand \selectlanguage [0]{\@gobble}%
\providecommand \bibinfo  [0]{\@secondoftwo}%
\providecommand \bibfield  [0]{\@secondoftwo}%
\providecommand \translation [1]{[#1]}%
\providecommand \BibitemOpen [0]{}%
\providecommand \bibitemStop [0]{}%
\providecommand \bibitemNoStop [0]{.\EOS\space}%
\providecommand \EOS [0]{\spacefactor3000\relax}%
\providecommand \BibitemShut  [1]{\csname bibitem#1\endcsname}%
\let\auto@bib@innerbib\@empty
\bibitem [{\citenamefont {Moore}(1994)}]{Moore:1994yx}%
  \BibitemOpen
  \bibfield  {author} {\bibinfo {author} {\bibfnamefont {B.}~\bibnamefont
  {Moore}},\ }\href {\doibase 10.1038/370629a0} {\bibfield  {journal} {\bibinfo
   {journal} {Nature}\ }\textbf {\bibinfo {volume} {370}},\ \bibinfo {pages}
  {629} (\bibinfo {year} {1994})}\BibitemShut {NoStop}%
\bibitem [{\citenamefont {Flores}\ and\ \citenamefont
  {Primack}(1994)}]{Flores:1994gz}%
  \BibitemOpen
  \bibfield  {author} {\bibinfo {author} {\bibfnamefont {R.~A.}\ \bibnamefont
  {Flores}}\ and\ \bibinfo {author} {\bibfnamefont {J.~R.}\ \bibnamefont
  {Primack}},\ }\href {\doibase 10.1086/187350} {\bibfield  {journal} {\bibinfo
   {journal} {Astrophys. J.}\ }\textbf {\bibinfo {volume} {427}},\ \bibinfo
  {pages} {L1} (\bibinfo {year} {1994})},\ \Eprint
  {http://arxiv.org/abs/astro-ph/9402004} {arXiv:astro-ph/9402004 [astro-ph]}
  \BibitemShut {NoStop}%
\bibitem [{\citenamefont {Navarro}\ \emph {et~al.}(1997)\citenamefont
  {Navarro}, \citenamefont {Frenk},\ and\ \citenamefont
  {White}}]{Navarro:1996gj}%
  \BibitemOpen
  \bibfield  {author} {\bibinfo {author} {\bibfnamefont {J.~F.}\ \bibnamefont
  {Navarro}}, \bibinfo {author} {\bibfnamefont {C.~S.}\ \bibnamefont {Frenk}},
  \ and\ \bibinfo {author} {\bibfnamefont {S.~D.~M.}\ \bibnamefont {White}},\
  }\href {\doibase 10.1086/304888} {\bibfield  {journal} {\bibinfo  {journal}
  {Astrophys. J.}\ }\textbf {\bibinfo {volume} {490}},\ \bibinfo {pages} {493}
  (\bibinfo {year} {1997})},\ \Eprint {http://arxiv.org/abs/astro-ph/9611107}
  {arXiv:astro-ph/9611107 [astro-ph]} \BibitemShut {NoStop}%
\bibitem [{\citenamefont {Klypin}\ \emph {et~al.}(1999)\citenamefont {Klypin},
  \citenamefont {Kravtsov}, \citenamefont {Valenzuela},\ and\ \citenamefont
  {Prada}}]{Klypin:1999uc}%
  \BibitemOpen
  \bibfield  {author} {\bibinfo {author} {\bibfnamefont {A.~A.}\ \bibnamefont
  {Klypin}}, \bibinfo {author} {\bibfnamefont {A.~V.}\ \bibnamefont
  {Kravtsov}}, \bibinfo {author} {\bibfnamefont {O.}~\bibnamefont
  {Valenzuela}}, \ and\ \bibinfo {author} {\bibfnamefont {F.}~\bibnamefont
  {Prada}},\ }\href {\doibase 10.1086/307643} {\bibfield  {journal} {\bibinfo
  {journal} {Astrophys. J.}\ }\textbf {\bibinfo {volume} {522}},\ \bibinfo
  {pages} {82} (\bibinfo {year} {1999})},\ \Eprint
  {http://arxiv.org/abs/astro-ph/9901240} {arXiv:astro-ph/9901240 [astro-ph]}
  \BibitemShut {NoStop}%
\bibitem [{\citenamefont {Moore}\ \emph {et~al.}(1999)\citenamefont {Moore},
  \citenamefont {Ghigna}, \citenamefont {Governato}, \citenamefont {Lake},
  \citenamefont {Quinn}, \citenamefont {Stadel},\ and\ \citenamefont
  {Tozzi}}]{Moore:1999nt}%
  \BibitemOpen
  \bibfield  {author} {\bibinfo {author} {\bibfnamefont {B.}~\bibnamefont
  {Moore}}, \bibinfo {author} {\bibfnamefont {S.}~\bibnamefont {Ghigna}},
  \bibinfo {author} {\bibfnamefont {F.}~\bibnamefont {Governato}}, \bibinfo
  {author} {\bibfnamefont {G.}~\bibnamefont {Lake}}, \bibinfo {author}
  {\bibfnamefont {T.~R.}\ \bibnamefont {Quinn}}, \bibinfo {author}
  {\bibfnamefont {J.}~\bibnamefont {Stadel}}, \ and\ \bibinfo {author}
  {\bibfnamefont {P.}~\bibnamefont {Tozzi}},\ }\href {\doibase 10.1086/312287}
  {\bibfield  {journal} {\bibinfo  {journal} {Astrophys. J.}\ }\textbf
  {\bibinfo {volume} {524}},\ \bibinfo {pages} {L19} (\bibinfo {year}
  {1999})},\ \Eprint {http://arxiv.org/abs/astro-ph/9907411}
  {arXiv:astro-ph/9907411 [astro-ph]} \BibitemShut {NoStop}%
\bibitem [{\citenamefont {Kauffmann}\ \emph {et~al.}(1993)\citenamefont
  {Kauffmann}, \citenamefont {White},\ and\ \citenamefont
  {Guiderdoni}}]{Kauffmann:1993gv}%
  \BibitemOpen
  \bibfield  {author} {\bibinfo {author} {\bibfnamefont {G.}~\bibnamefont
  {Kauffmann}}, \bibinfo {author} {\bibfnamefont {S.~D.~M.}\ \bibnamefont
  {White}}, \ and\ \bibinfo {author} {\bibfnamefont {B.}~\bibnamefont
  {Guiderdoni}},\ }\href@noop {} {\bibfield  {journal} {\bibinfo  {journal}
  {Mon. Not. Roy. Astron. Soc.}\ }\textbf {\bibinfo {volume} {264}},\ \bibinfo
  {pages} {201} (\bibinfo {year} {1993})}\BibitemShut {NoStop}%
\bibitem [{\citenamefont {Boylan-Kolchin}\ \emph {et~al.}(2011)\citenamefont
  {Boylan-Kolchin}, \citenamefont {Bullock},\ and\ \citenamefont
  {Kaplinghat}}]{BoylanKolchin:2011de}%
  \BibitemOpen
  \bibfield  {author} {\bibinfo {author} {\bibfnamefont {M.}~\bibnamefont
  {Boylan-Kolchin}}, \bibinfo {author} {\bibfnamefont {J.~S.}\ \bibnamefont
  {Bullock}}, \ and\ \bibinfo {author} {\bibfnamefont {M.}~\bibnamefont
  {Kaplinghat}},\ }\href {\doibase 10.1111/j.1745-3933.2011.01074.x} {\bibfield
   {journal} {\bibinfo  {journal} {Mon. Not. Roy. Astron. Soc.}\ }\textbf
  {\bibinfo {volume} {415}},\ \bibinfo {pages} {L40} (\bibinfo {year}
  {2011})},\ \Eprint {http://arxiv.org/abs/1103.0007} {arXiv:1103.0007
  [astro-ph.CO]} \BibitemShut {NoStop}%
\bibitem [{\citenamefont {Oh}\ \emph {et~al.}(2011)\citenamefont {Oh},
  \citenamefont {Brook}, \citenamefont {Governato}, \citenamefont {Brinks},
  \citenamefont {Mayer}, \citenamefont {de~Blok}, \citenamefont {Brooks},\ and\
  \citenamefont {Walter}}]{Oh:2010mc}%
  \BibitemOpen
  \bibfield  {author} {\bibinfo {author} {\bibfnamefont {S.-H.}\ \bibnamefont
  {Oh}}, \bibinfo {author} {\bibfnamefont {C.}~\bibnamefont {Brook}}, \bibinfo
  {author} {\bibfnamefont {F.}~\bibnamefont {Governato}}, \bibinfo {author}
  {\bibfnamefont {E.}~\bibnamefont {Brinks}}, \bibinfo {author} {\bibfnamefont
  {L.}~\bibnamefont {Mayer}}, \bibinfo {author} {\bibfnamefont {W.~J.~G.}\
  \bibnamefont {de~Blok}}, \bibinfo {author} {\bibfnamefont {A.}~\bibnamefont
  {Brooks}}, \ and\ \bibinfo {author} {\bibfnamefont {F.}~\bibnamefont
  {Walter}},\ }\href {\doibase 10.1088/0004-6256/142/1/24} {\bibfield
  {journal} {\bibinfo  {journal} {Astron. J.}\ }\textbf {\bibinfo {volume}
  {142}},\ \bibinfo {pages} {24} (\bibinfo {year} {2011})},\ \Eprint
  {http://arxiv.org/abs/1011.2777} {arXiv:1011.2777 [astro-ph.CO]} \BibitemShut
  {NoStop}%
\bibitem [{\citenamefont {Brook}\ \emph {et~al.}(2012)\citenamefont {Brook},
  \citenamefont {Stinson}, \citenamefont {Gibson}, \citenamefont {Roskar},
  \citenamefont {Wadsley},\ and\ \citenamefont {Quinn}}]{Brook:2011nz}%
  \BibitemOpen
  \bibfield  {author} {\bibinfo {author} {\bibfnamefont {C.~B.}\ \bibnamefont
  {Brook}}, \bibinfo {author} {\bibfnamefont {G.}~\bibnamefont {Stinson}},
  \bibinfo {author} {\bibfnamefont {B.~K.}\ \bibnamefont {Gibson}}, \bibinfo
  {author} {\bibfnamefont {R.}~\bibnamefont {Roskar}}, \bibinfo {author}
  {\bibfnamefont {J.}~\bibnamefont {Wadsley}}, \ and\ \bibinfo {author}
  {\bibfnamefont {T.}~\bibnamefont {Quinn}},\ }\href {\doibase
  10.1111/j.1365-2966.2011.19740.x} {\bibfield  {journal} {\bibinfo  {journal}
  {Mon. Not. Roy. Astron. Soc.}\ }\textbf {\bibinfo {volume} {419}},\ \bibinfo
  {pages} {771} (\bibinfo {year} {2012})},\ \Eprint
  {http://arxiv.org/abs/1105.2562} {arXiv:1105.2562 [astro-ph.CO]} \BibitemShut
  {NoStop}%
\bibitem [{\citenamefont {Pontzen}\ and\ \citenamefont
  {Governato}(2012)}]{Pontzen:2011ty}%
  \BibitemOpen
  \bibfield  {author} {\bibinfo {author} {\bibfnamefont {A.}~\bibnamefont
  {Pontzen}}\ and\ \bibinfo {author} {\bibfnamefont {F.}~\bibnamefont
  {Governato}},\ }\href {\doibase 10.1111/j.1365-2966.2012.20571.x} {\bibfield
  {journal} {\bibinfo  {journal} {Mon. Not. Roy. Astron. Soc.}\ }\textbf
  {\bibinfo {volume} {421}},\ \bibinfo {pages} {3464} (\bibinfo {year}
  {2012})},\ \Eprint {http://arxiv.org/abs/1106.0499} {arXiv:1106.0499
  [astro-ph.CO]} \BibitemShut {NoStop}%
\bibitem [{\citenamefont {Governato}\ \emph {et~al.}(2012)\citenamefont
  {Governato}, \citenamefont {Zolotov}, \citenamefont {Pontzen}, \citenamefont
  {Christensen}, \citenamefont {Oh}, \citenamefont {Brooks}, \citenamefont
  {Quinn}, \citenamefont {Shen},\ and\ \citenamefont
  {Wadsley}}]{Governato:2012fa}%
  \BibitemOpen
  \bibfield  {author} {\bibinfo {author} {\bibfnamefont {F.}~\bibnamefont
  {Governato}}, \bibinfo {author} {\bibfnamefont {A.}~\bibnamefont {Zolotov}},
  \bibinfo {author} {\bibfnamefont {A.}~\bibnamefont {Pontzen}}, \bibinfo
  {author} {\bibfnamefont {C.}~\bibnamefont {Christensen}}, \bibinfo {author}
  {\bibfnamefont {S.~H.}\ \bibnamefont {Oh}}, \bibinfo {author} {\bibfnamefont
  {A.~M.}\ \bibnamefont {Brooks}}, \bibinfo {author} {\bibfnamefont
  {T.}~\bibnamefont {Quinn}}, \bibinfo {author} {\bibfnamefont
  {S.}~\bibnamefont {Shen}}, \ and\ \bibinfo {author} {\bibfnamefont
  {J.}~\bibnamefont {Wadsley}},\ }\href {\doibase
  10.1111/j.1365-2966.2012.20696.x} {\bibfield  {journal} {\bibinfo  {journal}
  {Mon. Not. Roy. Astron. Soc.}\ }\textbf {\bibinfo {volume} {422}},\ \bibinfo
  {pages} {1231} (\bibinfo {year} {2012})},\ \Eprint
  {http://arxiv.org/abs/1202.0554} {arXiv:1202.0554 [astro-ph.CO]} \BibitemShut
  {NoStop}%
\bibitem [{\citenamefont {Liu}\ \emph {et~al.}(2011)\citenamefont {Liu},
  \citenamefont {Gerke}, \citenamefont {Wechsler}, \citenamefont {Behroozi},\
  and\ \citenamefont {Busha}}]{Liu:2010tn}%
  \BibitemOpen
  \bibfield  {author} {\bibinfo {author} {\bibfnamefont {L.}~\bibnamefont
  {Liu}}, \bibinfo {author} {\bibfnamefont {B.~F.}\ \bibnamefont {Gerke}},
  \bibinfo {author} {\bibfnamefont {R.~H.}\ \bibnamefont {Wechsler}}, \bibinfo
  {author} {\bibfnamefont {P.~S.}\ \bibnamefont {Behroozi}}, \ and\ \bibinfo
  {author} {\bibfnamefont {M.~T.}\ \bibnamefont {Busha}},\ }\href {\doibase
  10.1088/0004-637X/733/1/62} {\bibfield  {journal} {\bibinfo  {journal}
  {Astrophys. J.}\ }\textbf {\bibinfo {volume} {733}},\ \bibinfo {pages} {62}
  (\bibinfo {year} {2011})},\ \Eprint {http://arxiv.org/abs/1011.2255}
  {arXiv:1011.2255 [astro-ph.CO]} \BibitemShut {NoStop}%
\bibitem [{\citenamefont {Tollerud}\ \emph {et~al.}(2011)\citenamefont
  {Tollerud}, \citenamefont {Boylan-Kolchin}, \citenamefont {Barton},
  \citenamefont {Bullock},\ and\ \citenamefont {Trinh}}]{Tollerud:2011wt}%
  \BibitemOpen
  \bibfield  {author} {\bibinfo {author} {\bibfnamefont {E.~J.}\ \bibnamefont
  {Tollerud}}, \bibinfo {author} {\bibfnamefont {M.}~\bibnamefont
  {Boylan-Kolchin}}, \bibinfo {author} {\bibfnamefont {E.~J.}\ \bibnamefont
  {Barton}}, \bibinfo {author} {\bibfnamefont {J.~S.}\ \bibnamefont {Bullock}},
  \ and\ \bibinfo {author} {\bibfnamefont {C.~Q.}\ \bibnamefont {Trinh}},\
  }\href {\doibase 10.1088/0004-637X/738/1/102} {\bibfield  {journal} {\bibinfo
   {journal} {Astrophys. J.}\ }\textbf {\bibinfo {volume} {738}},\ \bibinfo
  {pages} {102} (\bibinfo {year} {2011})},\ \Eprint
  {http://arxiv.org/abs/1103.1875} {arXiv:1103.1875 [astro-ph.CO]} \BibitemShut
  {NoStop}%
\bibitem [{\citenamefont {Strigari}\ and\ \citenamefont
  {Wechsler}(2012)}]{Strigari:2011ps}%
  \BibitemOpen
  \bibfield  {author} {\bibinfo {author} {\bibfnamefont {L.~E.}\ \bibnamefont
  {Strigari}}\ and\ \bibinfo {author} {\bibfnamefont {R.~H.}\ \bibnamefont
  {Wechsler}},\ }\href {\doibase 10.1088/0004-637X/749/1/75} {\bibfield
  {journal} {\bibinfo  {journal} {Astrophys. J.}\ }\textbf {\bibinfo {volume}
  {749}},\ \bibinfo {pages} {75} (\bibinfo {year} {2012})},\ \Eprint
  {http://arxiv.org/abs/1111.2611} {arXiv:1111.2611 [astro-ph.CO]} \BibitemShut
  {NoStop}%
\bibitem [{\citenamefont {Vogelsberger}\ \emph {et~al.}(2012)\citenamefont
  {Vogelsberger}, \citenamefont {Zavala},\ and\ \citenamefont
  {Loeb}}]{Vogelsberger:2012ku}%
  \BibitemOpen
  \bibfield  {author} {\bibinfo {author} {\bibfnamefont {M.}~\bibnamefont
  {Vogelsberger}}, \bibinfo {author} {\bibfnamefont {J.}~\bibnamefont
  {Zavala}}, \ and\ \bibinfo {author} {\bibfnamefont {A.}~\bibnamefont
  {Loeb}},\ }\href {\doibase 10.1111/j.1365-2966.2012.21182.x} {\bibfield
  {journal} {\bibinfo  {journal} {Mon. Not. Roy. Astron. Soc.}\ }\textbf
  {\bibinfo {volume} {423}},\ \bibinfo {pages} {3740} (\bibinfo {year}
  {2012})},\ \Eprint {http://arxiv.org/abs/1201.5892} {arXiv:1201.5892
  [astro-ph.CO]} \BibitemShut {NoStop}%
\bibitem [{\citenamefont {Rocha}\ \emph {et~al.}(2013)\citenamefont {Rocha},
  \citenamefont {Peter}, \citenamefont {Bullock}, \citenamefont {Kaplinghat},
  \citenamefont {Garrison-Kimmel}, \citenamefont {Onorbe},\ and\ \citenamefont
  {Moustakas}}]{Rocha:2012jg}%
  \BibitemOpen
  \bibfield  {author} {\bibinfo {author} {\bibfnamefont {M.}~\bibnamefont
  {Rocha}}, \bibinfo {author} {\bibfnamefont {A.~H.~G.}\ \bibnamefont {Peter}},
  \bibinfo {author} {\bibfnamefont {J.~S.}\ \bibnamefont {Bullock}}, \bibinfo
  {author} {\bibfnamefont {M.}~\bibnamefont {Kaplinghat}}, \bibinfo {author}
  {\bibfnamefont {S.}~\bibnamefont {Garrison-Kimmel}}, \bibinfo {author}
  {\bibfnamefont {J.}~\bibnamefont {Onorbe}}, \ and\ \bibinfo {author}
  {\bibfnamefont {L.~A.}\ \bibnamefont {Moustakas}},\ }\href {\doibase
  10.1093/mnras/sts514} {\bibfield  {journal} {\bibinfo  {journal} {Mon. Not.
  Roy. Astron. Soc.}\ }\textbf {\bibinfo {volume} {430}},\ \bibinfo {pages}
  {81} (\bibinfo {year} {2013})},\ \Eprint {http://arxiv.org/abs/1208.3025}
  {arXiv:1208.3025 [astro-ph.CO]} \BibitemShut {NoStop}%
\bibitem [{\citenamefont {Zavala}\ \emph {et~al.}(2013)\citenamefont {Zavala},
  \citenamefont {Vogelsberger},\ and\ \citenamefont {Walker}}]{Zavala:2012us}%
  \BibitemOpen
  \bibfield  {author} {\bibinfo {author} {\bibfnamefont {J.}~\bibnamefont
  {Zavala}}, \bibinfo {author} {\bibfnamefont {M.}~\bibnamefont
  {Vogelsberger}}, \ and\ \bibinfo {author} {\bibfnamefont {M.~G.}\
  \bibnamefont {Walker}},\ }\href {\doibase 10.1093/mnrasl/sls053} {\bibfield
  {journal} {\bibinfo  {journal} {Monthly Notices of the Royal Astronomical
  Society: Letters}\ }\textbf {\bibinfo {volume} {431}},\ \bibinfo {pages}
  {L20} (\bibinfo {year} {2013})},\ \Eprint {http://arxiv.org/abs/1211.6426}
  {arXiv:1211.6426 [astro-ph.CO]} \BibitemShut {NoStop}%
\bibitem [{\citenamefont {Peter}\ \emph {et~al.}(2013)\citenamefont {Peter},
  \citenamefont {Rocha}, \citenamefont {Bullock},\ and\ \citenamefont
  {Kaplinghat}}]{Peter:2012jh}%
  \BibitemOpen
  \bibfield  {author} {\bibinfo {author} {\bibfnamefont {A.~H.~G.}\
  \bibnamefont {Peter}}, \bibinfo {author} {\bibfnamefont {M.}~\bibnamefont
  {Rocha}}, \bibinfo {author} {\bibfnamefont {J.~S.}\ \bibnamefont {Bullock}},
  \ and\ \bibinfo {author} {\bibfnamefont {M.}~\bibnamefont {Kaplinghat}},\
  }\href {\doibase 10.1093/mnras/sts535} {\bibfield  {journal} {\bibinfo
  {journal} {Mon. Not. Roy. Astron. Soc.}\ }\textbf {\bibinfo {volume} {430}},\
  \bibinfo {pages} {105} (\bibinfo {year} {2013})},\ \Eprint
  {http://arxiv.org/abs/1208.3026} {arXiv:1208.3026 [astro-ph.CO]} \BibitemShut
  {NoStop}%
\bibitem [{\citenamefont {Spergel}\ and\ \citenamefont
  {Steinhardt}(2000)}]{Spergel:1999mh}%
  \BibitemOpen
  \bibfield  {author} {\bibinfo {author} {\bibfnamefont {D.~N.}\ \bibnamefont
  {Spergel}}\ and\ \bibinfo {author} {\bibfnamefont {P.~J.}\ \bibnamefont
  {Steinhardt}},\ }\href {\doibase 10.1103/PhysRevLett.84.3760} {\bibfield
  {journal} {\bibinfo  {journal} {Phys. Rev. Lett.}\ }\textbf {\bibinfo
  {volume} {84}},\ \bibinfo {pages} {3760} (\bibinfo {year} {2000})},\ \Eprint
  {http://arxiv.org/abs/astro-ph/9909386} {arXiv:astro-ph/9909386 [astro-ph]}
  \BibitemShut {NoStop}%
\bibitem [{\citenamefont {Wandelt}\ \emph {et~al.}(2000)\citenamefont
  {Wandelt}, \citenamefont {Dave}, \citenamefont {Farrar}, \citenamefont
  {McGuire}, \citenamefont {Spergel},\ and\ \citenamefont
  {Steinhardt}}]{Wandelt:2000ad}%
  \BibitemOpen
  \bibfield  {author} {\bibinfo {author} {\bibfnamefont {B.~D.}\ \bibnamefont
  {Wandelt}}, \bibinfo {author} {\bibfnamefont {R.}~\bibnamefont {Dave}},
  \bibinfo {author} {\bibfnamefont {G.~R.}\ \bibnamefont {Farrar}}, \bibinfo
  {author} {\bibfnamefont {P.~C.}\ \bibnamefont {McGuire}}, \bibinfo {author}
  {\bibfnamefont {D.~N.}\ \bibnamefont {Spergel}}, \ and\ \bibinfo {author}
  {\bibfnamefont {P.~J.}\ \bibnamefont {Steinhardt}},\ }in\ \href
  {http://www.slac.stanford.edu/spires/find/books/www?cl=QB461:I57:2000} {\emph
  {\bibinfo {booktitle} {{Sources and detection of dark matter and dark energy
  in the universe. Proceedings, 4th International Symposium, DM 2000, Marina
  del Rey, USA, February 23-25, 2000}}}}\ (\bibinfo {year} {2000})\ pp.\
  \bibinfo {pages} {263--274},\ \Eprint {http://arxiv.org/abs/astro-ph/0006344}
  {arXiv:astro-ph/0006344 [astro-ph]} \BibitemShut {NoStop}%
\bibitem [{\citenamefont {Faraggi}\ and\ \citenamefont
  {Pospelov}(2002)}]{Faraggi:2000pv}%
  \BibitemOpen
  \bibfield  {author} {\bibinfo {author} {\bibfnamefont {A.~E.}\ \bibnamefont
  {Faraggi}}\ and\ \bibinfo {author} {\bibfnamefont {M.}~\bibnamefont
  {Pospelov}},\ }\href {\doibase 10.1016/S0927-6505(01)00121-9} {\bibfield
  {journal} {\bibinfo  {journal} {Astropart. Phys.}\ }\textbf {\bibinfo
  {volume} {16}},\ \bibinfo {pages} {451} (\bibinfo {year} {2002})},\ \Eprint
  {http://arxiv.org/abs/hep-ph/0008223} {arXiv:hep-ph/0008223 [hep-ph]}
  \BibitemShut {NoStop}%
\bibitem [{\citenamefont {Mohapatra}\ \emph {et~al.}(2002)\citenamefont
  {Mohapatra}, \citenamefont {Nussinov},\ and\ \citenamefont
  {Teplitz}}]{Mohapatra:2001sx}%
  \BibitemOpen
  \bibfield  {author} {\bibinfo {author} {\bibfnamefont {R.~N.}\ \bibnamefont
  {Mohapatra}}, \bibinfo {author} {\bibfnamefont {S.}~\bibnamefont {Nussinov}},
  \ and\ \bibinfo {author} {\bibfnamefont {V.~L.}\ \bibnamefont {Teplitz}},\
  }\href {\doibase 10.1103/PhysRevD.66.063002} {\bibfield  {journal} {\bibinfo
  {journal} {Phys. Rev.}\ }\textbf {\bibinfo {volume} {D66}},\ \bibinfo {pages}
  {063002} (\bibinfo {year} {2002})},\ \Eprint
  {http://arxiv.org/abs/hep-ph/0111381} {arXiv:hep-ph/0111381 [hep-ph]}
  \BibitemShut {NoStop}%
\bibitem [{\citenamefont {Kusenko}\ and\ \citenamefont
  {Steinhardt}(2001)}]{Kusenko:2001vu}%
  \BibitemOpen
  \bibfield  {author} {\bibinfo {author} {\bibfnamefont {A.}~\bibnamefont
  {Kusenko}}\ and\ \bibinfo {author} {\bibfnamefont {P.~J.}\ \bibnamefont
  {Steinhardt}},\ }\href {\doibase 10.1103/PhysRevLett.87.141301} {\bibfield
  {journal} {\bibinfo  {journal} {Phys. Rev. Lett.}\ }\textbf {\bibinfo
  {volume} {87}},\ \bibinfo {pages} {141301} (\bibinfo {year} {2001})},\
  \Eprint {http://arxiv.org/abs/astro-ph/0106008} {arXiv:astro-ph/0106008
  [astro-ph]} \BibitemShut {NoStop}%
\bibitem [{\citenamefont {Loeb}\ and\ \citenamefont
  {Weiner}(2011)}]{Loeb:2010gj}%
  \BibitemOpen
  \bibfield  {author} {\bibinfo {author} {\bibfnamefont {A.}~\bibnamefont
  {Loeb}}\ and\ \bibinfo {author} {\bibfnamefont {N.}~\bibnamefont {Weiner}},\
  }\href {\doibase 10.1103/PhysRevLett.106.171302} {\bibfield  {journal}
  {\bibinfo  {journal} {Phys. Rev. Lett.}\ }\textbf {\bibinfo {volume} {106}},\
  \bibinfo {pages} {171302} (\bibinfo {year} {2011})},\ \Eprint
  {http://arxiv.org/abs/1011.6374} {arXiv:1011.6374 [astro-ph.CO]} \BibitemShut
  {NoStop}%
\bibitem [{\citenamefont {Kouvaris}(2012)}]{Kouvaris:2011gb}%
  \BibitemOpen
  \bibfield  {author} {\bibinfo {author} {\bibfnamefont {C.}~\bibnamefont
  {Kouvaris}},\ }\href {\doibase 10.1103/PhysRevLett.108.191301} {\bibfield
  {journal} {\bibinfo  {journal} {Phys. Rev. Lett.}\ }\textbf {\bibinfo
  {volume} {108}},\ \bibinfo {pages} {191301} (\bibinfo {year} {2012})},\
  \Eprint {http://arxiv.org/abs/1111.4364} {arXiv:1111.4364 [astro-ph.CO]}
  \BibitemShut {NoStop}%
\bibitem [{\citenamefont {Vogelsberger}\ and\ \citenamefont
  {Zavala}(2013)}]{Vogelsberger:2012sa}%
  \BibitemOpen
  \bibfield  {author} {\bibinfo {author} {\bibfnamefont {M.}~\bibnamefont
  {Vogelsberger}}\ and\ \bibinfo {author} {\bibfnamefont {J.}~\bibnamefont
  {Zavala}},\ }\href {\doibase 10.1093/mnras/sts712} {\bibfield  {journal}
  {\bibinfo  {journal} {Mon. Not. Roy. Astron. Soc.}\ }\textbf {\bibinfo
  {volume} {430}},\ \bibinfo {pages} {1722} (\bibinfo {year} {2013})},\ \Eprint
  {http://arxiv.org/abs/1211.1377} {arXiv:1211.1377 [astro-ph.CO]} \BibitemShut
  {NoStop}%
\bibitem [{\citenamefont {Tulin}\ \emph {et~al.}(2013)\citenamefont {Tulin},
  \citenamefont {Yu},\ and\ \citenamefont {Zurek}}]{Tulin:2013teo}%
  \BibitemOpen
  \bibfield  {author} {\bibinfo {author} {\bibfnamefont {S.}~\bibnamefont
  {Tulin}}, \bibinfo {author} {\bibfnamefont {H.-B.}\ \bibnamefont {Yu}}, \
  and\ \bibinfo {author} {\bibfnamefont {K.~M.}\ \bibnamefont {Zurek}},\ }\href
  {\doibase 10.1103/PhysRevD.87.115007} {\bibfield  {journal} {\bibinfo
  {journal} {Phys. Rev.}\ }\textbf {\bibinfo {volume} {D87}},\ \bibinfo {pages}
  {115007} (\bibinfo {year} {2013})},\ \Eprint {http://arxiv.org/abs/1302.3898}
  {arXiv:1302.3898 [hep-ph]} \BibitemShut {NoStop}%
\bibitem [{\citenamefont {Kaplinghat}\ \emph
  {et~al.}(2014{\natexlab{a}})\citenamefont {Kaplinghat}, \citenamefont
  {Keeley}, \citenamefont {Linden},\ and\ \citenamefont
  {Yu}}]{Kaplinghat:2013xca}%
  \BibitemOpen
  \bibfield  {author} {\bibinfo {author} {\bibfnamefont {M.}~\bibnamefont
  {Kaplinghat}}, \bibinfo {author} {\bibfnamefont {R.~E.}\ \bibnamefont
  {Keeley}}, \bibinfo {author} {\bibfnamefont {T.}~\bibnamefont {Linden}}, \
  and\ \bibinfo {author} {\bibfnamefont {H.-B.}\ \bibnamefont {Yu}},\ }\href
  {\doibase 10.1103/PhysRevLett.113.021302} {\bibfield  {journal} {\bibinfo
  {journal} {Phys. Rev. Lett.}\ }\textbf {\bibinfo {volume} {113}},\ \bibinfo
  {pages} {021302} (\bibinfo {year} {2014}{\natexlab{a}})},\ \Eprint
  {http://arxiv.org/abs/1311.6524} {arXiv:1311.6524 [astro-ph.CO]} \BibitemShut
  {NoStop}%
\bibitem [{\citenamefont {Kaplinghat}\ \emph
  {et~al.}(2014{\natexlab{b}})\citenamefont {Kaplinghat}, \citenamefont
  {Tulin},\ and\ \citenamefont {Yu}}]{Kaplinghat:2013yxa}%
  \BibitemOpen
  \bibfield  {author} {\bibinfo {author} {\bibfnamefont {M.}~\bibnamefont
  {Kaplinghat}}, \bibinfo {author} {\bibfnamefont {S.}~\bibnamefont {Tulin}}, \
  and\ \bibinfo {author} {\bibfnamefont {H.-B.}\ \bibnamefont {Yu}},\ }\href
  {\doibase 10.1103/PhysRevD.89.035009} {\bibfield  {journal} {\bibinfo
  {journal} {Phys. Rev.}\ }\textbf {\bibinfo {volume} {D89}},\ \bibinfo {pages}
  {035009} (\bibinfo {year} {2014}{\natexlab{b}})},\ \Eprint
  {http://arxiv.org/abs/1310.7945} {arXiv:1310.7945 [hep-ph]} \BibitemShut
  {NoStop}%
\bibitem [{\citenamefont {Cline}\ \emph
  {et~al.}(2014{\natexlab{a}})\citenamefont {Cline}, \citenamefont {Liu},
  \citenamefont {Moore},\ and\ \citenamefont {Xue}}]{Cline:2013pca}%
  \BibitemOpen
  \bibfield  {author} {\bibinfo {author} {\bibfnamefont {J.~M.}\ \bibnamefont
  {Cline}}, \bibinfo {author} {\bibfnamefont {Z.}~\bibnamefont {Liu}}, \bibinfo
  {author} {\bibfnamefont {G.}~\bibnamefont {Moore}}, \ and\ \bibinfo {author}
  {\bibfnamefont {W.}~\bibnamefont {Xue}},\ }\href {\doibase
  10.1103/PhysRevD.89.043514} {\bibfield  {journal} {\bibinfo  {journal} {Phys.
  Rev.}\ }\textbf {\bibinfo {volume} {D89}},\ \bibinfo {pages} {043514}
  (\bibinfo {year} {2014}{\natexlab{a}})},\ \Eprint
  {http://arxiv.org/abs/1311.6468} {arXiv:1311.6468 [hep-ph]} \BibitemShut
  {NoStop}%
\bibitem [{\citenamefont {Cline}\ \emph
  {et~al.}(2014{\natexlab{b}})\citenamefont {Cline}, \citenamefont {Liu},
  \citenamefont {Moore},\ and\ \citenamefont {Xue}}]{Cline:2013zca}%
  \BibitemOpen
  \bibfield  {author} {\bibinfo {author} {\bibfnamefont {J.~M.}\ \bibnamefont
  {Cline}}, \bibinfo {author} {\bibfnamefont {Z.}~\bibnamefont {Liu}}, \bibinfo
  {author} {\bibfnamefont {G.}~\bibnamefont {Moore}}, \ and\ \bibinfo {author}
  {\bibfnamefont {W.}~\bibnamefont {Xue}},\ }\href {\doibase
  10.1103/PhysRevD.90.015023} {\bibfield  {journal} {\bibinfo  {journal} {Phys.
  Rev.}\ }\textbf {\bibinfo {volume} {D90}},\ \bibinfo {pages} {015023}
  (\bibinfo {year} {2014}{\natexlab{b}})},\ \Eprint
  {http://arxiv.org/abs/1312.3325} {arXiv:1312.3325 [hep-ph]} \BibitemShut
  {NoStop}%
\bibitem [{\citenamefont {Petraki}\ \emph {et~al.}(2014)\citenamefont
  {Petraki}, \citenamefont {Pearce},\ and\ \citenamefont
  {Kusenko}}]{Petraki:2014uza}%
  \BibitemOpen
  \bibfield  {author} {\bibinfo {author} {\bibfnamefont {K.}~\bibnamefont
  {Petraki}}, \bibinfo {author} {\bibfnamefont {L.}~\bibnamefont {Pearce}}, \
  and\ \bibinfo {author} {\bibfnamefont {A.}~\bibnamefont {Kusenko}},\ }\href
  {\doibase 10.1088/1475-7516/2014/07/039} {\bibfield  {journal} {\bibinfo
  {journal} {JCAP}\ }\textbf {\bibinfo {volume} {1407}},\ \bibinfo {pages}
  {039} (\bibinfo {year} {2014})},\ \Eprint {http://arxiv.org/abs/1403.1077}
  {arXiv:1403.1077 [hep-ph]} \BibitemShut {NoStop}%
\bibitem [{\citenamefont {Buckley}\ \emph {et~al.}(2014)\citenamefont
  {Buckley}, \citenamefont {Zavala}, \citenamefont {Cyr-Racine}, \citenamefont
  {Sigurdson},\ and\ \citenamefont {Vogelsberger}}]{Buckley:2014hja}%
  \BibitemOpen
  \bibfield  {author} {\bibinfo {author} {\bibfnamefont {M.~R.}\ \bibnamefont
  {Buckley}}, \bibinfo {author} {\bibfnamefont {J.}~\bibnamefont {Zavala}},
  \bibinfo {author} {\bibfnamefont {F.-Y.}\ \bibnamefont {Cyr-Racine}},
  \bibinfo {author} {\bibfnamefont {K.}~\bibnamefont {Sigurdson}}, \ and\
  \bibinfo {author} {\bibfnamefont {M.}~\bibnamefont {Vogelsberger}},\ }\href
  {\doibase 10.1103/PhysRevD.90.043524} {\bibfield  {journal} {\bibinfo
  {journal} {Phys. Rev.}\ }\textbf {\bibinfo {volume} {D90}},\ \bibinfo {pages}
  {043524} (\bibinfo {year} {2014})},\ \Eprint {http://arxiv.org/abs/1405.2075}
  {arXiv:1405.2075 [astro-ph.CO]} \BibitemShut {NoStop}%
\bibitem [{\citenamefont {Boddy}\ \emph {et~al.}(2014)\citenamefont {Boddy},
  \citenamefont {Feng}, \citenamefont {Kaplinghat},\ and\ \citenamefont
  {Tait}}]{Boddy:2014yra}%
  \BibitemOpen
  \bibfield  {author} {\bibinfo {author} {\bibfnamefont {K.~K.}\ \bibnamefont
  {Boddy}}, \bibinfo {author} {\bibfnamefont {J.~L.}\ \bibnamefont {Feng}},
  \bibinfo {author} {\bibfnamefont {M.}~\bibnamefont {Kaplinghat}}, \ and\
  \bibinfo {author} {\bibfnamefont {T.~M.~P.}\ \bibnamefont {Tait}},\ }\href
  {\doibase 10.1103/PhysRevD.89.115017} {\bibfield  {journal} {\bibinfo
  {journal} {Phys. Rev.}\ }\textbf {\bibinfo {volume} {D89}},\ \bibinfo {pages}
  {115017} (\bibinfo {year} {2014})},\ \Eprint {http://arxiv.org/abs/1402.3629}
  {arXiv:1402.3629 [hep-ph]} \BibitemShut {NoStop}%
\bibitem [{\citenamefont {Schutz}\ and\ \citenamefont
  {Slatyer}(2015)}]{Schutz:2014nka}%
  \BibitemOpen
  \bibfield  {author} {\bibinfo {author} {\bibfnamefont {K.}~\bibnamefont
  {Schutz}}\ and\ \bibinfo {author} {\bibfnamefont {T.~R.}\ \bibnamefont
  {Slatyer}},\ }\href {\doibase 10.1088/1475-7516/2015/01/021} {\bibfield
  {journal} {\bibinfo  {journal} {JCAP}\ }\textbf {\bibinfo {volume} {1501}},\
  \bibinfo {pages} {021} (\bibinfo {year} {2015})},\ \Eprint
  {http://arxiv.org/abs/1409.2867} {arXiv:1409.2867 [hep-ph]} \BibitemShut
  {NoStop}%
\bibitem [{\citenamefont {Feng}\ \emph {et~al.}(2009)\citenamefont {Feng},
  \citenamefont {Kaplinghat}, \citenamefont {Tu},\ and\ \citenamefont
  {Yu}}]{Feng:2009mn}%
  \BibitemOpen
  \bibfield  {author} {\bibinfo {author} {\bibfnamefont {J.~L.}\ \bibnamefont
  {Feng}}, \bibinfo {author} {\bibfnamefont {M.}~\bibnamefont {Kaplinghat}},
  \bibinfo {author} {\bibfnamefont {H.}~\bibnamefont {Tu}}, \ and\ \bibinfo
  {author} {\bibfnamefont {H.-B.}\ \bibnamefont {Yu}},\ }\href {\doibase
  10.1088/1475-7516/2009/07/004} {\bibfield  {journal} {\bibinfo  {journal}
  {JCAP}\ }\textbf {\bibinfo {volume} {0907}},\ \bibinfo {pages} {004}
  (\bibinfo {year} {2009})},\ \Eprint {http://arxiv.org/abs/0905.3039}
  {arXiv:0905.3039 [hep-ph]} \BibitemShut {NoStop}%
\bibitem [{\citenamefont {Feng}\ \emph {et~al.}(2010)\citenamefont {Feng},
  \citenamefont {Kaplinghat},\ and\ \citenamefont {Yu}}]{Feng:2009hw}%
  \BibitemOpen
  \bibfield  {author} {\bibinfo {author} {\bibfnamefont {J.~L.}\ \bibnamefont
  {Feng}}, \bibinfo {author} {\bibfnamefont {M.}~\bibnamefont {Kaplinghat}}, \
  and\ \bibinfo {author} {\bibfnamefont {H.-B.}\ \bibnamefont {Yu}},\ }\href
  {\doibase 10.1103/PhysRevLett.104.151301} {\bibfield  {journal} {\bibinfo
  {journal} {Phys. Rev. Lett.}\ }\textbf {\bibinfo {volume} {104}},\ \bibinfo
  {pages} {151301} (\bibinfo {year} {2010})},\ \Eprint
  {http://arxiv.org/abs/0911.0422} {arXiv:0911.0422 [hep-ph]} \BibitemShut
  {NoStop}%
\bibitem [{\citenamefont {Markevitch}\ \emph {et~al.}(2004)\citenamefont
  {Markevitch}, \citenamefont {Gonzalez}, \citenamefont {Clowe}, \citenamefont
  {Vikhlinin}, \citenamefont {David}, \citenamefont {Forman}, \citenamefont
  {Jones}, \citenamefont {Murray},\ and\ \citenamefont
  {Tucker}}]{Markevitch:2003at}%
  \BibitemOpen
  \bibfield  {author} {\bibinfo {author} {\bibfnamefont {M.}~\bibnamefont
  {Markevitch}}, \bibinfo {author} {\bibfnamefont {A.~H.}\ \bibnamefont
  {Gonzalez}}, \bibinfo {author} {\bibfnamefont {D.}~\bibnamefont {Clowe}},
  \bibinfo {author} {\bibfnamefont {A.}~\bibnamefont {Vikhlinin}}, \bibinfo
  {author} {\bibfnamefont {L.}~\bibnamefont {David}}, \bibinfo {author}
  {\bibfnamefont {W.}~\bibnamefont {Forman}}, \bibinfo {author} {\bibfnamefont
  {C.}~\bibnamefont {Jones}}, \bibinfo {author} {\bibfnamefont
  {S.}~\bibnamefont {Murray}}, \ and\ \bibinfo {author} {\bibfnamefont
  {W.}~\bibnamefont {Tucker}},\ }\href {\doibase 10.1086/383178} {\bibfield
  {journal} {\bibinfo  {journal} {Astrophys. J.}\ }\textbf {\bibinfo {volume}
  {606}},\ \bibinfo {pages} {819} (\bibinfo {year} {2004})},\ \Eprint
  {http://arxiv.org/abs/astro-ph/0309303} {arXiv:astro-ph/0309303 [astro-ph]}
  \BibitemShut {NoStop}%
\bibitem [{\citenamefont {Kouvaris}\ \emph {et~al.}(2015)\citenamefont
  {Kouvaris}, \citenamefont {Shoemaker},\ and\ \citenamefont
  {Tuominen}}]{Kouvaris:2014uoa}%
  \BibitemOpen
  \bibfield  {author} {\bibinfo {author} {\bibfnamefont {C.}~\bibnamefont
  {Kouvaris}}, \bibinfo {author} {\bibfnamefont {I.~M.}\ \bibnamefont
  {Shoemaker}}, \ and\ \bibinfo {author} {\bibfnamefont {K.}~\bibnamefont
  {Tuominen}},\ }\href {\doibase 10.1103/PhysRevD.91.043519} {\bibfield
  {journal} {\bibinfo  {journal} {Phys. Rev.}\ }\textbf {\bibinfo {volume}
  {D91}},\ \bibinfo {pages} {043519} (\bibinfo {year} {2015})},\ \Eprint
  {http://arxiv.org/abs/1411.3730} {arXiv:1411.3730 [hep-ph]} \BibitemShut
  {NoStop}%
\bibitem [{\citenamefont {Pollack}\ \emph {et~al.}(2015)\citenamefont
  {Pollack}, \citenamefont {Spergel},\ and\ \citenamefont
  {Steinhardt}}]{Pollack:2014rja}%
  \BibitemOpen
  \bibfield  {author} {\bibinfo {author} {\bibfnamefont {J.}~\bibnamefont
  {Pollack}}, \bibinfo {author} {\bibfnamefont {D.~N.}\ \bibnamefont
  {Spergel}}, \ and\ \bibinfo {author} {\bibfnamefont {P.~J.}\ \bibnamefont
  {Steinhardt}},\ }\href {\doibase 10.1088/0004-637X/804/2/131} {\bibfield
  {journal} {\bibinfo  {journal} {Astrophys. J.}\ }\textbf {\bibinfo {volume}
  {804}},\ \bibinfo {pages} {131} (\bibinfo {year} {2015})},\ \Eprint
  {http://arxiv.org/abs/1501.00017} {arXiv:1501.00017 [astro-ph.CO]}
  \BibitemShut {NoStop}%
\bibitem [{\citenamefont {Balberg}\ \emph {et~al.}(2002)\citenamefont
  {Balberg}, \citenamefont {Shapiro},\ and\ \citenamefont
  {Inagaki}}]{Balberg:2002ue}%
  \BibitemOpen
  \bibfield  {author} {\bibinfo {author} {\bibfnamefont {S.}~\bibnamefont
  {Balberg}}, \bibinfo {author} {\bibfnamefont {S.~L.}\ \bibnamefont
  {Shapiro}}, \ and\ \bibinfo {author} {\bibfnamefont {S.}~\bibnamefont
  {Inagaki}},\ }\href {\doibase 10.1086/339038} {\bibfield  {journal} {\bibinfo
   {journal} {Astrophys. J.}\ }\textbf {\bibinfo {volume} {568}},\ \bibinfo
  {pages} {475} (\bibinfo {year} {2002})},\ \Eprint
  {http://arxiv.org/abs/astro-ph/0110561} {arXiv:astro-ph/0110561 [astro-ph]}
  \BibitemShut {NoStop}%
\bibitem [{\citenamefont {Spolyar}\ \emph {et~al.}(2008)\citenamefont
  {Spolyar}, \citenamefont {Freese},\ and\ \citenamefont
  {Gondolo}}]{Spolyar:2007qv}%
  \BibitemOpen
  \bibfield  {author} {\bibinfo {author} {\bibfnamefont {D.}~\bibnamefont
  {Spolyar}}, \bibinfo {author} {\bibfnamefont {K.}~\bibnamefont {Freese}}, \
  and\ \bibinfo {author} {\bibfnamefont {P.}~\bibnamefont {Gondolo}},\ }\href
  {\doibase 10.1103/PhysRevLett.100.051101} {\bibfield  {journal} {\bibinfo
  {journal} {Phys. Rev. Lett.}\ }\textbf {\bibinfo {volume} {100}},\ \bibinfo
  {pages} {051101} (\bibinfo {year} {2008})},\ \Eprint
  {http://arxiv.org/abs/0705.0521} {arXiv:0705.0521 [astro-ph]} \BibitemShut
  {NoStop}%
\bibitem [{\citenamefont {Freese}\ \emph {et~al.}(2009)\citenamefont {Freese},
  \citenamefont {Gondolo}, \citenamefont {Sellwood},\ and\ \citenamefont
  {Spolyar}}]{Freese:2008hb}%
  \BibitemOpen
  \bibfield  {author} {\bibinfo {author} {\bibfnamefont {K.}~\bibnamefont
  {Freese}}, \bibinfo {author} {\bibfnamefont {P.}~\bibnamefont {Gondolo}},
  \bibinfo {author} {\bibfnamefont {J.~A.}\ \bibnamefont {Sellwood}}, \ and\
  \bibinfo {author} {\bibfnamefont {D.}~\bibnamefont {Spolyar}},\ }\href
  {\doibase 10.1088/0004-637X/693/2/1563} {\bibfield  {journal} {\bibinfo
  {journal} {Astrophys. J.}\ }\textbf {\bibinfo {volume} {693}},\ \bibinfo
  {pages} {1563} (\bibinfo {year} {2009})},\ \Eprint
  {http://arxiv.org/abs/0805.3540} {arXiv:0805.3540 [astro-ph]} \BibitemShut
  {NoStop}%
\bibitem [{\citenamefont {Freese}\ \emph {et~al.}(2008)\citenamefont {Freese},
  \citenamefont {Bodenheimer}, \citenamefont {Spolyar},\ and\ \citenamefont
  {Gondolo}}]{Freese:2008wh}%
  \BibitemOpen
  \bibfield  {author} {\bibinfo {author} {\bibfnamefont {K.}~\bibnamefont
  {Freese}}, \bibinfo {author} {\bibfnamefont {P.}~\bibnamefont {Bodenheimer}},
  \bibinfo {author} {\bibfnamefont {D.}~\bibnamefont {Spolyar}}, \ and\
  \bibinfo {author} {\bibfnamefont {P.}~\bibnamefont {Gondolo}},\ }\href
  {\doibase 10.1086/592685} {\bibfield  {journal} {\bibinfo  {journal}
  {Astrophys. J.}\ }\textbf {\bibinfo {volume} {685}},\ \bibinfo {pages} {L101}
  (\bibinfo {year} {2008})},\ \Eprint {http://arxiv.org/abs/0806.0617}
  {arXiv:0806.0617 [astro-ph]} \BibitemShut {NoStop}%
\bibitem [{\citenamefont {Nussinov}(1985)}]{Nussinov:1985xr}%
  \BibitemOpen
  \bibfield  {author} {\bibinfo {author} {\bibfnamefont {S.}~\bibnamefont
  {Nussinov}},\ }\href {\doibase 10.1016/0370-2693(85)90689-6} {\bibfield
  {journal} {\bibinfo  {journal} {Phys. Lett.}\ }\textbf {\bibinfo {volume}
  {B165}},\ \bibinfo {pages} {55} (\bibinfo {year} {1985})}\BibitemShut
  {NoStop}%
\bibitem [{\citenamefont {Barr}\ \emph {et~al.}(1990)\citenamefont {Barr},
  \citenamefont {Chivukula},\ and\ \citenamefont {Farhi}}]{Barr:1990ca}%
  \BibitemOpen
  \bibfield  {author} {\bibinfo {author} {\bibfnamefont {S.~M.}\ \bibnamefont
  {Barr}}, \bibinfo {author} {\bibfnamefont {R.~S.}\ \bibnamefont {Chivukula}},
  \ and\ \bibinfo {author} {\bibfnamefont {E.}~\bibnamefont {Farhi}},\ }\href
  {\doibase 10.1016/0370-2693(90)91661-T} {\bibfield  {journal} {\bibinfo
  {journal} {Phys. Lett.}\ }\textbf {\bibinfo {volume} {B241}},\ \bibinfo
  {pages} {387} (\bibinfo {year} {1990})}\BibitemShut {NoStop}%
\bibitem [{\citenamefont {Gudnason}\ \emph {et~al.}(2006)\citenamefont
  {Gudnason}, \citenamefont {Kouvaris},\ and\ \citenamefont
  {Sannino}}]{Gudnason:2006yj}%
  \BibitemOpen
  \bibfield  {author} {\bibinfo {author} {\bibfnamefont {S.~B.}\ \bibnamefont
  {Gudnason}}, \bibinfo {author} {\bibfnamefont {C.}~\bibnamefont {Kouvaris}},
  \ and\ \bibinfo {author} {\bibfnamefont {F.}~\bibnamefont {Sannino}},\ }\href
  {\doibase 10.1103/PhysRevD.74.095008} {\bibfield  {journal} {\bibinfo
  {journal} {Phys. Rev.}\ }\textbf {\bibinfo {volume} {D74}},\ \bibinfo {pages}
  {095008} (\bibinfo {year} {2006})},\ \Eprint
  {http://arxiv.org/abs/hep-ph/0608055} {arXiv:hep-ph/0608055 [hep-ph]}
  \BibitemShut {NoStop}%
\bibitem [{\citenamefont {Foadi}\ \emph {et~al.}(2009)\citenamefont {Foadi},
  \citenamefont {Frandsen},\ and\ \citenamefont {Sannino}}]{Foadi:2008qv}%
  \BibitemOpen
  \bibfield  {author} {\bibinfo {author} {\bibfnamefont {R.}~\bibnamefont
  {Foadi}}, \bibinfo {author} {\bibfnamefont {M.~T.}\ \bibnamefont {Frandsen}},
  \ and\ \bibinfo {author} {\bibfnamefont {F.}~\bibnamefont {Sannino}},\ }\href
  {\doibase 10.1103/PhysRevD.80.037702} {\bibfield  {journal} {\bibinfo
  {journal} {Phys. Rev.}\ }\textbf {\bibinfo {volume} {D80}},\ \bibinfo {pages}
  {037702} (\bibinfo {year} {2009})},\ \Eprint {http://arxiv.org/abs/0812.3406}
  {arXiv:0812.3406 [hep-ph]} \BibitemShut {NoStop}%
\bibitem [{\citenamefont {Dietrich}\ and\ \citenamefont
  {Sannino}(2007)}]{Dietrich:2006cm}%
  \BibitemOpen
  \bibfield  {author} {\bibinfo {author} {\bibfnamefont {D.~D.}\ \bibnamefont
  {Dietrich}}\ and\ \bibinfo {author} {\bibfnamefont {F.}~\bibnamefont
  {Sannino}},\ }\href {\doibase 10.1103/PhysRevD.75.085018} {\bibfield
  {journal} {\bibinfo  {journal} {Phys. Rev.}\ }\textbf {\bibinfo {volume}
  {D75}},\ \bibinfo {pages} {085018} (\bibinfo {year} {2007})},\ \Eprint
  {http://arxiv.org/abs/hep-ph/0611341} {arXiv:hep-ph/0611341 [hep-ph]}
  \BibitemShut {NoStop}%
\bibitem [{\citenamefont {Sannino}(2009)}]{Sannino:2009za}%
  \BibitemOpen
  \bibfield  {author} {\bibinfo {author} {\bibfnamefont {F.}~\bibnamefont
  {Sannino}},\ }\href@noop {} {\bibfield  {journal} {\bibinfo  {journal} {Acta
  Phys. Polon.}\ }\textbf {\bibinfo {volume} {B40}},\ \bibinfo {pages} {3533}
  (\bibinfo {year} {2009})},\ \Eprint {http://arxiv.org/abs/0911.0931}
  {arXiv:0911.0931 [hep-ph]} \BibitemShut {NoStop}%
\bibitem [{\citenamefont {Ryttov}\ and\ \citenamefont
  {Sannino}(2008)}]{Ryttov:2008xe}%
  \BibitemOpen
  \bibfield  {author} {\bibinfo {author} {\bibfnamefont {T.~A.}\ \bibnamefont
  {Ryttov}}\ and\ \bibinfo {author} {\bibfnamefont {F.}~\bibnamefont
  {Sannino}},\ }\href {\doibase 10.1103/PhysRevD.78.115010} {\bibfield
  {journal} {\bibinfo  {journal} {Phys. Rev.}\ }\textbf {\bibinfo {volume}
  {D78}},\ \bibinfo {pages} {115010} (\bibinfo {year} {2008})},\ \Eprint
  {http://arxiv.org/abs/0809.0713} {arXiv:0809.0713 [hep-ph]} \BibitemShut
  {NoStop}%
\bibitem [{\citenamefont {Sannino}\ and\ \citenamefont
  {Zwicky}(2009)}]{Sannino:2008nv}%
  \BibitemOpen
  \bibfield  {author} {\bibinfo {author} {\bibfnamefont {F.}~\bibnamefont
  {Sannino}}\ and\ \bibinfo {author} {\bibfnamefont {R.}~\bibnamefont
  {Zwicky}},\ }\href {\doibase 10.1103/PhysRevD.79.015016} {\bibfield
  {journal} {\bibinfo  {journal} {Phys. Rev.}\ }\textbf {\bibinfo {volume}
  {D79}},\ \bibinfo {pages} {015016} (\bibinfo {year} {2009})},\ \Eprint
  {http://arxiv.org/abs/0810.2686} {arXiv:0810.2686 [hep-ph]} \BibitemShut
  {NoStop}%
\bibitem [{\citenamefont {Kaplan}\ \emph {et~al.}(2009)\citenamefont {Kaplan},
  \citenamefont {Luty},\ and\ \citenamefont {Zurek}}]{Kaplan:2009ag}%
  \BibitemOpen
  \bibfield  {author} {\bibinfo {author} {\bibfnamefont {D.~E.}\ \bibnamefont
  {Kaplan}}, \bibinfo {author} {\bibfnamefont {M.~A.}\ \bibnamefont {Luty}}, \
  and\ \bibinfo {author} {\bibfnamefont {K.~M.}\ \bibnamefont {Zurek}},\ }\href
  {\doibase 10.1103/PhysRevD.79.115016} {\bibfield  {journal} {\bibinfo
  {journal} {Phys. Rev.}\ }\textbf {\bibinfo {volume} {D79}},\ \bibinfo {pages}
  {115016} (\bibinfo {year} {2009})},\ \Eprint {http://arxiv.org/abs/0901.4117}
  {arXiv:0901.4117 [hep-ph]} \BibitemShut {NoStop}%
\bibitem [{\citenamefont {Frandsen}\ and\ \citenamefont
  {Sannino}(2010)}]{Frandsen:2009mi}%
  \BibitemOpen
  \bibfield  {author} {\bibinfo {author} {\bibfnamefont {M.~T.}\ \bibnamefont
  {Frandsen}}\ and\ \bibinfo {author} {\bibfnamefont {F.}~\bibnamefont
  {Sannino}},\ }\href {\doibase 10.1103/PhysRevD.81.097704} {\bibfield
  {journal} {\bibinfo  {journal} {Phys. Rev.}\ }\textbf {\bibinfo {volume}
  {D81}},\ \bibinfo {pages} {097704} (\bibinfo {year} {2010})},\ \Eprint
  {http://arxiv.org/abs/0911.1570} {arXiv:0911.1570 [hep-ph]} \BibitemShut
  {NoStop}%
\bibitem [{\citenamefont {March-Russell}\ and\ \citenamefont
  {McCullough}(2012)}]{MarchRussell:2011fi}%
  \BibitemOpen
  \bibfield  {author} {\bibinfo {author} {\bibfnamefont {J.}~\bibnamefont
  {March-Russell}}\ and\ \bibinfo {author} {\bibfnamefont {M.}~\bibnamefont
  {McCullough}},\ }\href {\doibase 10.1088/1475-7516/2012/03/019} {\bibfield
  {journal} {\bibinfo  {journal} {JCAP}\ }\textbf {\bibinfo {volume} {1203}},\
  \bibinfo {pages} {019} (\bibinfo {year} {2012})},\ \Eprint
  {http://arxiv.org/abs/1106.4319} {arXiv:1106.4319 [hep-ph]} \BibitemShut
  {NoStop}%
\bibitem [{\citenamefont {Frandsen}\ \emph {et~al.}(2011)\citenamefont
  {Frandsen}, \citenamefont {Kahlhoefer}, \citenamefont {Sarkar},\ and\
  \citenamefont {Schmidt-Hoberg}}]{Frandsen:2011cg}%
  \BibitemOpen
  \bibfield  {author} {\bibinfo {author} {\bibfnamefont {M.~T.}\ \bibnamefont
  {Frandsen}}, \bibinfo {author} {\bibfnamefont {F.}~\bibnamefont
  {Kahlhoefer}}, \bibinfo {author} {\bibfnamefont {S.}~\bibnamefont {Sarkar}},
  \ and\ \bibinfo {author} {\bibfnamefont {K.}~\bibnamefont {Schmidt-Hoberg}},\
  }\href {\doibase 10.1007/JHEP09(2011)128} {\bibfield  {journal} {\bibinfo
  {journal} {JHEP}\ }\textbf {\bibinfo {volume} {09}},\ \bibinfo {pages} {128}
  (\bibinfo {year} {2011})},\ \Eprint {http://arxiv.org/abs/1107.2118}
  {arXiv:1107.2118 [hep-ph]} \BibitemShut {NoStop}%
\bibitem [{\citenamefont {Gao}\ \emph {et~al.}(2013)\citenamefont {Gao},
  \citenamefont {Kang},\ and\ \citenamefont {Li}}]{Gao:2011ka}%
  \BibitemOpen
  \bibfield  {author} {\bibinfo {author} {\bibfnamefont {X.}~\bibnamefont
  {Gao}}, \bibinfo {author} {\bibfnamefont {Z.}~\bibnamefont {Kang}}, \ and\
  \bibinfo {author} {\bibfnamefont {T.}~\bibnamefont {Li}},\ }\href {\doibase
  10.1088/1475-7516/2013/01/021} {\bibfield  {journal} {\bibinfo  {journal}
  {JCAP}\ }\textbf {\bibinfo {volume} {1301}},\ \bibinfo {pages} {021}
  (\bibinfo {year} {2013})},\ \Eprint {http://arxiv.org/abs/1107.3529}
  {arXiv:1107.3529 [hep-ph]} \BibitemShut {NoStop}%
\bibitem [{\citenamefont {Arina}\ and\ \citenamefont
  {Sahu}(2012)}]{Arina:2011cu}%
  \BibitemOpen
  \bibfield  {author} {\bibinfo {author} {\bibfnamefont {C.}~\bibnamefont
  {Arina}}\ and\ \bibinfo {author} {\bibfnamefont {N.}~\bibnamefont {Sahu}},\
  }\href {\doibase 10.1016/j.nuclphysb.2011.09.014} {\bibfield  {journal}
  {\bibinfo  {journal} {Nucl. Phys.}\ }\textbf {\bibinfo {volume} {B854}},\
  \bibinfo {pages} {666} (\bibinfo {year} {2012})},\ \Eprint
  {http://arxiv.org/abs/1108.3967} {arXiv:1108.3967 [hep-ph]} \BibitemShut
  {NoStop}%
\bibitem [{\citenamefont {Buckley}\ and\ \citenamefont
  {Profumo}(2012)}]{Buckley:2011ye}%
  \BibitemOpen
  \bibfield  {author} {\bibinfo {author} {\bibfnamefont {M.~R.}\ \bibnamefont
  {Buckley}}\ and\ \bibinfo {author} {\bibfnamefont {S.}~\bibnamefont
  {Profumo}},\ }\href {\doibase 10.1103/PhysRevLett.108.011301} {\bibfield
  {journal} {\bibinfo  {journal} {Phys. Rev. Lett.}\ }\textbf {\bibinfo
  {volume} {108}},\ \bibinfo {pages} {011301} (\bibinfo {year} {2012})},\
  \Eprint {http://arxiv.org/abs/1109.2164} {arXiv:1109.2164 [hep-ph]}
  \BibitemShut {NoStop}%
\bibitem [{\citenamefont {Lewis}\ \emph {et~al.}(2012)\citenamefont {Lewis},
  \citenamefont {Pica},\ and\ \citenamefont {Sannino}}]{Lewis:2011zb}%
  \BibitemOpen
  \bibfield  {author} {\bibinfo {author} {\bibfnamefont {R.}~\bibnamefont
  {Lewis}}, \bibinfo {author} {\bibfnamefont {C.}~\bibnamefont {Pica}}, \ and\
  \bibinfo {author} {\bibfnamefont {F.}~\bibnamefont {Sannino}},\ }\href
  {\doibase 10.1103/PhysRevD.85.014504} {\bibfield  {journal} {\bibinfo
  {journal} {Phys. Rev.}\ }\textbf {\bibinfo {volume} {D85}},\ \bibinfo {pages}
  {014504} (\bibinfo {year} {2012})},\ \Eprint {http://arxiv.org/abs/1109.3513}
  {arXiv:1109.3513 [hep-ph]} \BibitemShut {NoStop}%
\bibitem [{\citenamefont {Davoudiasl}\ \emph {et~al.}(2011)\citenamefont
  {Davoudiasl}, \citenamefont {Morrissey}, \citenamefont {Sigurdson},\ and\
  \citenamefont {Tulin}}]{Davoudiasl:2011fj}%
  \BibitemOpen
  \bibfield  {author} {\bibinfo {author} {\bibfnamefont {H.}~\bibnamefont
  {Davoudiasl}}, \bibinfo {author} {\bibfnamefont {D.~E.}\ \bibnamefont
  {Morrissey}}, \bibinfo {author} {\bibfnamefont {K.}~\bibnamefont
  {Sigurdson}}, \ and\ \bibinfo {author} {\bibfnamefont {S.}~\bibnamefont
  {Tulin}},\ }\href {\doibase 10.1103/PhysRevD.84.096008} {\bibfield  {journal}
  {\bibinfo  {journal} {Phys. Rev.}\ }\textbf {\bibinfo {volume} {D84}},\
  \bibinfo {pages} {096008} (\bibinfo {year} {2011})},\ \Eprint
  {http://arxiv.org/abs/1106.4320} {arXiv:1106.4320 [hep-ph]} \BibitemShut
  {NoStop}%
\bibitem [{\citenamefont {Graesser}\ \emph {et~al.}(2011)\citenamefont
  {Graesser}, \citenamefont {Shoemaker},\ and\ \citenamefont
  {Vecchi}}]{Graesser:2011wi}%
  \BibitemOpen
  \bibfield  {author} {\bibinfo {author} {\bibfnamefont {M.~L.}\ \bibnamefont
  {Graesser}}, \bibinfo {author} {\bibfnamefont {I.~M.}\ \bibnamefont
  {Shoemaker}}, \ and\ \bibinfo {author} {\bibfnamefont {L.}~\bibnamefont
  {Vecchi}},\ }\href {\doibase 10.1007/JHEP10(2011)110} {\bibfield  {journal}
  {\bibinfo  {journal} {JHEP}\ }\textbf {\bibinfo {volume} {10}},\ \bibinfo
  {pages} {110} (\bibinfo {year} {2011})},\ \Eprint
  {http://arxiv.org/abs/1103.2771} {arXiv:1103.2771 [hep-ph]} \BibitemShut
  {NoStop}%
\bibitem [{\citenamefont {Bell}\ \emph {et~al.}(2011)\citenamefont {Bell},
  \citenamefont {Petraki}, \citenamefont {Shoemaker},\ and\ \citenamefont
  {Volkas}}]{Bell:2011tn}%
  \BibitemOpen
  \bibfield  {author} {\bibinfo {author} {\bibfnamefont {N.~F.}\ \bibnamefont
  {Bell}}, \bibinfo {author} {\bibfnamefont {K.}~\bibnamefont {Petraki}},
  \bibinfo {author} {\bibfnamefont {I.~M.}\ \bibnamefont {Shoemaker}}, \ and\
  \bibinfo {author} {\bibfnamefont {R.~R.}\ \bibnamefont {Volkas}},\ }\href
  {\doibase 10.1103/PhysRevD.84.123505} {\bibfield  {journal} {\bibinfo
  {journal} {Phys. Rev.}\ }\textbf {\bibinfo {volume} {D84}},\ \bibinfo {pages}
  {123505} (\bibinfo {year} {2011})},\ \Eprint {http://arxiv.org/abs/1105.3730}
  {arXiv:1105.3730 [hep-ph]} \BibitemShut {NoStop}%
\bibitem [{\citenamefont {Cheung}\ and\ \citenamefont
  {Zurek}(2011)}]{Cheung:2011if}%
  \BibitemOpen
  \bibfield  {author} {\bibinfo {author} {\bibfnamefont {C.}~\bibnamefont
  {Cheung}}\ and\ \bibinfo {author} {\bibfnamefont {K.~M.}\ \bibnamefont
  {Zurek}},\ }\href {\doibase 10.1103/PhysRevD.84.035007} {\bibfield  {journal}
  {\bibinfo  {journal} {Phys. Rev.}\ }\textbf {\bibinfo {volume} {D84}},\
  \bibinfo {pages} {035007} (\bibinfo {year} {2011})},\ \Eprint
  {http://arxiv.org/abs/1105.4612} {arXiv:1105.4612 [hep-ph]} \BibitemShut
  {NoStop}%
\bibitem [{\citenamefont {Narain}\ \emph {et~al.}(2006)\citenamefont {Narain},
  \citenamefont {Schaffner-Bielich},\ and\ \citenamefont
  {Mishustin}}]{Narain:2006kx}%
  \BibitemOpen
  \bibfield  {author} {\bibinfo {author} {\bibfnamefont {G.}~\bibnamefont
  {Narain}}, \bibinfo {author} {\bibfnamefont {J.}~\bibnamefont
  {Schaffner-Bielich}}, \ and\ \bibinfo {author} {\bibfnamefont {I.~N.}\
  \bibnamefont {Mishustin}},\ }\href {\doibase 10.1103/PhysRevD.74.063003}
  {\bibfield  {journal} {\bibinfo  {journal} {Phys. Rev.}\ }\textbf {\bibinfo
  {volume} {D74}},\ \bibinfo {pages} {063003} (\bibinfo {year} {2006})},\
  \Eprint {http://arxiv.org/abs/astro-ph/0605724} {arXiv:astro-ph/0605724
  [astro-ph]} \BibitemShut {NoStop}%
\bibitem [{\citenamefont {Kouvaris}\ and\ \citenamefont
  {Nielsen}(2015)}]{Kouvaris:2015rea}%
  \BibitemOpen
  \bibfield  {author} {\bibinfo {author} {\bibfnamefont {C.}~\bibnamefont
  {Kouvaris}}\ and\ \bibinfo {author} {\bibfnamefont {N.~G.}\ \bibnamefont
  {Nielsen}},\ }\href {\doibase 10.1103/PhysRevD.92.063526} {\bibfield
  {journal} {\bibinfo  {journal} {Phys. Rev.}\ }\textbf {\bibinfo {volume}
  {D92}},\ \bibinfo {pages} {063526} (\bibinfo {year} {2015})},\ \Eprint
  {http://arxiv.org/abs/1507.00959} {arXiv:1507.00959 [hep-ph]} \BibitemShut
  {NoStop}%
\bibitem [{\citenamefont {Eby}\ \emph {et~al.}(2016{\natexlab{a}})\citenamefont
  {Eby}, \citenamefont {Kouvaris}, \citenamefont {Nielsen},\ and\ \citenamefont
  {Wijewardhana}}]{Eby:2015hsq}%
  \BibitemOpen
  \bibfield  {author} {\bibinfo {author} {\bibfnamefont {J.}~\bibnamefont
  {Eby}}, \bibinfo {author} {\bibfnamefont {C.}~\bibnamefont {Kouvaris}},
  \bibinfo {author} {\bibfnamefont {N.~G.}\ \bibnamefont {Nielsen}}, \ and\
  \bibinfo {author} {\bibfnamefont {L.~C.~R.}\ \bibnamefont {Wijewardhana}},\
  }\href {\doibase 10.1007/JHEP02(2016)028} {\bibfield  {journal} {\bibinfo
  {journal} {JHEP}\ }\textbf {\bibinfo {volume} {02}},\ \bibinfo {pages} {028}
  (\bibinfo {year} {2016}{\natexlab{a}})},\ \Eprint
  {http://arxiv.org/abs/1511.04474} {arXiv:1511.04474 [hep-ph]} \BibitemShut
  {NoStop}%
\bibitem [{\citenamefont {Kolb}\ and\ \citenamefont
  {Tkachev}(1993)}]{Kolb:1993zz}%
  \BibitemOpen
  \bibfield  {author} {\bibinfo {author} {\bibfnamefont {E.~W.}\ \bibnamefont
  {Kolb}}\ and\ \bibinfo {author} {\bibfnamefont {I.~I.}\ \bibnamefont
  {Tkachev}},\ }\href {\doibase 10.1103/PhysRevLett.71.3051} {\bibfield
  {journal} {\bibinfo  {journal} {Phys. Rev. Lett.}\ }\textbf {\bibinfo
  {volume} {71}},\ \bibinfo {pages} {3051} (\bibinfo {year} {1993})},\ \Eprint
  {http://arxiv.org/abs/hep-ph/9303313} {arXiv:hep-ph/9303313 [hep-ph]}
  \BibitemShut {NoStop}%
\bibitem [{\citenamefont {Kolb}\ and\ \citenamefont
  {Tkachev}(1994)}]{Kolb:1993hw}%
  \BibitemOpen
  \bibfield  {author} {\bibinfo {author} {\bibfnamefont {E.~W.}\ \bibnamefont
  {Kolb}}\ and\ \bibinfo {author} {\bibfnamefont {I.~I.}\ \bibnamefont
  {Tkachev}},\ }\href {\doibase 10.1103/PhysRevD.49.5040} {\bibfield  {journal}
  {\bibinfo  {journal} {Phys. Rev.}\ }\textbf {\bibinfo {volume} {D49}},\
  \bibinfo {pages} {5040} (\bibinfo {year} {1994})},\ \Eprint
  {http://arxiv.org/abs/astro-ph/9311037} {arXiv:astro-ph/9311037 [astro-ph]}
  \BibitemShut {NoStop}%
\bibitem [{\citenamefont {Chavanis}(2011)}]{Chavanis:2011zi}%
  \BibitemOpen
  \bibfield  {author} {\bibinfo {author} {\bibfnamefont {P.-H.}\ \bibnamefont
  {Chavanis}},\ }\href {\doibase 10.1103/PhysRevD.84.043531} {\bibfield
  {journal} {\bibinfo  {journal} {Phys. Rev.}\ }\textbf {\bibinfo {volume}
  {D84}},\ \bibinfo {pages} {043531} (\bibinfo {year} {2011})},\ \Eprint
  {http://arxiv.org/abs/1103.2050} {arXiv:1103.2050 [astro-ph.CO]} \BibitemShut
  {NoStop}%
\bibitem [{\citenamefont {Chavanis}\ and\ \citenamefont
  {Delfini}(2011)}]{Chavanis:2011zm}%
  \BibitemOpen
  \bibfield  {author} {\bibinfo {author} {\bibfnamefont {P.~H.}\ \bibnamefont
  {Chavanis}}\ and\ \bibinfo {author} {\bibfnamefont {L.}~\bibnamefont
  {Delfini}},\ }\href {\doibase 10.1103/PhysRevD.84.043532} {\bibfield
  {journal} {\bibinfo  {journal} {Phys. Rev.}\ }\textbf {\bibinfo {volume}
  {D84}},\ \bibinfo {pages} {043532} (\bibinfo {year} {2011})},\ \Eprint
  {http://arxiv.org/abs/1103.2054} {arXiv:1103.2054 [astro-ph.CO]} \BibitemShut
  {NoStop}%
\bibitem [{\citenamefont {Eby}\ \emph {et~al.}(2015)\citenamefont {Eby},
  \citenamefont {Suranyi}, \citenamefont {Vaz},\ and\ \citenamefont
  {Wijewardhana}}]{Eby:2014fya}%
  \BibitemOpen
  \bibfield  {author} {\bibinfo {author} {\bibfnamefont {J.}~\bibnamefont
  {Eby}}, \bibinfo {author} {\bibfnamefont {P.}~\bibnamefont {Suranyi}},
  \bibinfo {author} {\bibfnamefont {C.}~\bibnamefont {Vaz}}, \ and\ \bibinfo
  {author} {\bibfnamefont {L.~C.~R.}\ \bibnamefont {Wijewardhana}},\ }\href
  {\doibase 10.1007/JHEP11(2016)134, 10.1007/JHEP03(2015)080} {\bibfield
  {journal} {\bibinfo  {journal} {JHEP}\ }\textbf {\bibinfo {volume} {03}},\
  \bibinfo {pages} {080} (\bibinfo {year} {2015})},\ \bibinfo {note} {[Erratum:
  JHEP11,134(2016)]},\ \Eprint {http://arxiv.org/abs/1412.3430}
  {arXiv:1412.3430 [hep-th]} \BibitemShut {NoStop}%
\bibitem [{\citenamefont {Brito}\ \emph
  {et~al.}(2016{\natexlab{a}})\citenamefont {Brito}, \citenamefont {Cardoso},
  \citenamefont {Macedo}, \citenamefont {Okawa},\ and\ \citenamefont
  {Palenzuela}}]{Brito:2015yfh}%
  \BibitemOpen
  \bibfield  {author} {\bibinfo {author} {\bibfnamefont {R.}~\bibnamefont
  {Brito}}, \bibinfo {author} {\bibfnamefont {V.}~\bibnamefont {Cardoso}},
  \bibinfo {author} {\bibfnamefont {C.~F.~B.}\ \bibnamefont {Macedo}}, \bibinfo
  {author} {\bibfnamefont {H.}~\bibnamefont {Okawa}}, \ and\ \bibinfo {author}
  {\bibfnamefont {C.}~\bibnamefont {Palenzuela}},\ }\href {\doibase
  10.1103/PhysRevD.93.044045} {\bibfield  {journal} {\bibinfo  {journal} {Phys.
  Rev.}\ }\textbf {\bibinfo {volume} {D93}},\ \bibinfo {pages} {044045}
  (\bibinfo {year} {2016}{\natexlab{a}})},\ \Eprint
  {http://arxiv.org/abs/1512.00466} {arXiv:1512.00466 [astro-ph.SR]}
  \BibitemShut {NoStop}%
\bibitem [{\citenamefont {Eby}\ \emph {et~al.}(2016{\natexlab{b}})\citenamefont
  {Eby}, \citenamefont {Suranyi},\ and\ \citenamefont
  {Wijewardhana}}]{Eby:2015hyx}%
  \BibitemOpen
  \bibfield  {author} {\bibinfo {author} {\bibfnamefont {J.}~\bibnamefont
  {Eby}}, \bibinfo {author} {\bibfnamefont {P.}~\bibnamefont {Suranyi}}, \ and\
  \bibinfo {author} {\bibfnamefont {L.~C.~R.}\ \bibnamefont {Wijewardhana}},\
  }\href {\doibase 10.1142/S0217732316500905} {\bibfield  {journal} {\bibinfo
  {journal} {Mod. Phys. Lett.}\ }\textbf {\bibinfo {volume} {A31}},\ \bibinfo
  {pages} {1650090} (\bibinfo {year} {2016}{\natexlab{b}})},\ \Eprint
  {http://arxiv.org/abs/1512.01709} {arXiv:1512.01709 [hep-ph]} \BibitemShut
  {NoStop}%
\bibitem [{\citenamefont {Eby}\ \emph {et~al.}(2016{\natexlab{c}})\citenamefont
  {Eby}, \citenamefont {Leembruggen}, \citenamefont {Suranyi},\ and\
  \citenamefont {Wijewardhana}}]{Eby:2016cnq}%
  \BibitemOpen
  \bibfield  {author} {\bibinfo {author} {\bibfnamefont {J.}~\bibnamefont
  {Eby}}, \bibinfo {author} {\bibfnamefont {M.}~\bibnamefont {Leembruggen}},
  \bibinfo {author} {\bibfnamefont {P.}~\bibnamefont {Suranyi}}, \ and\
  \bibinfo {author} {\bibfnamefont {L.~C.~R.}\ \bibnamefont {Wijewardhana}},\
  }\href {\doibase 10.1007/JHEP12(2016)066} {\bibfield  {journal} {\bibinfo
  {journal} {JHEP}\ }\textbf {\bibinfo {volume} {12}},\ \bibinfo {pages} {066}
  (\bibinfo {year} {2016}{\natexlab{c}})},\ \Eprint
  {http://arxiv.org/abs/1608.06911} {arXiv:1608.06911 [astro-ph.CO]}
  \BibitemShut {NoStop}%
\bibitem [{\citenamefont {Cotner}(2016)}]{Cotner:2016aaq}%
  \BibitemOpen
  \bibfield  {author} {\bibinfo {author} {\bibfnamefont {E.}~\bibnamefont
  {Cotner}},\ }\href {\doibase 10.1103/PhysRevD.94.063503} {\bibfield
  {journal} {\bibinfo  {journal} {Phys. Rev.}\ }\textbf {\bibinfo {volume}
  {D94}},\ \bibinfo {pages} {063503} (\bibinfo {year} {2016})},\ \Eprint
  {http://arxiv.org/abs/1608.00547} {arXiv:1608.00547 [astro-ph.CO]}
  \BibitemShut {NoStop}%
\bibitem [{\citenamefont {Davidson}\ and\ \citenamefont
  {Schwetz}(2016)}]{Davidson:2016uok}%
  \BibitemOpen
  \bibfield  {author} {\bibinfo {author} {\bibfnamefont {S.}~\bibnamefont
  {Davidson}}\ and\ \bibinfo {author} {\bibfnamefont {T.}~\bibnamefont
  {Schwetz}},\ }\href {\doibase 10.1103/PhysRevD.93.123509} {\bibfield
  {journal} {\bibinfo  {journal} {Phys. Rev.}\ }\textbf {\bibinfo {volume}
  {D93}},\ \bibinfo {pages} {123509} (\bibinfo {year} {2016})},\ \Eprint
  {http://arxiv.org/abs/1603.04249} {arXiv:1603.04249 [astro-ph.CO]}
  \BibitemShut {NoStop}%
\bibitem [{\citenamefont {Chavanis}(2016)}]{Chavanis:2016dab}%
  \BibitemOpen
  \bibfield  {author} {\bibinfo {author} {\bibfnamefont {P.-H.}\ \bibnamefont
  {Chavanis}},\ }\href {\doibase 10.1103/PhysRevD.94.083007} {\bibfield
  {journal} {\bibinfo  {journal} {Phys. Rev.}\ }\textbf {\bibinfo {volume}
  {D94}},\ \bibinfo {pages} {083007} (\bibinfo {year} {2016})},\ \Eprint
  {http://arxiv.org/abs/1604.05904} {arXiv:1604.05904 [astro-ph.CO]}
  \BibitemShut {NoStop}%
\bibitem [{\citenamefont {Levkov}\ \emph {et~al.}(2017)\citenamefont {Levkov},
  \citenamefont {Panin},\ and\ \citenamefont {Tkachev}}]{Levkov:2016rkk}%
  \BibitemOpen
  \bibfield  {author} {\bibinfo {author} {\bibfnamefont {D.~G.}\ \bibnamefont
  {Levkov}}, \bibinfo {author} {\bibfnamefont {A.~G.}\ \bibnamefont {Panin}}, \
  and\ \bibinfo {author} {\bibfnamefont {I.~I.}\ \bibnamefont {Tkachev}},\
  }\href {\doibase 10.1103/PhysRevLett.118.011301} {\bibfield  {journal}
  {\bibinfo  {journal} {Phys. Rev. Lett.}\ }\textbf {\bibinfo {volume} {118}},\
  \bibinfo {pages} {011301} (\bibinfo {year} {2017})},\ \Eprint
  {http://arxiv.org/abs/1609.03611} {arXiv:1609.03611 [astro-ph.CO]}
  \BibitemShut {NoStop}%
\bibitem [{\citenamefont {Hui}\ \emph {et~al.}(2017)\citenamefont {Hui},
  \citenamefont {Ostriker}, \citenamefont {Tremaine},\ and\ \citenamefont
  {Witten}}]{Hui:2016ltb}%
  \BibitemOpen
  \bibfield  {author} {\bibinfo {author} {\bibfnamefont {L.}~\bibnamefont
  {Hui}}, \bibinfo {author} {\bibfnamefont {J.~P.}\ \bibnamefont {Ostriker}},
  \bibinfo {author} {\bibfnamefont {S.}~\bibnamefont {Tremaine}}, \ and\
  \bibinfo {author} {\bibfnamefont {E.}~\bibnamefont {Witten}},\ }\href
  {\doibase 10.1103/PhysRevD.95.043541} {\bibfield  {journal} {\bibinfo
  {journal} {Phys. Rev.}\ }\textbf {\bibinfo {volume} {D95}},\ \bibinfo {pages}
  {043541} (\bibinfo {year} {2017})},\ \Eprint
  {http://arxiv.org/abs/1610.08297} {arXiv:1610.08297 [astro-ph.CO]}
  \BibitemShut {NoStop}%
\bibitem [{\citenamefont {Bai}\ \emph {et~al.}(2016)\citenamefont {Bai},
  \citenamefont {Barger},\ and\ \citenamefont {Berger}}]{Bai:2016wpg}%
  \BibitemOpen
  \bibfield  {author} {\bibinfo {author} {\bibfnamefont {Y.}~\bibnamefont
  {Bai}}, \bibinfo {author} {\bibfnamefont {V.}~\bibnamefont {Barger}}, \ and\
  \bibinfo {author} {\bibfnamefont {J.}~\bibnamefont {Berger}},\ }\href
  {\doibase 10.1007/JHEP12(2016)127} {\bibfield  {journal} {\bibinfo  {journal}
  {JHEP}\ }\textbf {\bibinfo {volume} {12}},\ \bibinfo {pages} {127} (\bibinfo
  {year} {2016})},\ \Eprint {http://arxiv.org/abs/1612.00438} {arXiv:1612.00438
  [hep-ph]} \BibitemShut {NoStop}%
\bibitem [{\citenamefont {Eby}\ \emph {et~al.}(2017)\citenamefont {Eby},
  \citenamefont {Leembruggen}, \citenamefont {Leeney}, \citenamefont
  {Suranyi},\ and\ \citenamefont {Wijewardhana}}]{Eby:2017xaw}%
  \BibitemOpen
  \bibfield  {author} {\bibinfo {author} {\bibfnamefont {J.}~\bibnamefont
  {Eby}}, \bibinfo {author} {\bibfnamefont {M.}~\bibnamefont {Leembruggen}},
  \bibinfo {author} {\bibfnamefont {J.}~\bibnamefont {Leeney}}, \bibinfo
  {author} {\bibfnamefont {P.}~\bibnamefont {Suranyi}}, \ and\ \bibinfo
  {author} {\bibfnamefont {L.~C.~R.}\ \bibnamefont {Wijewardhana}},\
  }\href@noop {} {\  (\bibinfo {year} {2017})},\ \Eprint
  {http://arxiv.org/abs/1701.01476} {arXiv:1701.01476 [astro-ph.CO]}
  \BibitemShut {NoStop}%
\bibitem [{\citenamefont {Soni}\ and\ \citenamefont
  {Zhang}(2016)}]{Soni:2016gzf}%
  \BibitemOpen
  \bibfield  {author} {\bibinfo {author} {\bibfnamefont {A.}~\bibnamefont
  {Soni}}\ and\ \bibinfo {author} {\bibfnamefont {Y.}~\bibnamefont {Zhang}},\
  }\href {\doibase 10.1103/PhysRevD.93.115025} {\bibfield  {journal} {\bibinfo
  {journal} {Phys. Rev.}\ }\textbf {\bibinfo {volume} {D93}},\ \bibinfo {pages}
  {115025} (\bibinfo {year} {2016})},\ \Eprint
  {http://arxiv.org/abs/1602.00714} {arXiv:1602.00714 [hep-ph]} \BibitemShut
  {NoStop}%
\bibitem [{\citenamefont {Giudice}\ \emph {et~al.}(2016)\citenamefont
  {Giudice}, \citenamefont {McCullough},\ and\ \citenamefont
  {Urbano}}]{Giudice:2016zpa}%
  \BibitemOpen
  \bibfield  {author} {\bibinfo {author} {\bibfnamefont {G.~F.}\ \bibnamefont
  {Giudice}}, \bibinfo {author} {\bibfnamefont {M.}~\bibnamefont {McCullough}},
  \ and\ \bibinfo {author} {\bibfnamefont {A.}~\bibnamefont {Urbano}},\ }\href
  {\doibase 10.1088/1475-7516/2016/10/001} {\bibfield  {journal} {\bibinfo
  {journal} {JCAP}\ }\textbf {\bibinfo {volume} {1610}},\ \bibinfo {pages}
  {001} (\bibinfo {year} {2016})},\ \Eprint {http://arxiv.org/abs/1605.01209}
  {arXiv:1605.01209 [hep-ph]} \BibitemShut {NoStop}%
\bibitem [{\citenamefont {Cardoso}\ \emph {et~al.}(2017)\citenamefont
  {Cardoso}, \citenamefont {Franzin}, \citenamefont {Maselli}, \citenamefont
  {Pani},\ and\ \citenamefont {Raposo}}]{Cardoso:2017cfl}%
  \BibitemOpen
  \bibfield  {author} {\bibinfo {author} {\bibfnamefont {V.}~\bibnamefont
  {Cardoso}}, \bibinfo {author} {\bibfnamefont {E.}~\bibnamefont {Franzin}},
  \bibinfo {author} {\bibfnamefont {A.}~\bibnamefont {Maselli}}, \bibinfo
  {author} {\bibfnamefont {P.}~\bibnamefont {Pani}}, \ and\ \bibinfo {author}
  {\bibfnamefont {G.}~\bibnamefont {Raposo}},\ }\href@noop {} {\  (\bibinfo
  {year} {2017})},\ \Eprint {http://arxiv.org/abs/1701.01116} {arXiv:1701.01116
  [gr-qc]} \BibitemShut {NoStop}%
\bibitem [{\citenamefont {Dev}\ \emph {et~al.}(2016)\citenamefont {Dev},
  \citenamefont {Lindner},\ and\ \citenamefont {Ohmer}}]{Dev:2016hxv}%
  \BibitemOpen
  \bibfield  {author} {\bibinfo {author} {\bibfnamefont {P.~S.~B.}\
  \bibnamefont {Dev}}, \bibinfo {author} {\bibfnamefont {M.}~\bibnamefont
  {Lindner}}, \ and\ \bibinfo {author} {\bibfnamefont {S.}~\bibnamefont
  {Ohmer}},\ }\href@noop {} {\  (\bibinfo {year} {2016})},\ \Eprint
  {http://arxiv.org/abs/1609.03939} {arXiv:1609.03939 [hep-ph]} \BibitemShut
  {NoStop}%
\bibitem [{\citenamefont {Leung}\ \emph {et~al.}(2011)\citenamefont {Leung},
  \citenamefont {Chu},\ and\ \citenamefont {Lin}}]{Leung:2011zz}%
  \BibitemOpen
  \bibfield  {author} {\bibinfo {author} {\bibfnamefont {S.~C.}\ \bibnamefont
  {Leung}}, \bibinfo {author} {\bibfnamefont {M.~C.}\ \bibnamefont {Chu}}, \
  and\ \bibinfo {author} {\bibfnamefont {L.~M.}\ \bibnamefont {Lin}},\ }\href
  {\doibase 10.1103/PhysRevD.84.107301} {\bibfield  {journal} {\bibinfo
  {journal} {Phys. Rev.}\ }\textbf {\bibinfo {volume} {D84}},\ \bibinfo {pages}
  {107301} (\bibinfo {year} {2011})},\ \Eprint {http://arxiv.org/abs/1111.1787}
  {arXiv:1111.1787 [astro-ph.CO]} \BibitemShut {NoStop}%
\bibitem [{\citenamefont {Leung}\ \emph {et~al.}(2013)\citenamefont {Leung},
  \citenamefont {Chu}, \citenamefont {Lin},\ and\ \citenamefont
  {Wong}}]{Leung:2013pra}%
  \BibitemOpen
  \bibfield  {author} {\bibinfo {author} {\bibfnamefont {S.~C.}\ \bibnamefont
  {Leung}}, \bibinfo {author} {\bibfnamefont {M.~C.}\ \bibnamefont {Chu}},
  \bibinfo {author} {\bibfnamefont {L.~M.}\ \bibnamefont {Lin}}, \ and\
  \bibinfo {author} {\bibfnamefont {K.~W.}\ \bibnamefont {Wong}},\ }\href
  {\doibase 10.1103/PhysRevD.87.123506} {\bibfield  {journal} {\bibinfo
  {journal} {Phys. Rev.}\ }\textbf {\bibinfo {volume} {D87}},\ \bibinfo {pages}
  {123506} (\bibinfo {year} {2013})},\ \Eprint {http://arxiv.org/abs/1305.6142}
  {arXiv:1305.6142 [astro-ph.CO]} \BibitemShut {NoStop}%
\bibitem [{\citenamefont {Tolos}\ and\ \citenamefont
  {Schaffner-Bielich}(2015)}]{Tolos:2015qra}%
  \BibitemOpen
  \bibfield  {author} {\bibinfo {author} {\bibfnamefont {L.}~\bibnamefont
  {Tolos}}\ and\ \bibinfo {author} {\bibfnamefont {J.}~\bibnamefont
  {Schaffner-Bielich}},\ }\href {\doibase 10.1103/PhysRevD.92.123002}
  {\bibfield  {journal} {\bibinfo  {journal} {Phys. Rev.}\ }\textbf {\bibinfo
  {volume} {D92}},\ \bibinfo {pages} {123002} (\bibinfo {year} {2015})},\
  \Eprint {http://arxiv.org/abs/1507.08197} {arXiv:1507.08197 [astro-ph.HE]}
  \BibitemShut {NoStop}%
\bibitem [{\citenamefont {Mukhopadhyay}\ and\ \citenamefont
  {Schaffner-Bielich}(2016)}]{Mukhopadhyay:2015xhs}%
  \BibitemOpen
  \bibfield  {author} {\bibinfo {author} {\bibfnamefont {P.}~\bibnamefont
  {Mukhopadhyay}}\ and\ \bibinfo {author} {\bibfnamefont {J.}~\bibnamefont
  {Schaffner-Bielich}},\ }\href {\doibase 10.1103/PhysRevD.93.083009}
  {\bibfield  {journal} {\bibinfo  {journal} {Phys. Rev.}\ }\textbf {\bibinfo
  {volume} {D93}},\ \bibinfo {pages} {083009} (\bibinfo {year} {2016})},\
  \Eprint {http://arxiv.org/abs/1511.00238} {arXiv:1511.00238 [astro-ph.HE]}
  \BibitemShut {NoStop}%
\bibitem [{\citenamefont {Lynden-Bell}\ and\ \citenamefont
  {Wood}(1968)}]{LyndenBell:1968yw}%
  \BibitemOpen
  \bibfield  {author} {\bibinfo {author} {\bibfnamefont {D.}~\bibnamefont
  {Lynden-Bell}}\ and\ \bibinfo {author} {\bibfnamefont {R.}~\bibnamefont
  {Wood}},\ }\href@noop {} {\bibfield  {journal} {\bibinfo  {journal} {Mon.
  Not. Roy. Astron. Soc.}\ }\textbf {\bibinfo {volume} {138}},\ \bibinfo
  {pages} {495} (\bibinfo {year} {1968})}\BibitemShut {NoStop}%
\bibitem [{\citenamefont {Kouvaris}\ and\ \citenamefont
  {Tinyakov}(2010)}]{Kouvaris:2010vv}%
  \BibitemOpen
  \bibfield  {author} {\bibinfo {author} {\bibfnamefont {C.}~\bibnamefont
  {Kouvaris}}\ and\ \bibinfo {author} {\bibfnamefont {P.}~\bibnamefont
  {Tinyakov}},\ }\href {\doibase 10.1103/PhysRevD.82.063531} {\bibfield
  {journal} {\bibinfo  {journal} {Phys. Rev.}\ }\textbf {\bibinfo {volume}
  {D82}},\ \bibinfo {pages} {063531} (\bibinfo {year} {2010})},\ \Eprint
  {http://arxiv.org/abs/1004.0586} {arXiv:1004.0586 [astro-ph.GA]} \BibitemShut
  {NoStop}%
\bibitem [{\citenamefont {Fan}\ \emph {et~al.}(2013)\citenamefont {Fan},
  \citenamefont {Katz}, \citenamefont {Randall},\ and\ \citenamefont
  {Reece}}]{Fan:2013yva}%
  \BibitemOpen
  \bibfield  {author} {\bibinfo {author} {\bibfnamefont {J.}~\bibnamefont
  {Fan}}, \bibinfo {author} {\bibfnamefont {A.}~\bibnamefont {Katz}}, \bibinfo
  {author} {\bibfnamefont {L.}~\bibnamefont {Randall}}, \ and\ \bibinfo
  {author} {\bibfnamefont {M.}~\bibnamefont {Reece}},\ }\href {\doibase
  10.1016/j.dark.2013.07.001} {\bibfield  {journal} {\bibinfo  {journal} {Phys.
  Dark Univ.}\ }\textbf {\bibinfo {volume} {2}},\ \bibinfo {pages} {139}
  (\bibinfo {year} {2013})},\ \Eprint {http://arxiv.org/abs/1303.1521}
  {arXiv:1303.1521 [astro-ph.CO]} \BibitemShut {NoStop}%
\bibitem [{\citenamefont {Hartle}(1967)}]{Hartle:1967he}%
  \BibitemOpen
  \bibfield  {author} {\bibinfo {author} {\bibfnamefont {J.~B.}\ \bibnamefont
  {Hartle}},\ }\href {\doibase 10.1086/149400} {\bibfield  {journal} {\bibinfo
  {journal} {Astrophys. J.}\ }\textbf {\bibinfo {volume} {150}},\ \bibinfo
  {pages} {1005} (\bibinfo {year} {1967})}\BibitemShut {NoStop}%
\bibitem [{\citenamefont {{Hartle}}\ and\ \citenamefont
  {{Thorne}}(1968)}]{1968ApJ...153..807H}%
  \BibitemOpen
  \bibfield  {author} {\bibinfo {author} {\bibfnamefont {J.~B.}\ \bibnamefont
  {{Hartle}}}\ and\ \bibinfo {author} {\bibfnamefont {K.~S.}\ \bibnamefont
  {{Thorne}}},\ }\href {\doibase 10.1086/149707} {\bibfield  {journal}
  {\bibinfo  {journal} {\apj}\ }\textbf {\bibinfo {volume} {153}},\ \bibinfo
  {pages} {807} (\bibinfo {year} {1968})}\BibitemShut {NoStop}%
\bibitem [{\citenamefont {Hinderer}(2008)}]{Hinderer:2007mb}%
  \BibitemOpen
  \bibfield  {author} {\bibinfo {author} {\bibfnamefont {T.}~\bibnamefont
  {Hinderer}},\ }\href {\doibase 10.1086/533487} {\bibfield  {journal}
  {\bibinfo  {journal} {Astrophys. J.}\ }\textbf {\bibinfo {volume} {677}},\
  \bibinfo {pages} {1216} (\bibinfo {year} {2008})},\ \bibinfo {note}
  {{Erratum: {\it ibid.}
  \href{https://dx.doi.org/10.1088/0004-637X/697/1/964}{{\bf 697}, 964
  (2009)}}},\ \Eprint {http://arxiv.org/abs/0711.2420} {arXiv:0711.2420
  [astro-ph]} \BibitemShut {NoStop}%
\bibitem [{\citenamefont {Akmal}\ \emph {et~al.}(1998)\citenamefont {Akmal},
  \citenamefont {Pandharipande},\ and\ \citenamefont
  {Ravenhall}}]{Akmal:1998cf}%
  \BibitemOpen
  \bibfield  {author} {\bibinfo {author} {\bibfnamefont {A.}~\bibnamefont
  {Akmal}}, \bibinfo {author} {\bibfnamefont {V.~R.}\ \bibnamefont
  {Pandharipande}}, \ and\ \bibinfo {author} {\bibfnamefont {D.~G.}\
  \bibnamefont {Ravenhall}},\ }\href {\doibase 10.1103/PhysRevC.58.1804}
  {\bibfield  {journal} {\bibinfo  {journal} {Phys. Rev.}\ }\textbf {\bibinfo
  {volume} {C58}},\ \bibinfo {pages} {1804} (\bibinfo {year} {1998})},\ \Eprint
  {http://arxiv.org/abs/nucl-th/9804027} {arXiv:nucl-th/9804027 [nucl-th]}
  \BibitemShut {NoStop}%
\bibitem [{\citenamefont {Muller}\ and\ \citenamefont
  {Serot}(1995)}]{Muller:1995ji}%
  \BibitemOpen
  \bibfield  {author} {\bibinfo {author} {\bibfnamefont {H.}~\bibnamefont
  {Muller}}\ and\ \bibinfo {author} {\bibfnamefont {B.~D.}\ \bibnamefont
  {Serot}},\ }\href {\doibase 10.1103/PhysRevC.52.2072} {\bibfield  {journal}
  {\bibinfo  {journal} {Phys. Rev.}\ }\textbf {\bibinfo {volume} {C52}},\
  \bibinfo {pages} {2072} (\bibinfo {year} {1995})},\ \Eprint
  {http://arxiv.org/abs/nucl-th/9505013} {arXiv:nucl-th/9505013 [nucl-th]}
  \BibitemShut {NoStop}%
\bibitem [{\citenamefont {{Shapiro}}\ and\ \citenamefont
  {{Teukolsky}}(1983)}]{1983bhwd.book.....S}%
  \BibitemOpen
  \bibfield  {author} {\bibinfo {author} {\bibfnamefont {S.~L.}\ \bibnamefont
  {{Shapiro}}}\ and\ \bibinfo {author} {\bibfnamefont {S.~A.}\ \bibnamefont
  {{Teukolsky}}},\ }\href@noop {} {\emph {\bibinfo {title} {{Black holes, white
  dwarfs, and neutron stars: The physics of compact objects}}}}\ (\bibinfo
  {year} {1983})\BibitemShut {NoStop}%
\bibitem [{\citenamefont {Kaup}(1968)}]{Kaup:1968zz}%
  \BibitemOpen
  \bibfield  {author} {\bibinfo {author} {\bibfnamefont {D.~J.}\ \bibnamefont
  {Kaup}},\ }\href {\doibase 10.1103/PhysRev.172.1331} {\bibfield  {journal}
  {\bibinfo  {journal} {Phys. Rev.}\ }\textbf {\bibinfo {volume} {172}},\
  \bibinfo {pages} {1331} (\bibinfo {year} {1968})}\BibitemShut {NoStop}%
\bibitem [{\citenamefont {Ruffini}\ and\ \citenamefont
  {Bonazzola}(1969)}]{Ruffini:1969qy}%
  \BibitemOpen
  \bibfield  {author} {\bibinfo {author} {\bibfnamefont {R.}~\bibnamefont
  {Ruffini}}\ and\ \bibinfo {author} {\bibfnamefont {S.}~\bibnamefont
  {Bonazzola}},\ }\href {\doibase 10.1103/PhysRev.187.1767} {\bibfield
  {journal} {\bibinfo  {journal} {Phys. Rev.}\ }\textbf {\bibinfo {volume}
  {187}},\ \bibinfo {pages} {1767} (\bibinfo {year} {1969})}\BibitemShut
  {NoStop}%
\bibitem [{\citenamefont {Colpi}\ \emph {et~al.}(1986)\citenamefont {Colpi},
  \citenamefont {Shapiro},\ and\ \citenamefont {Wasserman}}]{Colpi:1986ye}%
  \BibitemOpen
  \bibfield  {author} {\bibinfo {author} {\bibfnamefont {M.}~\bibnamefont
  {Colpi}}, \bibinfo {author} {\bibfnamefont {S.~L.}\ \bibnamefont {Shapiro}},
  \ and\ \bibinfo {author} {\bibfnamefont {I.}~\bibnamefont {Wasserman}},\
  }\href {\doibase 10.1103/PhysRevLett.57.2485} {\bibfield  {journal} {\bibinfo
   {journal} {Phys. Rev. Lett.}\ }\textbf {\bibinfo {volume} {57}},\ \bibinfo
  {pages} {2485} (\bibinfo {year} {1986})}\BibitemShut {NoStop}%
\bibitem [{\citenamefont {Liebling}\ and\ \citenamefont
  {Palenzuela}(2012)}]{Liebling:2012fv}%
  \BibitemOpen
  \bibfield  {author} {\bibinfo {author} {\bibfnamefont {S.~L.}\ \bibnamefont
  {Liebling}}\ and\ \bibinfo {author} {\bibfnamefont {C.}~\bibnamefont
  {Palenzuela}},\ }\href {\doibase 10.12942/lrr-2012-6} {\bibfield  {journal}
  {\bibinfo  {journal} {Living Rev. Rel.}\ }\textbf {\bibinfo {volume} {15}},\
  \bibinfo {pages} {6} (\bibinfo {year} {2012})},\ \Eprint
  {http://arxiv.org/abs/1202.5809} {arXiv:1202.5809 [gr-qc]} \BibitemShut
  {NoStop}%
\bibitem [{\citenamefont {Seidel}\ and\ \citenamefont
  {Suen}(1991)}]{Seidel:1991zh}%
  \BibitemOpen
  \bibfield  {author} {\bibinfo {author} {\bibfnamefont {E.}~\bibnamefont
  {Seidel}}\ and\ \bibinfo {author} {\bibfnamefont {W.~M.}\ \bibnamefont
  {Suen}},\ }\href {\doibase 10.1103/PhysRevLett.66.1659} {\bibfield  {journal}
  {\bibinfo  {journal} {Phys. Rev. Lett.}\ }\textbf {\bibinfo {volume} {66}},\
  \bibinfo {pages} {1659} (\bibinfo {year} {1991})}\BibitemShut {NoStop}%
\bibitem [{\citenamefont {Brito}\ \emph
  {et~al.}(2016{\natexlab{b}})\citenamefont {Brito}, \citenamefont {Cardoso},
  \citenamefont {Herdeiro},\ and\ \citenamefont {Radu}}]{Brito:2015pxa}%
  \BibitemOpen
  \bibfield  {author} {\bibinfo {author} {\bibfnamefont {R.}~\bibnamefont
  {Brito}}, \bibinfo {author} {\bibfnamefont {V.}~\bibnamefont {Cardoso}},
  \bibinfo {author} {\bibfnamefont {C.~A.~R.}\ \bibnamefont {Herdeiro}}, \ and\
  \bibinfo {author} {\bibfnamefont {E.}~\bibnamefont {Radu}},\ }\href {\doibase
  10.1016/j.physletb.2015.11.051} {\bibfield  {journal} {\bibinfo  {journal}
  {Phys. Lett.}\ }\textbf {\bibinfo {volume} {B752}},\ \bibinfo {pages} {291}
  (\bibinfo {year} {2016}{\natexlab{b}})},\ \Eprint
  {http://arxiv.org/abs/1508.05395} {arXiv:1508.05395 [gr-qc]} \BibitemShut
  {NoStop}%
\bibitem [{\citenamefont {Sikivie}\ and\ \citenamefont
  {Yang}(2009)}]{Sikivie:2009qn}%
  \BibitemOpen
  \bibfield  {author} {\bibinfo {author} {\bibfnamefont {P.}~\bibnamefont
  {Sikivie}}\ and\ \bibinfo {author} {\bibfnamefont {Q.}~\bibnamefont {Yang}},\
  }\href {\doibase 10.1103/PhysRevLett.103.111301} {\bibfield  {journal}
  {\bibinfo  {journal} {Phys. Rev. Lett.}\ }\textbf {\bibinfo {volume} {103}},\
  \bibinfo {pages} {111301} (\bibinfo {year} {2009})},\ \Eprint
  {http://arxiv.org/abs/0901.1106} {arXiv:0901.1106 [hep-ph]} \BibitemShut
  {NoStop}%
\bibitem [{\citenamefont {Will}(2014)}]{Will:2014kxa}%
  \BibitemOpen
  \bibfield  {author} {\bibinfo {author} {\bibfnamefont {C.~M.}\ \bibnamefont
  {Will}},\ }\href {\doibase 10.12942/lrr-2014-4} {\bibfield  {journal}
  {\bibinfo  {journal} {Living Rev. Rel.}\ }\textbf {\bibinfo {volume} {17}},\
  \bibinfo {pages} {4} (\bibinfo {year} {2014})},\ \Eprint
  {http://arxiv.org/abs/1403.7377} {arXiv:1403.7377 [gr-qc]} \BibitemShut
  {NoStop}%
\bibitem [{\citenamefont {Lattimer}\ and\ \citenamefont
  {Prakash}(2007)}]{Lattimer:2006xb}%
  \BibitemOpen
  \bibfield  {author} {\bibinfo {author} {\bibfnamefont {J.~M.}\ \bibnamefont
  {Lattimer}}\ and\ \bibinfo {author} {\bibfnamefont {M.}~\bibnamefont
  {Prakash}},\ }\href {\doibase 10.1016/j.physrep.2007.02.003} {\bibfield
  {journal} {\bibinfo  {journal} {Phys. Rept.}\ }\textbf {\bibinfo {volume}
  {442}},\ \bibinfo {pages} {109} (\bibinfo {year} {2007})},\ \Eprint
  {http://arxiv.org/abs/astro-ph/0612440} {arXiv:astro-ph/0612440 [astro-ph]}
  \BibitemShut {NoStop}%
\bibitem [{\citenamefont {Aasi}\ \emph {et~al.}(2015)\citenamefont {Aasi} \emph
  {et~al.}}]{0264-9381-32-11-115012}%
  \BibitemOpen
  \bibfield  {author} {\bibinfo {author} {\bibfnamefont {J.}~\bibnamefont
  {Aasi}} \emph {et~al.},\ }\href@noop {} {\bibfield  {journal} {\bibinfo
  {journal} {Classical and Quantum Gravity}\ }\textbf {\bibinfo {volume}
  {32}},\ \bibinfo {pages} {115012} (\bibinfo {year} {2015})}\BibitemShut
  {NoStop}%
\bibitem [{\citenamefont {Acernese}\ \emph {et~al.}(2015)\citenamefont
  {Acernese} \emph {et~al.}}]{0264-9381-32-2-024001}%
  \BibitemOpen
  \bibfield  {author} {\bibinfo {author} {\bibfnamefont {F.}~\bibnamefont
  {Acernese}} \emph {et~al.},\ }\href@noop {} {\bibfield  {journal} {\bibinfo
  {journal} {Classical and Quantum Gravity}\ }\textbf {\bibinfo {volume}
  {32}},\ \bibinfo {pages} {024001} (\bibinfo {year} {2015})}\BibitemShut
  {NoStop}%
\bibitem [{\citenamefont {Sathyaprakash}\ and\ \citenamefont
  {Schutz}(2009)}]{Sathyaprakash:2009xs}%
  \BibitemOpen
  \bibfield  {author} {\bibinfo {author} {\bibfnamefont {B.~S.}\ \bibnamefont
  {Sathyaprakash}}\ and\ \bibinfo {author} {\bibfnamefont {B.~F.}\ \bibnamefont
  {Schutz}},\ }\href {\doibase 10.12942/lrr-2009-2} {\bibfield  {journal}
  {\bibinfo  {journal} {Living Rev. Rel.}\ }\textbf {\bibinfo {volume} {12}},\
  \bibinfo {pages} {2} (\bibinfo {year} {2009})},\ \Eprint
  {http://arxiv.org/abs/0903.0338} {arXiv:0903.0338 [gr-qc]} \BibitemShut
  {NoStop}%
\bibitem [{\citenamefont {Flanagan}\ and\ \citenamefont
  {Hinderer}(2008)}]{Flanagan:2007ix}%
  \BibitemOpen
  \bibfield  {author} {\bibinfo {author} {\bibfnamefont {E.~E.}\ \bibnamefont
  {Flanagan}}\ and\ \bibinfo {author} {\bibfnamefont {T.}~\bibnamefont
  {Hinderer}},\ }\href {\doibase 10.1103/PhysRevD.77.021502} {\bibfield
  {journal} {\bibinfo  {journal} {Phys. Rev.}\ }\textbf {\bibinfo {volume}
  {D77}},\ \bibinfo {pages} {021502} (\bibinfo {year} {2008})},\ \Eprint
  {http://arxiv.org/abs/0709.1915} {arXiv:0709.1915 [astro-ph]} \BibitemShut
  {NoStop}%
\bibitem [{\citenamefont {Cottam}\ \emph {et~al.}(2002)\citenamefont {Cottam},
  \citenamefont {Paerels},\ and\ \citenamefont {Mendez}}]{Cottam:2002cu}%
  \BibitemOpen
  \bibfield  {author} {\bibinfo {author} {\bibfnamefont {J.}~\bibnamefont
  {Cottam}}, \bibinfo {author} {\bibfnamefont {F.}~\bibnamefont {Paerels}}, \
  and\ \bibinfo {author} {\bibfnamefont {M.}~\bibnamefont {Mendez}},\ }\href
  {\doibase 10.1038/nature01159} {\bibfield  {journal} {\bibinfo  {journal}
  {Nature}\ }\textbf {\bibinfo {volume} {420}},\ \bibinfo {pages} {51}
  (\bibinfo {year} {2002})},\ \Eprint {http://arxiv.org/abs/astro-ph/0211126}
  {arXiv:astro-ph/0211126 [astro-ph]} \BibitemShut {NoStop}%
\bibitem [{\citenamefont {{V{\"o}lkel}}\ and\ \citenamefont
  {{Kokkotas}}(2017)}]{2017arXiv170308156V}%
  \BibitemOpen
  \bibfield  {author} {\bibinfo {author} {\bibfnamefont {S.~H.}\ \bibnamefont
  {{V{\"o}lkel}}}\ and\ \bibinfo {author} {\bibfnamefont {K.~D.}\ \bibnamefont
  {{Kokkotas}}},\ }\href@noop {} {\bibfield  {journal} {\bibinfo  {journal}
  {ArXiv e-prints}\ } (\bibinfo {year} {2017})},\ \Eprint
  {http://arxiv.org/abs/1703.08156} {arXiv:1703.08156 [gr-qc]} \BibitemShut
  {NoStop}%
\bibitem [{\citenamefont {Hessels}\ \emph {et~al.}(2006)\citenamefont
  {Hessels}, \citenamefont {Ransom}, \citenamefont {Stairs}, \citenamefont
  {Freire}, \citenamefont {Kaspi},\ and\ \citenamefont
  {Camilo}}]{Hessels:2006ze}%
  \BibitemOpen
  \bibfield  {author} {\bibinfo {author} {\bibfnamefont {J.~W.~T.}\
  \bibnamefont {Hessels}}, \bibinfo {author} {\bibfnamefont {S.~M.}\
  \bibnamefont {Ransom}}, \bibinfo {author} {\bibfnamefont {I.~H.}\
  \bibnamefont {Stairs}}, \bibinfo {author} {\bibfnamefont {P.~C.~C.}\
  \bibnamefont {Freire}}, \bibinfo {author} {\bibfnamefont {V.~M.}\
  \bibnamefont {Kaspi}}, \ and\ \bibinfo {author} {\bibfnamefont
  {F.}~\bibnamefont {Camilo}},\ }\href {\doibase 10.1126/science.1123430}
  {\bibfield  {journal} {\bibinfo  {journal} {Science}\ }\textbf {\bibinfo
  {volume} {311}},\ \bibinfo {pages} {1901} (\bibinfo {year} {2006})},\ \Eprint
  {http://arxiv.org/abs/astro-ph/0601337} {arXiv:astro-ph/0601337 [astro-ph]}
  \BibitemShut {NoStop}%
\bibitem [{\citenamefont {Lattimer}\ and\ \citenamefont
  {Prakash}(2001)}]{0004-637X-550-1-426}%
  \BibitemOpen
  \bibfield  {author} {\bibinfo {author} {\bibfnamefont {J.~M.}\ \bibnamefont
  {Lattimer}}\ and\ \bibinfo {author} {\bibfnamefont {M.}~\bibnamefont
  {Prakash}},\ }\href@noop {} {\bibfield  {journal} {\bibinfo  {journal} {The
  Astrophysical Journal}\ }\textbf {\bibinfo {volume} {550}},\ \bibinfo {pages}
  {426} (\bibinfo {year} {2001})}\BibitemShut {NoStop}%
\bibitem [{\citenamefont {Lattimer}\ and\ \citenamefont
  {Lim}(2013)}]{0004-637X-771-1-51}%
  \BibitemOpen
  \bibfield  {author} {\bibinfo {author} {\bibfnamefont {J.~M.}\ \bibnamefont
  {Lattimer}}\ and\ \bibinfo {author} {\bibfnamefont {Y.}~\bibnamefont {Lim}},\
  }\href@noop {} {\bibfield  {journal} {\bibinfo  {journal} {The Astrophysical
  Journal}\ }\textbf {\bibinfo {volume} {771}},\ \bibinfo {pages} {51}
  (\bibinfo {year} {2013})}\BibitemShut {NoStop}%
\bibitem [{\citenamefont {Lattimer}\ and\ \citenamefont
  {Schutz}(2005)}]{Lattimer:2004nj}%
  \BibitemOpen
  \bibfield  {author} {\bibinfo {author} {\bibfnamefont {J.~M.}\ \bibnamefont
  {Lattimer}}\ and\ \bibinfo {author} {\bibfnamefont {B.~F.}\ \bibnamefont
  {Schutz}},\ }\href {\doibase 10.1086/431543} {\bibfield  {journal} {\bibinfo
  {journal} {Astrophys. J.}\ }\textbf {\bibinfo {volume} {629}},\ \bibinfo
  {pages} {979} (\bibinfo {year} {2005})},\ \Eprint
  {http://arxiv.org/abs/astro-ph/0411470} {arXiv:astro-ph/0411470 [astro-ph]}
  \BibitemShut {NoStop}%
\bibitem [{\citenamefont {Binnington}\ and\ \citenamefont
  {Poisson}(2009)}]{Binnington:2009bb}%
  \BibitemOpen
  \bibfield  {author} {\bibinfo {author} {\bibfnamefont {T.}~\bibnamefont
  {Binnington}}\ and\ \bibinfo {author} {\bibfnamefont {E.}~\bibnamefont
  {Poisson}},\ }\href {\doibase 10.1103/PhysRevD.80.084018} {\bibfield
  {journal} {\bibinfo  {journal} {Phys. Rev.}\ }\textbf {\bibinfo {volume}
  {D80}},\ \bibinfo {pages} {084018} (\bibinfo {year} {2009})},\ \Eprint
  {http://arxiv.org/abs/0906.1366} {arXiv:0906.1366 [gr-qc]} \BibitemShut
  {NoStop}%
\bibitem [{\citenamefont {Damour}\ and\ \citenamefont
  {Nagar}(2009)}]{Damour:2009vw}%
  \BibitemOpen
  \bibfield  {author} {\bibinfo {author} {\bibfnamefont {T.}~\bibnamefont
  {Damour}}\ and\ \bibinfo {author} {\bibfnamefont {A.}~\bibnamefont {Nagar}},\
  }\href {\doibase 10.1103/PhysRevD.80.084035} {\bibfield  {journal} {\bibinfo
  {journal} {Phys. Rev.}\ }\textbf {\bibinfo {volume} {D80}},\ \bibinfo {pages}
  {084035} (\bibinfo {year} {2009})},\ \Eprint {http://arxiv.org/abs/0906.0096}
  {arXiv:0906.0096 [gr-qc]} \BibitemShut {NoStop}%
\bibitem [{\citenamefont {Damour}\ and\ \citenamefont
  {Nagar}(2010)}]{Damour:2009wj}%
  \BibitemOpen
  \bibfield  {author} {\bibinfo {author} {\bibfnamefont {T.}~\bibnamefont
  {Damour}}\ and\ \bibinfo {author} {\bibfnamefont {A.}~\bibnamefont {Nagar}},\
  }\href {\doibase 10.1103/PhysRevD.81.084016} {\bibfield  {journal} {\bibinfo
  {journal} {Phys. Rev.}\ }\textbf {\bibinfo {volume} {D81}},\ \bibinfo {pages}
  {084016} (\bibinfo {year} {2010})},\ \Eprint {http://arxiv.org/abs/0911.5041}
  {arXiv:0911.5041 [gr-qc]} \BibitemShut {NoStop}%
\bibitem [{\citenamefont {Maselli}\ \emph
  {et~al.}(2013{\natexlab{a}})\citenamefont {Maselli}, \citenamefont
  {Gualtieri},\ and\ \citenamefont {Ferrari}}]{Maselli:2013rza}%
  \BibitemOpen
  \bibfield  {author} {\bibinfo {author} {\bibfnamefont {A.}~\bibnamefont
  {Maselli}}, \bibinfo {author} {\bibfnamefont {L.}~\bibnamefont {Gualtieri}},
  \ and\ \bibinfo {author} {\bibfnamefont {V.}~\bibnamefont {Ferrari}},\ }\href
  {\doibase 10.1103/PhysRevD.88.104040} {\bibfield  {journal} {\bibinfo
  {journal} {Phys. Rev.}\ }\textbf {\bibinfo {volume} {D88}},\ \bibinfo {pages}
  {104040} (\bibinfo {year} {2013}{\natexlab{a}})},\ \Eprint
  {http://arxiv.org/abs/1310.5381} {arXiv:1310.5381 [gr-qc]} \BibitemShut
  {NoStop}%
\bibitem [{\citenamefont {Damour}\ \emph {et~al.}(2012)\citenamefont {Damour},
  \citenamefont {Nagar},\ and\ \citenamefont {Villain}}]{Damour:2012yf}%
  \BibitemOpen
  \bibfield  {author} {\bibinfo {author} {\bibfnamefont {T.}~\bibnamefont
  {Damour}}, \bibinfo {author} {\bibfnamefont {A.}~\bibnamefont {Nagar}}, \
  and\ \bibinfo {author} {\bibfnamefont {L.}~\bibnamefont {Villain}},\ }\href
  {\doibase 10.1103/PhysRevD.85.123007} {\bibfield  {journal} {\bibinfo
  {journal} {Phys. Rev.}\ }\textbf {\bibinfo {volume} {D85}},\ \bibinfo {pages}
  {123007} (\bibinfo {year} {2012})},\ \Eprint {http://arxiv.org/abs/1203.4352}
  {arXiv:1203.4352 [gr-qc]} \BibitemShut {NoStop}%
\bibitem [{\citenamefont {Hinderer}\ \emph {et~al.}(2010)\citenamefont
  {Hinderer}, \citenamefont {Lackey}, \citenamefont {Lang},\ and\ \citenamefont
  {Read}}]{Hinderer:2009ca}%
  \BibitemOpen
  \bibfield  {author} {\bibinfo {author} {\bibfnamefont {T.}~\bibnamefont
  {Hinderer}}, \bibinfo {author} {\bibfnamefont {B.~D.}\ \bibnamefont
  {Lackey}}, \bibinfo {author} {\bibfnamefont {R.~N.}\ \bibnamefont {Lang}}, \
  and\ \bibinfo {author} {\bibfnamefont {J.~S.}\ \bibnamefont {Read}},\ }\href
  {\doibase 10.1103/PhysRevD.81.123016} {\bibfield  {journal} {\bibinfo
  {journal} {Phys. Rev.}\ }\textbf {\bibinfo {volume} {D81}},\ \bibinfo {pages}
  {123016} (\bibinfo {year} {2010})},\ \Eprint {http://arxiv.org/abs/0911.3535}
  {arXiv:0911.3535 [astro-ph.HE]} \BibitemShut {NoStop}%
\bibitem [{\citenamefont {Bernuzzi}\ \emph {et~al.}(2012)\citenamefont
  {Bernuzzi}, \citenamefont {Nagar}, \citenamefont {Thierfelder},\ and\
  \citenamefont {Brugmann}}]{Bernuzzi:2012ci}%
  \BibitemOpen
  \bibfield  {author} {\bibinfo {author} {\bibfnamefont {S.}~\bibnamefont
  {Bernuzzi}}, \bibinfo {author} {\bibfnamefont {A.}~\bibnamefont {Nagar}},
  \bibinfo {author} {\bibfnamefont {M.}~\bibnamefont {Thierfelder}}, \ and\
  \bibinfo {author} {\bibfnamefont {B.}~\bibnamefont {Brugmann}},\ }\href
  {\doibase 10.1103/PhysRevD.86.044030} {\bibfield  {journal} {\bibinfo
  {journal} {Phys. Rev.}\ }\textbf {\bibinfo {volume} {D86}},\ \bibinfo {pages}
  {044030} (\bibinfo {year} {2012})},\ \Eprint {http://arxiv.org/abs/1205.3403}
  {arXiv:1205.3403 [gr-qc]} \BibitemShut {NoStop}%
\bibitem [{\citenamefont {Read}\ \emph {et~al.}(2013)\citenamefont {Read},
  \citenamefont {Baiotti}, \citenamefont {Creighton}, \citenamefont {Friedman},
  \citenamefont {Giacomazzo}, \citenamefont {Kyutoku}, \citenamefont
  {Markakis}, \citenamefont {Rezzolla}, \citenamefont {Shibata},\ and\
  \citenamefont {Taniguchi}}]{Read:2013zra}%
  \BibitemOpen
  \bibfield  {author} {\bibinfo {author} {\bibfnamefont {J.~S.}\ \bibnamefont
  {Read}}, \bibinfo {author} {\bibfnamefont {L.}~\bibnamefont {Baiotti}},
  \bibinfo {author} {\bibfnamefont {J.~D.~E.}\ \bibnamefont {Creighton}},
  \bibinfo {author} {\bibfnamefont {J.~L.}\ \bibnamefont {Friedman}}, \bibinfo
  {author} {\bibfnamefont {B.}~\bibnamefont {Giacomazzo}}, \bibinfo {author}
  {\bibfnamefont {K.}~\bibnamefont {Kyutoku}}, \bibinfo {author} {\bibfnamefont
  {C.}~\bibnamefont {Markakis}}, \bibinfo {author} {\bibfnamefont
  {L.}~\bibnamefont {Rezzolla}}, \bibinfo {author} {\bibfnamefont
  {M.}~\bibnamefont {Shibata}}, \ and\ \bibinfo {author} {\bibfnamefont
  {K.}~\bibnamefont {Taniguchi}},\ }\href {\doibase 10.1103/PhysRevD.88.044042}
  {\bibfield  {journal} {\bibinfo  {journal} {Phys. Rev.}\ }\textbf {\bibinfo
  {volume} {D88}},\ \bibinfo {pages} {044042} (\bibinfo {year} {2013})},\
  \Eprint {http://arxiv.org/abs/1306.4065} {arXiv:1306.4065 [gr-qc]}
  \BibitemShut {NoStop}%
\bibitem [{\citenamefont {Baiotti}\ \emph {et~al.}(2010)\citenamefont
  {Baiotti}, \citenamefont {Damour}, \citenamefont {Giacomazzo}, \citenamefont
  {Nagar},\ and\ \citenamefont {Rezzolla}}]{Baiotti:2010xh}%
  \BibitemOpen
  \bibfield  {author} {\bibinfo {author} {\bibfnamefont {L.}~\bibnamefont
  {Baiotti}}, \bibinfo {author} {\bibfnamefont {T.}~\bibnamefont {Damour}},
  \bibinfo {author} {\bibfnamefont {B.}~\bibnamefont {Giacomazzo}}, \bibinfo
  {author} {\bibfnamefont {A.}~\bibnamefont {Nagar}}, \ and\ \bibinfo {author}
  {\bibfnamefont {L.}~\bibnamefont {Rezzolla}},\ }\href {\doibase
  10.1103/PhysRevLett.105.261101} {\bibfield  {journal} {\bibinfo  {journal}
  {Phys. Rev. Lett.}\ }\textbf {\bibinfo {volume} {105}},\ \bibinfo {pages}
  {261101} (\bibinfo {year} {2010})},\ \Eprint {http://arxiv.org/abs/1009.0521}
  {arXiv:1009.0521 [gr-qc]} \BibitemShut {NoStop}%
\bibitem [{\citenamefont {Baiotti}\ \emph {et~al.}(2011)\citenamefont
  {Baiotti}, \citenamefont {Damour}, \citenamefont {Giacomazzo}, \citenamefont
  {Nagar},\ and\ \citenamefont {Rezzolla}}]{Baiotti:2011am}%
  \BibitemOpen
  \bibfield  {author} {\bibinfo {author} {\bibfnamefont {L.}~\bibnamefont
  {Baiotti}}, \bibinfo {author} {\bibfnamefont {T.}~\bibnamefont {Damour}},
  \bibinfo {author} {\bibfnamefont {B.}~\bibnamefont {Giacomazzo}}, \bibinfo
  {author} {\bibfnamefont {A.}~\bibnamefont {Nagar}}, \ and\ \bibinfo {author}
  {\bibfnamefont {L.}~\bibnamefont {Rezzolla}},\ }\href {\doibase
  10.1103/PhysRevD.84.024017} {\bibfield  {journal} {\bibinfo  {journal} {Phys.
  Rev.}\ }\textbf {\bibinfo {volume} {D84}},\ \bibinfo {pages} {024017}
  (\bibinfo {year} {2011})},\ \Eprint {http://arxiv.org/abs/1103.3874}
  {arXiv:1103.3874 [gr-qc]} \BibitemShut {NoStop}%
\bibitem [{\citenamefont {Vines}\ \emph {et~al.}(2011)\citenamefont {Vines},
  \citenamefont {Flanagan},\ and\ \citenamefont {Hinderer}}]{Vines:2011ud}%
  \BibitemOpen
  \bibfield  {author} {\bibinfo {author} {\bibfnamefont {J.}~\bibnamefont
  {Vines}}, \bibinfo {author} {\bibfnamefont {E.~E.}\ \bibnamefont {Flanagan}},
  \ and\ \bibinfo {author} {\bibfnamefont {T.}~\bibnamefont {Hinderer}},\
  }\href {\doibase 10.1103/PhysRevD.83.084051} {\bibfield  {journal} {\bibinfo
  {journal} {Phys. Rev.}\ }\textbf {\bibinfo {volume} {D83}},\ \bibinfo {pages}
  {084051} (\bibinfo {year} {2011})},\ \Eprint {http://arxiv.org/abs/1101.1673}
  {arXiv:1101.1673 [gr-qc]} \BibitemShut {NoStop}%
\bibitem [{\citenamefont {Pannarale}\ \emph {et~al.}(2011)\citenamefont
  {Pannarale}, \citenamefont {Rezzolla}, \citenamefont {Ohme},\ and\
  \citenamefont {Read}}]{Pannarale:2011pk}%
  \BibitemOpen
  \bibfield  {author} {\bibinfo {author} {\bibfnamefont {F.}~\bibnamefont
  {Pannarale}}, \bibinfo {author} {\bibfnamefont {L.}~\bibnamefont {Rezzolla}},
  \bibinfo {author} {\bibfnamefont {F.}~\bibnamefont {Ohme}}, \ and\ \bibinfo
  {author} {\bibfnamefont {J.~S.}\ \bibnamefont {Read}},\ }\href {\doibase
  10.1103/PhysRevD.84.104017} {\bibfield  {journal} {\bibinfo  {journal} {Phys.
  Rev.}\ }\textbf {\bibinfo {volume} {D84}},\ \bibinfo {pages} {104017}
  (\bibinfo {year} {2011})},\ \Eprint {http://arxiv.org/abs/1103.3526}
  {arXiv:1103.3526 [astro-ph.HE]} \BibitemShut {NoStop}%
\bibitem [{\citenamefont {Vines}\ and\ \citenamefont
  {Flanagan}(2013)}]{Vines:2010ca}%
  \BibitemOpen
  \bibfield  {author} {\bibinfo {author} {\bibfnamefont {J.~E.}\ \bibnamefont
  {Vines}}\ and\ \bibinfo {author} {\bibfnamefont {E.~E.}\ \bibnamefont
  {Flanagan}},\ }\href {\doibase 10.1103/PhysRevD.88.024046} {\bibfield
  {journal} {\bibinfo  {journal} {Phys. Rev.}\ }\textbf {\bibinfo {volume}
  {D88}},\ \bibinfo {pages} {024046} (\bibinfo {year} {2013})},\ \Eprint
  {http://arxiv.org/abs/1009.4919} {arXiv:1009.4919 [gr-qc]} \BibitemShut
  {NoStop}%
\bibitem [{\citenamefont {Lackey}\ \emph {et~al.}(2012)\citenamefont {Lackey},
  \citenamefont {Kyutoku}, \citenamefont {Shibata}, \citenamefont {Brady},\
  and\ \citenamefont {Friedman}}]{Lackey:2011vz}%
  \BibitemOpen
  \bibfield  {author} {\bibinfo {author} {\bibfnamefont {B.~D.}\ \bibnamefont
  {Lackey}}, \bibinfo {author} {\bibfnamefont {K.}~\bibnamefont {Kyutoku}},
  \bibinfo {author} {\bibfnamefont {M.}~\bibnamefont {Shibata}}, \bibinfo
  {author} {\bibfnamefont {P.~R.}\ \bibnamefont {Brady}}, \ and\ \bibinfo
  {author} {\bibfnamefont {J.~L.}\ \bibnamefont {Friedman}},\ }\href {\doibase
  10.1103/PhysRevD.85.044061} {\bibfield  {journal} {\bibinfo  {journal} {Phys.
  Rev.}\ }\textbf {\bibinfo {volume} {D85}},\ \bibinfo {pages} {044061}
  (\bibinfo {year} {2012})},\ \Eprint {http://arxiv.org/abs/1109.3402}
  {arXiv:1109.3402 [astro-ph.HE]} \BibitemShut {NoStop}%
\bibitem [{\citenamefont {Lackey}\ \emph {et~al.}(2014)\citenamefont {Lackey},
  \citenamefont {Kyutoku}, \citenamefont {Shibata}, \citenamefont {Brady},\
  and\ \citenamefont {Friedman}}]{Lackey:2013axa}%
  \BibitemOpen
  \bibfield  {author} {\bibinfo {author} {\bibfnamefont {B.~D.}\ \bibnamefont
  {Lackey}}, \bibinfo {author} {\bibfnamefont {K.}~\bibnamefont {Kyutoku}},
  \bibinfo {author} {\bibfnamefont {M.}~\bibnamefont {Shibata}}, \bibinfo
  {author} {\bibfnamefont {P.~R.}\ \bibnamefont {Brady}}, \ and\ \bibinfo
  {author} {\bibfnamefont {J.~L.}\ \bibnamefont {Friedman}},\ }\href {\doibase
  10.1103/PhysRevD.89.043009} {\bibfield  {journal} {\bibinfo  {journal} {Phys.
  Rev.}\ }\textbf {\bibinfo {volume} {D89}},\ \bibinfo {pages} {043009}
  (\bibinfo {year} {2014})},\ \Eprint {http://arxiv.org/abs/1303.6298}
  {arXiv:1303.6298 [gr-qc]} \BibitemShut {NoStop}%
\bibitem [{\citenamefont {Yagi}\ and\ \citenamefont
  {Yunes}(2014)}]{Yagi:2013baa}%
  \BibitemOpen
  \bibfield  {author} {\bibinfo {author} {\bibfnamefont {K.}~\bibnamefont
  {Yagi}}\ and\ \bibinfo {author} {\bibfnamefont {N.}~\bibnamefont {Yunes}},\
  }\href {\doibase 10.1103/PhysRevD.89.021303} {\bibfield  {journal} {\bibinfo
  {journal} {Phys. Rev.}\ }\textbf {\bibinfo {volume} {D89}},\ \bibinfo {pages}
  {021303} (\bibinfo {year} {2014})},\ \Eprint {http://arxiv.org/abs/1310.8358}
  {arXiv:1310.8358 [gr-qc]} \BibitemShut {NoStop}%
\bibitem [{\citenamefont {Hotokezaka}\ \emph {et~al.}(2016)\citenamefont
  {Hotokezaka}, \citenamefont {Kyutoku}, \citenamefont {Sekiguchi},\ and\
  \citenamefont {Shibata}}]{Hotokezaka:2016bzh}%
  \BibitemOpen
  \bibfield  {author} {\bibinfo {author} {\bibfnamefont {K.}~\bibnamefont
  {Hotokezaka}}, \bibinfo {author} {\bibfnamefont {K.}~\bibnamefont {Kyutoku}},
  \bibinfo {author} {\bibfnamefont {Y.-i.}\ \bibnamefont {Sekiguchi}}, \ and\
  \bibinfo {author} {\bibfnamefont {M.}~\bibnamefont {Shibata}},\ }\href
  {\doibase 10.1103/PhysRevD.93.064082} {\bibfield  {journal} {\bibinfo
  {journal} {Phys. Rev.}\ }\textbf {\bibinfo {volume} {D93}},\ \bibinfo {pages}
  {064082} (\bibinfo {year} {2016})},\ \Eprint
  {http://arxiv.org/abs/1603.01286} {arXiv:1603.01286 [gr-qc]} \BibitemShut
  {NoStop}%
\bibitem [{\citenamefont {Hinderer}\ \emph {et~al.}(2016)\citenamefont
  {Hinderer} \emph {et~al.}}]{Hinderer:2016eia}%
  \BibitemOpen
  \bibfield  {author} {\bibinfo {author} {\bibfnamefont {T.}~\bibnamefont
  {Hinderer}} \emph {et~al.},\ }\href@noop {} {\  (\bibinfo {year} {2016})},\
  \Eprint {http://arxiv.org/abs/1602.00599} {arXiv:1602.00599 [gr-qc]}
  \BibitemShut {NoStop}%
\bibitem [{\citenamefont {Essick}\ \emph {et~al.}(2016)\citenamefont {Essick},
  \citenamefont {Vitale},\ and\ \citenamefont {Weinberg}}]{Essick:2016tkn}%
  \BibitemOpen
  \bibfield  {author} {\bibinfo {author} {\bibfnamefont {R.}~\bibnamefont
  {Essick}}, \bibinfo {author} {\bibfnamefont {S.}~\bibnamefont {Vitale}}, \
  and\ \bibinfo {author} {\bibfnamefont {N.~N.}\ \bibnamefont {Weinberg}},\
  }\href {\doibase 10.1103/PhysRevD.94.103012} {\bibfield  {journal} {\bibinfo
  {journal} {Phys. Rev.}\ }\textbf {\bibinfo {volume} {D94}},\ \bibinfo {pages}
  {103012} (\bibinfo {year} {2016})},\ \Eprint
  {http://arxiv.org/abs/1609.06362} {arXiv:1609.06362 [astro-ph.HE]}
  \BibitemShut {NoStop}%
\bibitem [{\citenamefont {Krishnendu}\ \emph {et~al.}(2017)\citenamefont
  {Krishnendu}, \citenamefont {Arun},\ and\ \citenamefont
  {Mishra}}]{Krishnendu:2017shb}%
  \BibitemOpen
  \bibfield  {author} {\bibinfo {author} {\bibfnamefont {N.~V.}\ \bibnamefont
  {Krishnendu}}, \bibinfo {author} {\bibfnamefont {K.~G.}\ \bibnamefont
  {Arun}}, \ and\ \bibinfo {author} {\bibfnamefont {C.~K.}\ \bibnamefont
  {Mishra}},\ }\href@noop {} {\  (\bibinfo {year} {2017})},\ \Eprint
  {http://arxiv.org/abs/1701.06318} {arXiv:1701.06318 [gr-qc]} \BibitemShut
  {NoStop}%
\bibitem [{\citenamefont {Pappas}(2012)}]{Pappas:2012nt}%
  \BibitemOpen
  \bibfield  {author} {\bibinfo {author} {\bibfnamefont {G.}~\bibnamefont
  {Pappas}},\ }\href {\doibase 10.1111/j.1365-2966.2012.20817.x} {\bibfield
  {journal} {\bibinfo  {journal} {Mon. Not. Roy. Astron. Soc.}\ }\textbf
  {\bibinfo {volume} {422}},\ \bibinfo {pages} {2581} (\bibinfo {year}
  {2012})},\ \Eprint {http://arxiv.org/abs/1201.6071} {arXiv:1201.6071
  [astro-ph.HE]} \BibitemShut {NoStop}%
\bibitem [{\citenamefont {Vallisneri}(2008)}]{Vallisneri:2007ev}%
  \BibitemOpen
  \bibfield  {author} {\bibinfo {author} {\bibfnamefont {M.}~\bibnamefont
  {Vallisneri}},\ }\href {\doibase 10.1103/PhysRevD.77.042001} {\bibfield
  {journal} {\bibinfo  {journal} {Phys. Rev.}\ }\textbf {\bibinfo {volume}
  {D77}},\ \bibinfo {pages} {042001} (\bibinfo {year} {2008})},\ \Eprint
  {http://arxiv.org/abs/gr-qc/0703086} {arXiv:gr-qc/0703086 [GR-QC]}
  \BibitemShut {NoStop}%
\bibitem [{\citenamefont {Shoemaker}(2010)}]{zerodet}%
  \BibitemOpen
  \bibfield  {author} {\bibinfo {author} {\bibfnamefont {D.}~\bibnamefont
  {Shoemaker}} (\bibinfo {collaboration} {LIGO}),\ }\href
  {https://dcc.ligo.org/LIGO-T0900288/public} {\emph {\bibinfo {title}
  {Advanced LIGO anticipated sensitivity curves}}},\ \bibinfo {type} {Tech.
  Rep.}\ \bibinfo {number} {T0900288-v3}\ (\bibinfo {year} {2010})\BibitemShut
  {NoStop}%
\bibitem [{\citenamefont {Yagi}\ and\ \citenamefont
  {Yunes}(2013{\natexlab{a}})}]{Yagi:2013bca}%
  \BibitemOpen
  \bibfield  {author} {\bibinfo {author} {\bibfnamefont {K.}~\bibnamefont
  {Yagi}}\ and\ \bibinfo {author} {\bibfnamefont {N.}~\bibnamefont {Yunes}},\
  }\href {\doibase 10.1126/science.1236462} {\bibfield  {journal} {\bibinfo
  {journal} {Science}\ }\textbf {\bibinfo {volume} {341}},\ \bibinfo {pages}
  {365} (\bibinfo {year} {2013}{\natexlab{a}})},\ \Eprint
  {http://arxiv.org/abs/1302.4499} {arXiv:1302.4499 [gr-qc]} \BibitemShut
  {NoStop}%
\bibitem [{\citenamefont {Yagi}\ and\ \citenamefont
  {Yunes}(2013{\natexlab{b}})}]{Yagi:2013awa}%
  \BibitemOpen
  \bibfield  {author} {\bibinfo {author} {\bibfnamefont {K.}~\bibnamefont
  {Yagi}}\ and\ \bibinfo {author} {\bibfnamefont {N.}~\bibnamefont {Yunes}},\
  }\href {\doibase 10.1103/PhysRevD.88.023009} {\bibfield  {journal} {\bibinfo
  {journal} {Phys. Rev.}\ }\textbf {\bibinfo {volume} {D88}},\ \bibinfo {pages}
  {023009} (\bibinfo {year} {2013}{\natexlab{b}})},\ \Eprint
  {http://arxiv.org/abs/1303.1528} {arXiv:1303.1528 [gr-qc]} \BibitemShut
  {NoStop}%
\bibitem [{\citenamefont {Baubock}\ \emph {et~al.}(2013)\citenamefont
  {Baubock}, \citenamefont {Berti}, \citenamefont {Psaltis},\ and\
  \citenamefont {Ozel}}]{Baubock:2013gna}%
  \BibitemOpen
  \bibfield  {author} {\bibinfo {author} {\bibfnamefont {M.}~\bibnamefont
  {Baubock}}, \bibinfo {author} {\bibfnamefont {E.}~\bibnamefont {Berti}},
  \bibinfo {author} {\bibfnamefont {D.}~\bibnamefont {Psaltis}}, \ and\
  \bibinfo {author} {\bibfnamefont {F.}~\bibnamefont {Ozel}},\ }\href {\doibase
  10.1088/0004-637X/777/1/68} {\bibfield  {journal} {\bibinfo  {journal}
  {Astrophys. J.}\ }\textbf {\bibinfo {volume} {777}},\ \bibinfo {pages} {68}
  (\bibinfo {year} {2013})},\ \Eprint {http://arxiv.org/abs/1306.0569}
  {arXiv:1306.0569 [astro-ph.HE]} \BibitemShut {NoStop}%
\bibitem [{\citenamefont {Psaltis}\ and\ \citenamefont
  {Ozel}(2014)}]{Psaltis:2013zja}%
  \BibitemOpen
  \bibfield  {author} {\bibinfo {author} {\bibfnamefont {D.}~\bibnamefont
  {Psaltis}}\ and\ \bibinfo {author} {\bibfnamefont {F.}~\bibnamefont {Ozel}},\
  }\href {\doibase 10.1088/0004-637X/792/2/87} {\bibfield  {journal} {\bibinfo
  {journal} {Astrophys. J.}\ }\textbf {\bibinfo {volume} {792}},\ \bibinfo
  {pages} {87} (\bibinfo {year} {2014})},\ \Eprint
  {http://arxiv.org/abs/1305.6615} {arXiv:1305.6615 [astro-ph.HE]} \BibitemShut
  {NoStop}%
\bibitem [{\citenamefont {Maselli}\ \emph
  {et~al.}(2013{\natexlab{b}})\citenamefont {Maselli}, \citenamefont {Cardoso},
  \citenamefont {Ferrari}, \citenamefont {Gualtieri},\ and\ \citenamefont
  {Pani}}]{Maselli:2013mva}%
  \BibitemOpen
  \bibfield  {author} {\bibinfo {author} {\bibfnamefont {A.}~\bibnamefont
  {Maselli}}, \bibinfo {author} {\bibfnamefont {V.}~\bibnamefont {Cardoso}},
  \bibinfo {author} {\bibfnamefont {V.}~\bibnamefont {Ferrari}}, \bibinfo
  {author} {\bibfnamefont {L.}~\bibnamefont {Gualtieri}}, \ and\ \bibinfo
  {author} {\bibfnamefont {P.}~\bibnamefont {Pani}},\ }\href {\doibase
  10.1103/PhysRevD.88.023007} {\bibfield  {journal} {\bibinfo  {journal} {Phys.
  Rev.}\ }\textbf {\bibinfo {volume} {D88}},\ \bibinfo {pages} {023007}
  (\bibinfo {year} {2013}{\natexlab{b}})},\ \Eprint
  {http://arxiv.org/abs/1304.2052} {arXiv:1304.2052 [gr-qc]} \BibitemShut
  {NoStop}%
\bibitem [{\citenamefont {Doneva}\ \emph {et~al.}(2013)\citenamefont {Doneva},
  \citenamefont {Yazadjiev}, \citenamefont {Stergioulas},\ and\ \citenamefont
  {Kokkotas}}]{Doneva:2013rha}%
  \BibitemOpen
  \bibfield  {author} {\bibinfo {author} {\bibfnamefont {D.~D.}\ \bibnamefont
  {Doneva}}, \bibinfo {author} {\bibfnamefont {S.~S.}\ \bibnamefont
  {Yazadjiev}}, \bibinfo {author} {\bibfnamefont {N.}~\bibnamefont
  {Stergioulas}}, \ and\ \bibinfo {author} {\bibfnamefont {K.~D.}\ \bibnamefont
  {Kokkotas}},\ }\href {\doibase 10.1088/2041-8205/781/1/L6} {\bibfield
  {journal} {\bibinfo  {journal} {Astrophys. J.}\ }\textbf {\bibinfo {volume}
  {781}},\ \bibinfo {pages} {L6} (\bibinfo {year} {2013})},\ \Eprint
  {http://arxiv.org/abs/1310.7436} {arXiv:1310.7436 [gr-qc]} \BibitemShut
  {NoStop}%
\bibitem [{\citenamefont {Chakrabarti}\ \emph {et~al.}(2014)\citenamefont
  {Chakrabarti}, \citenamefont {Delsate}, \citenamefont {Gurlebeck},\ and\
  \citenamefont {Steinhoff}}]{Chakrabarti:2013tca}%
  \BibitemOpen
  \bibfield  {author} {\bibinfo {author} {\bibfnamefont {S.}~\bibnamefont
  {Chakrabarti}}, \bibinfo {author} {\bibfnamefont {T.}~\bibnamefont
  {Delsate}}, \bibinfo {author} {\bibfnamefont {N.}~\bibnamefont {Gurlebeck}},
  \ and\ \bibinfo {author} {\bibfnamefont {J.}~\bibnamefont {Steinhoff}},\
  }\href {\doibase 10.1103/PhysRevLett.112.201102} {\bibfield  {journal}
  {\bibinfo  {journal} {Phys. Rev. Lett.}\ }\textbf {\bibinfo {volume} {112}},\
  \bibinfo {pages} {201102} (\bibinfo {year} {2014})},\ \Eprint
  {http://arxiv.org/abs/1311.6509} {arXiv:1311.6509 [gr-qc]} \BibitemShut
  {NoStop}%
\bibitem [{\citenamefont {Pappas}\ and\ \citenamefont
  {Apostolatos}(2014)}]{Pappas:2013naa}%
  \BibitemOpen
  \bibfield  {author} {\bibinfo {author} {\bibfnamefont {G.}~\bibnamefont
  {Pappas}}\ and\ \bibinfo {author} {\bibfnamefont {T.~A.}\ \bibnamefont
  {Apostolatos}},\ }\href {\doibase 10.1103/PhysRevLett.112.121101} {\bibfield
  {journal} {\bibinfo  {journal} {Phys. Rev. Lett.}\ }\textbf {\bibinfo
  {volume} {112}},\ \bibinfo {pages} {121101} (\bibinfo {year} {2014})},\
  \Eprint {http://arxiv.org/abs/1311.5508} {arXiv:1311.5508 [gr-qc]}
  \BibitemShut {NoStop}%
\bibitem [{\citenamefont {Stein}\ \emph {et~al.}(2014)\citenamefont {Stein},
  \citenamefont {Yagi},\ and\ \citenamefont {Yunes}}]{Stein:2013ofa}%
  \BibitemOpen
  \bibfield  {author} {\bibinfo {author} {\bibfnamefont {L.~C.}\ \bibnamefont
  {Stein}}, \bibinfo {author} {\bibfnamefont {K.}~\bibnamefont {Yagi}}, \ and\
  \bibinfo {author} {\bibfnamefont {N.}~\bibnamefont {Yunes}},\ }\href
  {\doibase 10.1088/0004-637X/788/1/15} {\bibfield  {journal} {\bibinfo
  {journal} {Astrophys. J.}\ }\textbf {\bibinfo {volume} {788}},\ \bibinfo
  {pages} {15} (\bibinfo {year} {2014})},\ \Eprint
  {http://arxiv.org/abs/1312.4532} {arXiv:1312.4532 [gr-qc]} \BibitemShut
  {NoStop}%
\bibitem [{\citenamefont {Haskell}\ \emph {et~al.}(2014)\citenamefont
  {Haskell}, \citenamefont {Ciolfi}, \citenamefont {Pannarale},\ and\
  \citenamefont {Rezzolla}}]{Haskell:2013vha}%
  \BibitemOpen
  \bibfield  {author} {\bibinfo {author} {\bibfnamefont {B.}~\bibnamefont
  {Haskell}}, \bibinfo {author} {\bibfnamefont {R.}~\bibnamefont {Ciolfi}},
  \bibinfo {author} {\bibfnamefont {F.}~\bibnamefont {Pannarale}}, \ and\
  \bibinfo {author} {\bibfnamefont {L.}~\bibnamefont {Rezzolla}},\ }\href
  {\doibase 10.1093/mnrasl/slt161} {\bibfield  {journal} {\bibinfo  {journal}
  {Mon. Not. Roy. Astron. Soc.}\ }\textbf {\bibinfo {volume} {438}},\ \bibinfo
  {pages} {L71} (\bibinfo {year} {2014})},\ \Eprint
  {http://arxiv.org/abs/13409.3885} {arXiv:13409.3885 [astro-ph.SR]}
  \BibitemShut {NoStop}%
\bibitem [{\citenamefont {Martinon}\ \emph {et~al.}(2014)\citenamefont
  {Martinon}, \citenamefont {Maselli}, \citenamefont {Gualtieri},\ and\
  \citenamefont {Ferrari}}]{PhysRevD.90.064026}%
  \BibitemOpen
  \bibfield  {author} {\bibinfo {author} {\bibfnamefont {G.}~\bibnamefont
  {Martinon}}, \bibinfo {author} {\bibfnamefont {A.}~\bibnamefont {Maselli}},
  \bibinfo {author} {\bibfnamefont {L.}~\bibnamefont {Gualtieri}}, \ and\
  \bibinfo {author} {\bibfnamefont {V.}~\bibnamefont {Ferrari}},\ }\href
  {\doibase 10.1103/PhysRevD.90.064026} {\bibfield  {journal} {\bibinfo
  {journal} {Phys. Rev. D}\ }\textbf {\bibinfo {volume} {90}},\ \bibinfo
  {pages} {064026} (\bibinfo {year} {2014})}\BibitemShut {NoStop}%
\bibitem [{\citenamefont {Maselli}\ and\ \citenamefont
  {Ferrari}(2014)}]{PhysRevD.89.064056}%
  \BibitemOpen
  \bibfield  {author} {\bibinfo {author} {\bibfnamefont {A.}~\bibnamefont
  {Maselli}}\ and\ \bibinfo {author} {\bibfnamefont {V.}~\bibnamefont
  {Ferrari}},\ }\href {\doibase 10.1103/PhysRevD.89.064056} {\bibfield
  {journal} {\bibinfo  {journal} {Phys. Rev. D}\ }\textbf {\bibinfo {volume}
  {89}},\ \bibinfo {pages} {064056} (\bibinfo {year} {2014})}\BibitemShut
  {NoStop}%
\bibitem [{\citenamefont {Yagi}\ \emph
  {et~al.}(2014{\natexlab{a}})\citenamefont {Yagi}, \citenamefont {Kyutoku},
  \citenamefont {Pappas}, \citenamefont {Yunes},\ and\ \citenamefont
  {Apostolatos}}]{Yagi:2014bxa}%
  \BibitemOpen
  \bibfield  {author} {\bibinfo {author} {\bibfnamefont {K.}~\bibnamefont
  {Yagi}}, \bibinfo {author} {\bibfnamefont {K.}~\bibnamefont {Kyutoku}},
  \bibinfo {author} {\bibfnamefont {G.}~\bibnamefont {Pappas}}, \bibinfo
  {author} {\bibfnamefont {N.}~\bibnamefont {Yunes}}, \ and\ \bibinfo {author}
  {\bibfnamefont {T.~A.}\ \bibnamefont {Apostolatos}},\ }\href {\doibase
  10.1103/PhysRevD.89.124013} {\bibfield  {journal} {\bibinfo  {journal} {Phys.
  Rev.}\ }\textbf {\bibinfo {volume} {D89}},\ \bibinfo {pages} {124013}
  (\bibinfo {year} {2014}{\natexlab{a}})},\ \Eprint
  {http://arxiv.org/abs/1403.6243} {arXiv:1403.6243 [gr-qc]} \BibitemShut
  {NoStop}%
\bibitem [{\citenamefont {Chatziioannou}\ \emph {et~al.}(2014)\citenamefont
  {Chatziioannou}, \citenamefont {Yagi},\ and\ \citenamefont
  {Yunes}}]{Chatziioannou:2014tha}%
  \BibitemOpen
  \bibfield  {author} {\bibinfo {author} {\bibfnamefont {K.}~\bibnamefont
  {Chatziioannou}}, \bibinfo {author} {\bibfnamefont {K.}~\bibnamefont {Yagi}},
  \ and\ \bibinfo {author} {\bibfnamefont {N.}~\bibnamefont {Yunes}},\ }\href
  {\doibase 10.1103/PhysRevD.90.064030} {\bibfield  {journal} {\bibinfo
  {journal} {Phys. Rev.}\ }\textbf {\bibinfo {volume} {D90}},\ \bibinfo {pages}
  {064030} (\bibinfo {year} {2014})},\ \Eprint {http://arxiv.org/abs/1406.7135}
  {arXiv:1406.7135 [gr-qc]} \BibitemShut {NoStop}%
\bibitem [{\citenamefont {Yagi}\ and\ \citenamefont
  {Yunes}(2015{\natexlab{a}})}]{Yagi:2015pkc}%
  \BibitemOpen
  \bibfield  {author} {\bibinfo {author} {\bibfnamefont {K.}~\bibnamefont
  {Yagi}}\ and\ \bibinfo {author} {\bibfnamefont {N.}~\bibnamefont {Yunes}},\
  }\href@noop {} {\  (\bibinfo {year} {2015}{\natexlab{a}})},\ \Eprint
  {http://arxiv.org/abs/1512.02639} {arXiv:1512.02639 [gr-qc]} \BibitemShut
  {NoStop}%
\bibitem [{\citenamefont {Majumder}\ \emph {et~al.}(2015)\citenamefont
  {Majumder}, \citenamefont {Yagi},\ and\ \citenamefont
  {Yunes}}]{Majumder:2015kfa}%
  \BibitemOpen
  \bibfield  {author} {\bibinfo {author} {\bibfnamefont {B.}~\bibnamefont
  {Majumder}}, \bibinfo {author} {\bibfnamefont {K.}~\bibnamefont {Yagi}}, \
  and\ \bibinfo {author} {\bibfnamefont {N.}~\bibnamefont {Yunes}},\ }\href
  {\doibase 10.1103/PhysRevD.92.024020} {\bibfield  {journal} {\bibinfo
  {journal} {Phys. Rev.}\ }\textbf {\bibinfo {volume} {D92}},\ \bibinfo {pages}
  {024020} (\bibinfo {year} {2015})},\ \Eprint
  {http://arxiv.org/abs/1504.02506} {arXiv:1504.02506 [gr-qc]} \BibitemShut
  {NoStop}%
\bibitem [{\citenamefont {Sham}\ \emph {et~al.}(2014)\citenamefont {Sham},
  \citenamefont {Lin},\ and\ \citenamefont {Leung}}]{Sham:2013cya}%
  \BibitemOpen
  \bibfield  {author} {\bibinfo {author} {\bibfnamefont {Y.~H.}\ \bibnamefont
  {Sham}}, \bibinfo {author} {\bibfnamefont {L.~M.}\ \bibnamefont {Lin}}, \
  and\ \bibinfo {author} {\bibfnamefont {P.~T.}\ \bibnamefont {Leung}},\ }\href
  {\doibase 10.1088/0004-637X/781/2/66} {\bibfield  {journal} {\bibinfo
  {journal} {Astrophys. J.}\ }\textbf {\bibinfo {volume} {781}},\ \bibinfo
  {pages} {66} (\bibinfo {year} {2014})},\ \Eprint
  {http://arxiv.org/abs/1312.1011} {arXiv:1312.1011 [gr-qc]} \BibitemShut
  {NoStop}%
\bibitem [{\citenamefont {Pani}(2015)}]{Pani:2015tga}%
  \BibitemOpen
  \bibfield  {author} {\bibinfo {author} {\bibfnamefont {P.}~\bibnamefont
  {Pani}},\ }\href {\doibase 10.1103/PhysRevD.92.124030} {\bibfield  {journal}
  {\bibinfo  {journal} {Phys. Rev.}\ }\textbf {\bibinfo {volume} {D92}},\
  \bibinfo {pages} {124030} (\bibinfo {year} {2015})},\ \Eprint
  {http://arxiv.org/abs/1506.06050} {arXiv:1506.06050 [gr-qc]} \BibitemShut
  {NoStop}%
\bibitem [{\citenamefont {Doneva}\ \emph {et~al.}(2014)\citenamefont {Doneva},
  \citenamefont {Yazadjiev}, \citenamefont {Staykov},\ and\ \citenamefont
  {Kokkotas}}]{Doneva:2014faa}%
  \BibitemOpen
  \bibfield  {author} {\bibinfo {author} {\bibfnamefont {D.~D.}\ \bibnamefont
  {Doneva}}, \bibinfo {author} {\bibfnamefont {S.~S.}\ \bibnamefont
  {Yazadjiev}}, \bibinfo {author} {\bibfnamefont {K.~V.}\ \bibnamefont
  {Staykov}}, \ and\ \bibinfo {author} {\bibfnamefont {K.~D.}\ \bibnamefont
  {Kokkotas}},\ }\href {\doibase 10.1103/PhysRevD.90.104021} {\bibfield
  {journal} {\bibinfo  {journal} {Phys. Rev.}\ }\textbf {\bibinfo {volume}
  {D90}},\ \bibinfo {pages} {104021} (\bibinfo {year} {2014})},\ \Eprint
  {http://arxiv.org/abs/1408.1641} {arXiv:1408.1641 [gr-qc]} \BibitemShut
  {NoStop}%
\bibitem [{\citenamefont {Doneva}\ \emph {et~al.}(2015)\citenamefont {Doneva},
  \citenamefont {Yazadjiev},\ and\ \citenamefont {Kokkotas}}]{Doneva:2015hsa}%
  \BibitemOpen
  \bibfield  {author} {\bibinfo {author} {\bibfnamefont {D.~D.}\ \bibnamefont
  {Doneva}}, \bibinfo {author} {\bibfnamefont {S.~S.}\ \bibnamefont
  {Yazadjiev}}, \ and\ \bibinfo {author} {\bibfnamefont {K.~D.}\ \bibnamefont
  {Kokkotas}},\ }\href {\doibase 10.1103/PhysRevD.92.064015} {\bibfield
  {journal} {\bibinfo  {journal} {Phys. Rev.}\ }\textbf {\bibinfo {volume}
  {D92}},\ \bibinfo {pages} {064015} (\bibinfo {year} {2015})},\ \Eprint
  {http://arxiv.org/abs/1507.00378} {arXiv:1507.00378 [gr-qc]} \BibitemShut
  {NoStop}%
\bibitem [{\citenamefont {Hawking}\ and\ \citenamefont
  {Ellis}(1973)}]{Hawking:1973uf}%
  \BibitemOpen
  \bibfield  {author} {\bibinfo {author} {\bibfnamefont {S.~W.}\ \bibnamefont
  {Hawking}}\ and\ \bibinfo {author} {\bibfnamefont {G.~F.~R.}\ \bibnamefont
  {Ellis}},\ }\href@noop {} {\emph {\bibinfo {title} {The Large Scale Structure
  of Space-Time}}},\ Cambridge Monographs on Mathematical Physics\ (\bibinfo
  {publisher} {Cambridge University Press},\ \bibinfo {address} {Cambridge},\
  \bibinfo {year} {1973})\BibitemShut {NoStop}%
\bibitem [{\citenamefont {Yagi}\ \emph
  {et~al.}(2014{\natexlab{b}})\citenamefont {Yagi}, \citenamefont {Stein},
  \citenamefont {Pappas}, \citenamefont {Yunes},\ and\ \citenamefont
  {Apostolatos}}]{PhysRevD.90.063010}%
  \BibitemOpen
  \bibfield  {author} {\bibinfo {author} {\bibfnamefont {K.}~\bibnamefont
  {Yagi}}, \bibinfo {author} {\bibfnamefont {L.~C.}\ \bibnamefont {Stein}},
  \bibinfo {author} {\bibfnamefont {G.}~\bibnamefont {Pappas}}, \bibinfo
  {author} {\bibfnamefont {N.}~\bibnamefont {Yunes}}, \ and\ \bibinfo {author}
  {\bibfnamefont {T.~A.}\ \bibnamefont {Apostolatos}},\ }\href {\doibase
  10.1103/PhysRevD.90.063010} {\bibfield  {journal} {\bibinfo  {journal} {Phys.
  Rev. D}\ }\textbf {\bibinfo {volume} {90}},\ \bibinfo {pages} {063010}
  (\bibinfo {year} {2014}{\natexlab{b}})}\BibitemShut {NoStop}%
\bibitem [{\citenamefont {Yagi}\ and\ \citenamefont
  {Yunes}(2015{\natexlab{b}})}]{Yagi:2015hda}%
  \BibitemOpen
  \bibfield  {author} {\bibinfo {author} {\bibfnamefont {K.}~\bibnamefont
  {Yagi}}\ and\ \bibinfo {author} {\bibfnamefont {N.}~\bibnamefont {Yunes}},\
  }\href {\doibase 10.1103/PhysRevD.91.123008} {\bibfield  {journal} {\bibinfo
  {journal} {Phys. Rev.}\ }\textbf {\bibinfo {volume} {D91}},\ \bibinfo {pages}
  {123008} (\bibinfo {year} {2015}{\natexlab{b}})},\ \Eprint
  {http://arxiv.org/abs/1503.02726} {arXiv:1503.02726 [gr-qc]} \BibitemShut
  {NoStop}%
\bibitem [{\citenamefont {Yagi}\ and\ \citenamefont
  {Yunes}(2016)}]{Yagi:2016ejg}%
  \BibitemOpen
  \bibfield  {author} {\bibinfo {author} {\bibfnamefont {K.}~\bibnamefont
  {Yagi}}\ and\ \bibinfo {author} {\bibfnamefont {N.}~\bibnamefont {Yunes}},\
  }\href@noop {} {\  (\bibinfo {year} {2016})},\ \Eprint
  {http://arxiv.org/abs/1601.02171} {arXiv:1601.02171 [gr-qc]} \BibitemShut
  {NoStop}%
\bibitem [{\citenamefont {Sham}\ \emph {et~al.}(2015)\citenamefont {Sham},
  \citenamefont {Chan}, \citenamefont {Lin},\ and\ \citenamefont
  {Leung}}]{Sham:2014kea}%
  \BibitemOpen
  \bibfield  {author} {\bibinfo {author} {\bibfnamefont {Y.~H.}\ \bibnamefont
  {Sham}}, \bibinfo {author} {\bibfnamefont {T.~K.}\ \bibnamefont {Chan}},
  \bibinfo {author} {\bibfnamefont {L.~M.}\ \bibnamefont {Lin}}, \ and\
  \bibinfo {author} {\bibfnamefont {P.~T.}\ \bibnamefont {Leung}},\ }\href
  {\doibase 10.1088/0004-637X/798/2/121} {\bibfield  {journal} {\bibinfo
  {journal} {Astrophys. J.}\ }\textbf {\bibinfo {volume} {798}},\ \bibinfo
  {pages} {121} (\bibinfo {year} {2015})},\ \Eprint
  {http://arxiv.org/abs/1410.8271} {arXiv:1410.8271 [gr-qc]} \BibitemShut
  {NoStop}%
\bibitem [{\citenamefont {Chan}\ \emph {et~al.}(2016)\citenamefont {Chan},
  \citenamefont {Chan},\ and\ \citenamefont {Leung}}]{Chan:2015iou}%
  \BibitemOpen
  \bibfield  {author} {\bibinfo {author} {\bibfnamefont {T.~K.}\ \bibnamefont
  {Chan}}, \bibinfo {author} {\bibfnamefont {A.~P.~O.}\ \bibnamefont {Chan}}, \
  and\ \bibinfo {author} {\bibfnamefont {P.~T.}\ \bibnamefont {Leung}},\ }\href
  {\doibase 10.1103/PhysRevD.93.024033} {\bibfield  {journal} {\bibinfo
  {journal} {Phys. Rev.}\ }\textbf {\bibinfo {volume} {D93}},\ \bibinfo {pages}
  {024033} (\bibinfo {year} {2016})},\ \Eprint
  {http://arxiv.org/abs/1511.08566} {arXiv:1511.08566 [gr-qc]} \BibitemShut
  {NoStop}%
\end{thebibliography}%
%
\end{document}